\documentclass[
    aip,
    amsmath,
    amssymb,
    reprint,
    nofootinbib
]{revtex4-2}

\usepackage{mathtools}
\usepackage{amsthm}
\usepackage{bm}
\usepackage{mathrsfs}
\usepackage{bbm}
\usepackage{tikz-cd}
\usepackage{quiver}

\newcommand{\R}{\mathbb{R}}
\newcommand{\Z}{\mathbb{Z}}
\newcommand{\calH}{\mathcal{H}}
\newcommand{\calO}{\mathcal{O}}

\newcommand{\calP}{\mathcal{P}}
\newcommand{\PhysProc}{\mathbf{PhysProc}}
\newcommand{\AbsProc}{\mathbf{AbsProc}}
\newcommand{\id}{\mathrm{id}}
\newcommand{\dist}{\operatorname{dist}}

\newtheorem{proposition}{Proposition}
\newtheorem{criterion}{Criterion}
\newtheorem{remark}{Remark}

\usepackage{hyperref}
\hypersetup{
    colorlinks=true,
    linkcolor=blue,
    citecolor=blue,
    urlcolor=blue
}

\usepackage{xcolor}

\makeatletter
\def\@fnsymbol#1{\ensuremath{\ifcase#1\or *\or \dagger\or \ddagger\or
   \mathsection\or \mathparagraph\or \|\or **\or \dagger\dagger
   \or \ddagger\ddagger \else\@ctrerr\fi}}

\makeatother

\begin{document}

\title[When Does Wave Memory Compute?]{Autonomous Physical Computation: A Categorical Closure Criterion for Physical and Neuromorphic Reservoirs}

\author{Nima Dehghani}
\email{nima.dehghani@mit.edu}
\affiliation{ 
McGovern Institute for Brain Research, MIT, Cambridge, MA, USA.
}%
\affiliation{NSF AI Institute for Artificial Intelligence and Fundamental Interactions, MIT, Cambridge, MA, USA.}

\date{\today}

\begin{abstract}
Physical reservoirs, neuromorphic devices, and wave-mediated systems often possess memory, feedback, and rich state-dependent dynamics, but these properties do not by themselves establish autonomous computation. Here we develop a closure criterion for autonomous physical computation, motivated by the wave--particle walker. We formulate the walker as a stroboscopic reservoir with state \(x_n=(\mathbf r_n,\mathbf v_n,\sigma_n,H_n)\), where the wave field \(H_n\) stores an exponentially decaying trace of previous droplet impacts and guides future motion through local slope coupling. This model separates physical writing, storage, reading, feedback, and externally triggered erasure. We then define computation as robust coarse-grained transition preservation: a physical map \(F:X\to X\) implements an abstract transition \(G:A\to A\) only when a coarse-graining \(\Pi:X\to A\) satisfies \(\Pi\circ F=G\circ\Pi\), with abstract states realized by separated physical basins and transitions stable under noise. Autonomous physical computation requires a further closure condition: an internal physical readout state \(y_n\) must select the next physical operation, \(x_{n+1}=F_{C(y_n)}(x_n)\). This criterion classifies the wave--particle walker as a wave-memory machine with genuine Turing-like primitives, but not as a closed autonomous physical computer, because the erasing phase shift is externally imposed. The framework turns this distinction into a design principle: memory becomes autonomous computation when physical readout basins are coupled back to operation selection. 
\footnotetext[2]{\label{note1}Context $\&$ Overview:\\ \url{https://neurovium.science/posts/pblog-Wave-compute}}
\end{abstract}


\maketitle
\section{Introduction}

Time reversal has long provided a sharp diagnostic for the difference between waves and particles. In conservative systems, microscopic dynamical laws may be invariant under reversal of time \cite{RobertsQuispel1992ChaosReversible,Devaney1976Reversible}. Yet in practice, particle trajectories in chaotic regimes rapidly lose reversibility because small errors in initial conditions are amplified \cite{SniederScales1998TimeReversal}. Waves occupy a different position. Through phase conjugation and time-reversal protocols, wave fields can refocus, reconstruct, and in certain settings effectively reverse the propagation of information even in complex media \cite{Zeldovich1985PhaseConjugation,Derode1995TimeReversal,DraegerFink1997ElasticTimeReversal,FinkPrada2001AcousticTimeReversal,Lerosey2004TimeReversalElectromagnetic,Bacot2016TimeReversal}. The contrast between particle irreversibility and wave reversibility becomes especially interesting in systems where a localized object is guided by a wave field that it has generated itself.

Walking droplets on vertically vibrated baths provide one of the cleanest physical examples of such a coupled wave--particle entity \cite{Couder2005WalkingDroplets,Protiere2006ParticleWave,Bush2015PilotWaveHydrodynamics}. A droplet bouncing near the Faraday threshold excites standing surface waves at each impact \cite{BenjaminUrsell1954Faraday,Douady1990Faraday}. Because these waves decay slowly, the present wave field contains a finite-memory trace of previous impact positions \cite{Eddi2011InformationStored}. The droplet is then propelled by the local slope of this self-generated wave field. The resulting object is neither a passive particle nor an externally prescribed wave packet, but a recurrent physical system in which the droplet writes to an extended wave reservoir, later reads that reservoir through the local slope, and is guided by the memory it has written.

In an elegant experiment, Perrard, Fort, and Couder used this system to demonstrate a striking form of memory-mediated reversal \cite{Perrard2016WaveTuring}. By imposing a controlled \(\pi\)-shift between the droplet bounce and the preexisting wave field, they reversed the effective horizontal wave-induced kick. In confined chaotic regimes, the droplet did not merely reverse its instantaneous velocity; it transiently retraced its previous complex trajectory \cite{Perrard2014Eigenstates,Perrard2014Chaos}. During this backtracking, newly emitted waves were written with opposite phase relative to the old field, destructively interfering with the stored memory and thereby erasing it. The walker could thus read its own wave memory backward while erasing that memory. This experiment motivates the description of the system in terms of writing, storing, reading, and erasing operations, with the wave field acting as a global information repository.

The question addressed here is what follows from this observation. \emph{Does a physical system that writes, stores, reads, and erases information thereby compute?} More specifically, \textit{when does a wave-memory system become an autonomous physical computer rather than a dynamical system to which an external observer assigns a computational interpretation?} This question is not semantic. It is a physical and mathematical question about state spaces, memory variables, robust readout, transition structure, and internal control.

The distinction matters because physical memory is not yet computation. A dissipative system may contain a recoverable trace of its past without implementing an abstract transition rule. A continuous dynamical system may admit many observer-defined labels without realizing a stable symbolic process. A reservoir may support rich state-dependent responses while still relying on an external readout to extract useful outputs. Conversely, engineered and biological physical computers do not require a one-to-one mapping between microscopic physical states and abstract symbols. What they require is a physically robust organization of state space such that coarse-grained states are distinguishable, transitions among them are preserved, and the relevant readout states causally affect subsequent physical evolution \cite{DehghaniCaterina2024PhysicalComputing, Dehghani2024hilbert}.

The motivation for distinguishing physical memory from autonomous computation is not limited to hydrodynamic walkers. Traveling waves are also a recurrent motif in cortical dynamics, where they have been observed and modeled across sensory, motor, sleep, and cognitive contexts. Their proposed roles range from coordinating excitability and spike timing to integrating spatial information over time and structuring neural representations \cite{ErmentroutKleinfeld2001wave,LeVanQuyen2016highfrq,Muller2018wave,KellerWelling2023wavemachine}. This raises a closely related question: when should a cortical traveling wave be regarded as part of a computation, rather than as a dynamical correlate or carrier of activity? The framework developed here is meant to make that question sharper. It does not identify cortical waves with hydrodynamic walkers. 
Instead, it asks whether wave-like activity participates in robust coarse-grained transitions and whether its readout is physically closed through downstream neural dynamics.

These questions---about the walker and about wave-mediated computation more broadly---motivate the framework developed below. We make the tripartite distinction precise in Sec.~\ref{sec:overview}, summarize the resulting closure criterion and its consequences for the walker, and then develop each ingredient in turn.

\section{Overview of the framework and main results}
\label{sec:overview}

We separate three levels that are often conflated:


\[
\begin{array}{l}
\text{physical memory} \\
\rightarrow\; 
\text{transition-preserving physical computation} \\
\rightarrow\; 
\text{closed autonomous physical computation}.
\end{array}
\]

Physical memory means that a physical state encodes information about previous states. Transition-preserving computation means that there exists a coarse-graining \(\Pi:X\to A\) from physical states to abstract states such that physical evolution descends to an abstract transition rule,

\[
\Pi\circ F = G\circ \Pi .
\]

Closed autonomous physical computation requires still more: the system must contain a physically realized readout or control state whose value selects the next physical operation. In this case the physical evolution has the form

\[
x_{n+1}=F_{C(y_n)}(x_n),
\]

where \(y_n\) is an internal physical readout state and \(C\) maps that readout state to a physical operation. The readout is not merely observed; it is coupled back into the dynamics.

The closure criterion is developed using the walking-droplet system as a motivating test case. It is deliberately formulated first in the language of dynamical systems and physical basins, and only afterward lifted into category-theoretic language. This ordering is important. The physics supplies the state space, memory field, readout variable, and physical operations. The categorical formulation then clarifies the compositional structure: physical processes form a category, abstract transitions form another, and computation corresponds to a coarse-grained functorial relation between them (Sec.~\ref{sec:CAT}). Autonomy appears as internal morphism selection, where the next physical process is selected by a state inside the physical system.

The analysis has three main contributions. First, we give a physics-level decomposition of the wave--particle walker into a stroboscopic reservoir model (Sec.~\ref{sec:model}). The physical state combines the droplet's position, velocity, and bouncing phase with a wave-memory field that stores an exponentially decaying trace of previous impacts and guides the droplet through its local slope. This makes explicit how writing, finite-time storage, local reading, feedback, and externally triggered erasure all arise from a single physical update rule.

Second, we define computation as robust coarse-grained transition preservation (Secs.~\ref{sec:coarse-grain} and \ref{sec:physSymb}). A physical system implements an abstract transition only when its dynamics respects the equivalence classes induced by the coarse-graining \(\Pi\). For finite-state abstractions this is realized by separated physical basins whose transitions remain stable under noise (Sec.~\ref{sec:physSymb}), which replaces both arbitrary observer labeling and overly strong bijective physical-symbol mappings with a physically meaningful notion of robust symbolization.

Third, we introduce the closure criterion for autonomous physical computation (Sec.~\ref{sec:closure}). A system may possess memory and even externally controlled computational primitives without being autonomous. Autonomy requires an internal physical readout-control layer, in which a dedicated readout subsystem maps the memory state to an operation that is then applied back to the system, so that the next operation is selected by the system's own physical state. Applying this criterion to the Perrard--Fort--Couder walker (Sec.~\ref{sec:classification}), we conclude that the walker is a wave-memory physical machine with genuine Turing-like primitives: writing, storage, reading, feedback, finite-time reversal, and erasure. However, as experimentally described, it is not a closed autonomous physical computer, because the \(\pi\)-shift that triggers erasure is externally imposed rather than selected by an internal physical readout state.

This classification is intended as a sharpening rather than a rejection of the wave-based computing interpretation. The walker demonstrates that memory, reversal, and erasure can be physically embodied in a continuous wave-mediated system. The closure criterion identifies the missing architectural ingredient needed to turn such a system into an autonomous physical computer: robust physical readout states that select subsequent operations. This leads naturally to constructive extensions (Sec.~\ref{sec:construct}), including multistable readout traps, thresholded wave detectors coupled to phase control, boundary-controlled Faraday reservoirs, and coupled-walker controllers.

The closure criterion applies well beyond walking droplets. Reservoir computing, neuromorphic hardware, and biological recurrent circuits all rely on high-dimensional physical or dynamical states whose useful computational content is extracted by a readout, and in each case one may ask whether that readout is merely external or physically internalized. In this sense the wave-memory walker is not only a hydrodynamic curiosity but a concrete model system for a broader question at the foundations of physical and neuromorphic computation: \emph{when does memory in matter become autonomous computation?}

\section{Physical model: wave memory as a stroboscopic reservoir}
\label{sec:model}

We begin with a physical system whose natural clock is supplied by the Faraday period. A droplet bouncing on a vertically vibrated bath generates standing Faraday waves at each impact. Because the bath is driven close to the Faraday instability threshold, these waves decay slowly and form a spatially distributed memory of previous impacts. The goal of this section is not to reproduce the full hydrodynamic model of the walking droplet, but to extract the minimal dynamical structure needed for analyzing physical memory, readout, erasure, and computation.

Let

\[
t_n=nT_F,
\qquad
n\in\Z_{\geq 0},
\]

denote the stroboscopic impact times, where \(T_F\) is the Faraday period. At each impact, define the physical state

\[
x_n=
(\mathbf r_n,\mathbf v_n,\sigma_n,H_n)\in X.
\]

Here

\[
\mathbf r_n\in\Omega\subset\R^2
\]

is the horizontal droplet position,

\[
\mathbf v_n\in\R^2
\]

is the horizontal velocity,

\[
\sigma_n\in\{+1,-1\}
\]

is the bouncing phase relative to a fixed stroboscopic Faraday phase, and

\[
H_n\in\calH
\]

is the surface-wave memory field over the bath domain \(\Omega\). We take \(\calH\) to be a function space such as \(H^1(\Omega)\), since the droplet responds to the local slope of the wave field.

The physical state space is therefore

\[
X=\Omega\times\R^2\times\{+1,-1\}\times\calH.
\]

At each bounce, the droplet writes a localized Faraday-wave source into the bath. In the Bessel approximation \cite{Eddi2011InformationStored,Oza2013TrajectoryEquation}, the source centered at position \(\mathbf r\) is

\[
\psi_{\mathbf r}(\rho)
=
h_0J_0(k_F|\rho-\mathbf r|),
\]

where \(\rho\in\Omega\), \(h_0\) is the single-impact amplitude, \(J_0\) is the zeroth-order Bessel function, and

\[
k_F=\frac{2\pi}{\lambda_F}
\]

is the Faraday wavenumber. If boundary effects are important, \(\psi_{\mathbf r}\) may be replaced by a domain-dependent response kernel \(h_0G_\Omega(\rho,\mathbf r)\). Nothing in the following depends on the exact Green's function.

The wave-memory update is

\[
H_{n+1}(\rho)
=
\lambda H_n(\rho)
+
\sigma_n\psi_{\mathbf r_n}(\rho),
\tag{1}
\]

with

\[
\lambda=e^{-1/M_e},
\qquad
M_e=\frac{\tau}{T_F}.
\]

The parameter \(M_e\) is the dimensionless memory parameter; it measures the wave decay time \(\tau\) in units of the Faraday period. Iterating Eq.~(1) gives

\[
H_n
=
\sum_{q=1}^{n}
\lambda^{q-1}
\sigma_{n-q}
\psi_{\mathbf r_{n-q}},
\tag{2}
\]

assuming the initial field has decayed. Thus \(H_n\) is an exponentially weighted trace of previous droplet impacts.

The droplet does not read the entire wave field. It samples the local slope at its impact position:

\[
\mathbf g_n
=
\nabla H_n(\mathbf r_n).
\tag{3}
\]



The wave-mediated horizontal force is proportional to this slope, sampled at the droplet's own bouncing phase,

\[
\mathbf F^{\mathrm{wave}}_n
=
-C\,\sigma_n\nabla H_n(\mathbf r_n).
\]

The factor \(\sigma_n\) encodes the phase relation between the bounce and the stored field. In the phase-locked regime it is inert, since \(\sigma_n\) squares against the \(\sigma\) factors carried by \(H_n\); its role is to make phase interventions act on the force law, as in Sec.~\ref{sec:erasure}.

A minimal stroboscopic impact map is therefore


\[
\mathbf v_{n+1}
=
a\mathbf v_n
-
\kappa\,\sigma_n\nabla H_n(\mathbf r_n)
-
\chi\nabla V(\mathbf r_n)
+
\boldsymbol\xi_n,
\tag{4}
\]

\[
\mathbf r_{n+1}
=
\mathbf r_n+T_F\mathbf v_{n+1}.
\tag{5}
\]

Here \(a\in[0,1]\) is a horizontal persistence parameter, \(\kappa>0\) is the effective wave-kick strength, \(V\) is an external confinement potential, \(\chi\) sets the confinement strength, and \(\boldsymbol\xi_n\) represents unresolved fluctuations or noise. Equations~(4)--(5) are a discrete-time reduction of the stroboscopic trajectory equation
of Oza, Rosales, and Bush \cite{Oza2013TrajectoryEquation}, retaining the wave-memory
kernel and the local-slope coupling while suppressing the within-bounce hydrodynamics.

Equations~(1)--(5) define the stroboscopic physical evolution

\[
x_{n+1}=F(x_n).
\tag{6}
\]

This map has the form of a physical reservoir system \cite{Jaeger2001EchoState,JaegerHaas2004esn,Maass2002LiquidState,Tanaka2019PhysicalReservoir}. The extended field \(H_n\) is the memory reservoir; the droplet is a localized probe and actuator; the readout \(\nabla H_n(\mathbf r_n)\) feeds back into future trajectory and future writing.

\section{Physical memory and erasure}
\label{sec:erasure}

The wave field stores a finite-lifetime trace of the recent path. We assume the bounce phase is constant over the memory window, $\sigma_{n-q}=\sigma$ for $1\leq q\leq L$, as holds for the unperturbed walker, which remains phase-locked; the stored field is then determined by the trajectory positions alone, with $\sigma$ a fixed overall sign. For a trajectory segment

\[
\mathcal T_n^{(L)}
=
(\mathbf r_{n-L},\ldots,\mathbf r_{n-1})\in\Omega^L,
\]

define the finite-memory writing map

\[
\mathcal W_L:\Omega^L\to\calH
\]

by

\[
\mathcal W_L(\mathcal T_n^{(L)})
=
\sum_{q=1}^{L}
\lambda^{q-1}
\sigma_{n-q}
\psi_{\mathbf r_{n-q}}.
\tag{7}
\]

The residual contribution from impacts older than \(L\) is bounded, for \(\|\psi_{\mathbf r}\|_{\calH}\leq \Psi_0\), by

\[
\|H_n-H_n^{(L)}\|_{\calH}
\leq
\Psi_0\frac{\lambda^L}{1-\lambda}.
\tag{8}
\]

Thus, at fixed field resolution, the effective memory horizon scales as

\[
L_{\mathrm{eff}}\sim M_e.
\tag{9}
\]

The stored field is dynamically readable because Eq.~(3) enters Eq.~(4). The memory is not merely an observer's reconstruction of the past; it is a physical variable that affects the future trajectory.

To quantify erasure, define the normalized wave energy

\[
E_W[n]
=
\frac{\|H_n\|_{L^2(\Omega)}^2}
{\|\psi_{\mathbf 0}\|_{L^2(\Omega)}^2}.
\tag{10}
\]

Consider a phase-shift operation \(P_\pi\) at bounce \(n_\pi\),

\[
P_\pi:
(\mathbf r,\mathbf v,\sigma,H)
\mapsto
(\mathbf r,\mathbf v,-\sigma,H).
\tag{11}
\]

The stored field is not instantaneously removed. Rather, subsequent impacts write sources of opposite phase relative to the preexisting field. For \(m\geq0\), decompose

\[
H_{n_\pi+m}
=
H_{\mathrm{old}}^{(m)}
+
H_{\mathrm{new}}^{(m)},
\]

where

\[
H_{\mathrm{old}}^{(m)}
=
\lambda^mH_{n_\pi},
\tag{12}
\]

and

\[
H_{\mathrm{new}}^{(m)}
=
-\sigma_+
\sum_{\ell=0}^{m-1}
\lambda^{m-1-\ell}
\psi_{\mathbf r_{n_\pi+\ell}}.
\tag{13}
\]

Wave erasure occurs when phase-opposed writing reduces the wave energy beyond passive decay:

\[
\|H_{\mathrm{old}}^{(m)}+H_{\mathrm{new}}^{(m)}\|_{L^2}^2
<
\|H_{\mathrm{old}}^{(m)}\|_{L^2}^2.
\tag{14}
\]

Equivalently,

\[
2\langle H_{\mathrm{old}}^{(m)},H_{\mathrm{new}}^{(m)}\rangle
+
\|H_{\mathrm{new}}^{(m)}\|^2
<0.
\tag{15}
\]

This inequality gives a physical definition of erasure: the new wave sources destructively interfere with the old memory field.



Because \(P_\pi\) reverses \(\sigma\) while leaving the stored field unchanged, the effective kick from the preexisting field reverses:

\[
-C\,\sigma_+\nabla H(\mathbf r)
\mapsto
+C\,\sigma_+\nabla H(\mathbf r).
\tag{16}
\]

The result is finite-time backtracking, not global time-reversal invariance. If

\[
d_{\mathrm{rev}}(m)
=
|\mathbf r_{n_\pi+m}-\mathbf r_{n_\pi-m}|,
\]

then memory-mediated reversal holds over a horizon \(m_\ast\) when

\[
d_{\mathrm{rev}}(m)<\varepsilon_{\mathrm{rev}},
\qquad
1\leq m\leq m_\ast.
\tag{17}
\]

For the walker, \(m_\ast\) is controlled by the memory lifetime and is of order \(M_e/2\). The system reads its wave memory backward while erasing it.

\section{Computation as coarse-grained transition preservation}
\label{sec:coarse-grain}

Physical memory is not yet computation. A system may store recoverable information about its past without implementing an abstract transition rule. We therefore introduce a first criterion for physical computation.
This criterion has a precise lineage. Abstraction/representation (AR) theory frames
computation as a commuting relation between physical dynamics and an evolving abstract
representation \cite{Horsman2014AR,Horsman2018AR}. AR deliberately places no semantic restriction on representations, which
protects it from mind‑centric accounts but leaves it vulnerable to the simple‑mapping
objection and pancomputationalist readings \cite{Putnam1960MindsMachines,Putnam1988RepresentationReality,Chalmers1996RockFSA}. 
The mechanistic account of physical computation meets this by demanding robust, functionally individuated structure
rather than arbitrary labeling
\cite{Piccinini2007ComputingMechanisms,PiccininiScarantino2010ComputationInformation,Piccinini2025PhysicalComputation}.
However, it too falls short of a concrete tie with physical systems and their dynamics and thus remains descriptive. 

We developed a categorical formulation of computation in order to address these specific
limitations of AR and the mechanistic account \cite{DehghaniCaterina2024PhysicalComputing}. Categorical formulation casts
that relation as a functor between a category of physical processes and one of abstract
processes \cite{DehghaniCaterina2024PhysicalComputing}. Here, we further specialize and extend that framework (see Sec.~\ref{sec:CAT}) to the present physical substrate,
imposing dynamical constraints (transition‑preservation and robust symbolization that rule 
out arbitrary labelings and give a physically meaningful criterion for when a
wave‑memory system should be regarded as computing rather than merely storing information).

The compute cycle becomes a compositional theory once physical systems and processes
are organized into a category $\PhysProc$, abstract data and programs into $\AbsProc$,
and computation into a functor pair $(\mathcal{R}_T,\widetilde{\mathcal{R}}_T)$ relating
them \cite{DehghaniCaterina2024PhysicalComputing}. Here we put that machinery to work on
a concrete substrate and push it one step further. Instantiating $\PhysProc$ on the
explicit state space of a wave-memory walker exposes a question the AR theory
never had to confront: at each step, what chooses the morphism? In AR theory and its
functorial form alike, the operation is handed to the diagram from outside. Closure (see Sec.~\ref{sec:CAT}) is
the condition that the system choose it---that the next physical process be selected by
the system's own readout state. This is the structural content of autonomy, and it is
what separates a wave field that merely remembers from one that computes.

The condition that separates
computation from arbitrary relabeling---and blocks the conclusion that any system
computes anything one can map onto it \cite{Putnam1960MindsMachines}---is that the
correspondence be preserved under the dynamics, which in a coarse-grained physical
setting is exactly Eq.~(19) below. That a coarse-graining of a dynamical system carries
intrinsic computational structure precisely when its states have well-defined transitions
is also the organizing idea of computational mechanics
\cite{ShaliziCrutchfield2001ComputationalMechanics,Crutchfield2012OrderChaos}.

Let \(X\) be a physical state space and

\[
F:X\to X
\]

a physical transition map. Let \(A\) be an abstract state space and

\[
G:A\to A
\]

an abstract transition map. A coarse-graining is a map

\[
\Pi:X\to A.
\]

The physical system implements \(G\) on a domain \(U\subseteq X\) when

\[
\Pi(F(x))=G(\Pi(x)),
\qquad x\in U.
\tag{18}
\]

Equivalently,

\[
\Pi\circ F=G\circ \Pi
\qquad
\text{on }U.
\tag{19}
\]

This is the commuting diagram

\[
\begin{tikzcd}
X \arrow[r,"F"] \arrow[d,"\Pi"'] & X \arrow[d,"\Pi"] \\
A \arrow[r,"G"'] & A .
\end{tikzcd}
\tag{20}
\]

The map \(\Pi\) is generally many-to-one. It induces an equivalence relation

\[
x\sim_\Pi x'
\quad\Longleftrightarrow\quad
\Pi(x)=\Pi(x').
\]

The physical transition \(F\) descends to a well-defined abstract transition on the quotient \(X/{\sim_\Pi}\) only if

\[
\Pi(x)=\Pi(x')
\quad\Longrightarrow\quad
\Pi(F(x))=\Pi(F(x')).
\tag{21}
\]

\begin{proposition}
Let \(F:X\to X\) be a physical transition map and \(\Pi:X\to A\) a surjective coarse-graining. If Eq.~(21) holds on \(U\subseteq X\), then there exists a unique map \(G:\Pi(U)\to A\) such that Eq.~(18) holds for all \(x\in U\).
\end{proposition}

This proposition is the formal replacement for one-to-one physical-symbol mapping. Abstract states are not microstates. They are equivalence classes, macrostates, or physical basins. Computation requires preservation of transition structure under coarse-graining.

\section{Robust physical symbolization}
\label{sec:physSymb}

Exact commutation is an idealization. Real physical systems are noisy, finite-resolution, and subject to perturbations. We therefore require physical symbols to be realized as separated basins.

Let

\[
A=\{a_1,\ldots,a_K\}
\]

be a finite abstract state space. A physical realization of \(A\) is a collection of regions

\[
\mathcal B=\{B_1,\ldots,B_K\},
\qquad
B_i\subset X,
\]

where \(B_i\) realizes \(a_i\). The coarse-graining is

\[
\Pi(x)=a_i
\qquad
\text{if }
x\in B_i.
\]

Let \(d_X\) be a physically meaningful metric on \(X\), and let \(\varepsilon_X\) be the uncertainty scale. Robust distinguishability requires

\[
\dist(B_i,B_j)>2\varepsilon_X,
\qquad
i\neq j.
\tag{22}
\]

If \(G(a_i)=a_{g(i)}\), deterministic robust transition preservation requires

\[
F(B_i)\subseteq B_{g(i)}.
\tag{23}
\]

At finite resolution,

\[
F(N_{\varepsilon_X}(B_i))
\subseteq
N_{\varepsilon_X}(B_{g(i)}).
\tag{24}
\]

For a stochastic transition kernel \(K(x,dy)\), the corresponding condition is

\[
K(x,B_{g(i)})\geq 1-\delta,
\qquad
x\in N_{\varepsilon_X}(B_i).
\tag{25}
\]

This defines a robust physical implementation of the abstract transition map.

The main failure modes are:

\[
\begin{aligned}
\text{aliasing}, \\
\text{representative dependence}, \\
\text{noise-induced basin crossing}, \\
\text{memory decay}, \\
\text{transition deformation}.
\end{aligned}
\]

Aliasing occurs when distinct intended abstract states cannot be physically separated:

\[
N_{\varepsilon_X}(B_i)\cap N_{\varepsilon_X}(B_j)\neq\varnothing.
\tag{26}
\]

Representative dependence occurs when two states in the same abstract basin evolve to different abstract basins:

\[
\Pi(x)=\Pi(x')
\quad
\text{but}
\quad
\Pi(F(x))\neq\Pi(F(x')).
\tag{27}
\]

Memory decay is especially relevant for wave reservoirs. If two wave-memory states have initial separation \(D_0\), passive decay gives

\[
D_m\approx \lambda^mD_0.
\]

They become indistinguishable when

\[
\lambda^mD_0\leq 2\varepsilon_H,
\]

or

\[
m\geq
M_e\log\left(\frac{D_0}{2\varepsilon_H}\right).
\tag{28}
\]

Thus the memory parameter \(M_e\) bounds the reliable lifetime of memory-dependent abstract distinctions.

\section{Closure and autonomy}
\label{sec:closure}

The criteria above define externally controlled or externally interpreted physical computation. The central step in this paper is the closure criterion.

Let \(\mathcal O\) be a set of physically realizable operations, with

\[
F_o:X\to X,
\qquad
o\in\mathcal O.
\]

If an external agent supplies the operation sequence \(o_0,o_1,\ldots\), then

\[
x_{n+1}=F_{o_n}(x_n).
\tag{29}
\]

This may implement a valid abstract transition structure, but the operation selection is external.

A closed physical computer must contain a physical readout/control state. Decompose

\[
X=X_R\times X_Y\times X_Z,
\]

where \(X_R\) is the reservoir or memory subsystem, \(X_Y\) is the physical readout-control subsystem, and \(X_Z\) contains the remaining degrees of freedom. Let

\[
p_Y:X\to X_Y
\]

be the projection onto the readout subsystem, and define

\[
y_n=p_Y(x_n).
\]

A physical operation-selection map is

\[
C:X_Y\to\mathcal O.
\]

The closed-loop physical evolution is

\[
x_{n+1}
=
F_{C(y_n)}(x_n)
=
F_{C(p_Y(x_n))}(x_n).
\tag{30}
\]

Define

\[
\Phi(x)=F_{C(p_Y(x))}(x).
\tag{31}
\]

Then the system implements a closed physical computation when

\[
\Pi(\Phi(x))=G(\Pi(x))
\tag{32}
\]

for a fixed abstract transition \(G\), or more generally,

\[
\Pi(F_{C(p_Y(x))}(x))
=
G_{\Theta(C(p_Y(x)))}(\Pi(x)),
\tag{33}
\]

where \(\Theta:\mathcal O\to\mathcal U\) maps physical operations to abstract operations.

\begin{criterion}[Closure]
A physical system satisfies autonomous closure when the operation applied at step \(n\) is selected by an internal physical readout state \(y_n\), rather than supplied externally:
\[
o_n=C(y_n),
\qquad
y_n=p_Y(x_n).
\]
\end{criterion}

\begin{remark}
Closure is a condition on \emph{autonomy}, not on computational power. A system can
satisfy the closure criterion and still compute nothing of interest: a two-state
controller that flips \(o_{\mathrm{normal}}\leftrightarrow o_\pi\) at every step is
closed, transition-preserving, and autonomous, yet realizes only a trivial period-two
cycle. Closure asserts that the system, rather than an external agent, selects its
operations; it does not assert that the selected abstract process is rich. Computational
nontriviality is a separate requirement on \(\mathcal O\), \(A\), and \(G\)---a sufficient
alphabet, programmable transitions, and composable operations---taken up in the
Limitations. Throughout, closure marks the boundary between externally driven and
autonomous physical computation, not between weak and powerful computation.
\end{remark}

This criterion distinguishes ordinary physical feedback from autonomous physical computation. The walker already has the feedback loop

\[
H_n
\to
\nabla H_n(\mathbf r_n)
\to
\mathbf v_{n+1}
\to
H_{n+1}.
\]

But feedback alone does not imply operation selection. Closure requires a physical readout state that selects among distinct operations, such as normal propagation, phase flip, erasure, branching, or confinement change.

\section{Categorical formulation}
\label{sec:CAT}
The preceding structure can be expressed categorically. 
We introduced the $\PhysProc$/$\AbsProc$ framework in Ref.~\cite{DehghaniCaterina2024PhysicalComputing},
drawing on ART's representation diagrams \cite{Horsman2014AR} and on categorical
treatments of information such as \cite{CoeckePaquette2010Categories}; here we specialize and extend that
framework to the present substrate in order to supply the functorial structure required to make closure explicit.

In that formulation,

\[
\PhysProc
\]

is a category whose objects are physical regions \(U,V\subseteq X\), and whose morphisms are physically realizable processes

\[
f:U\to V.
\]

Composition in $PhysProc$ is sequential composition of physical processes.

Meanwhile 

\[
\AbsProc
\]

is a category whose objects are abstract state spaces or abstract regions, and whose morphisms are abstract transitions.

Physical computation is expressed by a functor pair 
$(\mathcal{R}_T,\widetilde{\mathcal{R}}_T)$ between the $PhysProc$ and $AbsProc$.

A coarse-graining

\[
\Pi:X\to A
\]

plays the role of the representation functor $\mathcal{R}_T$ and maps physical regions to abstract regions:

\[
U\mapsto \Pi(U).
\]

Given a physical morphism \(f:U\to V\), ``coarse-graining'' induces an abstract morphism only if it respects \(\Pi\)-equivalence:

\[
\Pi(x)=\Pi(x')
\Longrightarrow
\Pi(f(x))=\Pi(f(x')).
\tag{34}
\]

When this holds, define

\[
Q_\Pi(f)(\Pi(x))=\Pi(f(x)).
\]

Thus \(Q_\Pi\) is a quotient functor on the subcategory \(\PhysProc_\Pi\) of processes that preserve the coarse-graining:

\[
Q_\Pi:\PhysProc_\Pi\to\AbsProc.
\tag{35}
\]

The condition

\[
Q_\Pi(F)=G
\]

is the categorical form of

\[
\Pi\circ F=G\circ\Pi.
\]

The coarse-graining also induces abstraction and concretization maps between regions:

\[
\alpha:\calP(X)\to\calP(A),
\qquad
\alpha(U)=\Pi(U),
\]

\[
\gamma:\calP(A)\to\calP(X),
\qquad
\gamma(S)=\Pi^{-1}(S).
\]

These form a Galois-style relation:

\[
\alpha(U)\subseteq S
\Longleftrightarrow
U\subseteq\gamma(S).
\tag{36}
\]

For a single abstract state \(a\),

\[
\gamma(\{a\})=B_a
\]

is its physical realization basin. Transition preservation becomes

\[
F(\gamma(S))\subseteq \gamma(G(S)),
\tag{37}
\]

or for a single state,

\[
F(B_a)\subseteq B_{G(a)}.
\tag{38}
\]
In realistic, noisy systems, this strict functoriality relaxes to an approximate quotient functor \(Q_\Pi(F) \approx_{\varepsilon,\delta} G\), where \(\varepsilon\) captures finite physical resolution and \(\delta\) bounds the stochastic transition failure probability (see Supplementary Section S6). Closure internalizes morphism selection. The closed-loop physical morphism is

\[
\Phi
=
\mathrm{Act}\circ
\langle C\circ p_Y,\id_X\rangle,
\tag{39}
\]

where

\[
\mathrm{Act}(o,x)=F_o(x).
\]

This is precisely the ingredient left open by the functorial formulation of Ref.~\cite{DehghaniCaterina2024PhysicalComputing}: there the representation functor
relates physical and abstract \emph{processes}, but which process is applied remains external to the diagram. Closure does not merely fill that gap; it changes the structure of the diagram itself by eliminating the external index. The index $o\in\mathcal{O}$ becomes a value of the system's own readout, $o = C(p_Y(x))$, so that the choice of morphism is generated internally rather than supplied from outside. Coalgebraically, $\Phi$ equips $X$ with the structure of a deterministic transition system whose next-step map is determined by an internal observation; autonomy is the statement that this structure map factors through the system's own state rather than through an external driver. Operation selection is therefore no longer an externally chosen label. It is a morphism produced by the physical state of the system itself.

The categorical compatibility condition is

\[
\Pi(\Phi(x))
=
G_{\Theta(C(p_Y(x)))}(\Pi(x)).
\tag{40}
\]

This is the categorical closure criterion for autonomous physical computation.

\section{Classification of the wave--particle walker}
\label{sec:classification}

We now classify the wave--particle walker using the preceding hierarchy.

First, the walker has physical memory. Its wave field stores an exponentially weighted trace of recent impacts:

\[
H_n=
\sum_{q=1}^{n}
\lambda^{q-1}
\sigma_{n-q}
\psi_{\mathbf r_{n-q}}.
\tag{41}
\]

Second, the walker writes into memory. Each droplet impact changes the bath state:

\[
H_{n+1}=\lambda H_n+\sigma_n\psi_{\mathbf r_n}.
\]

Third, it stores this memory for a finite time controlled by \(M_e\). Fourth, it physically reads the memory through

\[
\nabla H_n(\mathbf r_n),
\]

which affects the next velocity. Fifth, it has a closed analog feedback loop: the droplet writes the wave field, the wave field guides the droplet, and the guided droplet writes the future wave field.

Sixth, the walker realizes erasure when a \(\pi\)-shift is externally imposed. The phase shift reverses the sign of subsequently written sources, producing destructive interference with the old field:

\[
\|H_{\mathrm{old}}^{(m)}+H_{\mathrm{new}}^{(m)}\|^2
<
\|H_{\mathrm{old}}^{(m)}\|^2.
\tag{42}
\]

Seventh, this same process supports finite-time backtracking: the droplet reads its stored wave memory backward while erasing it.

These are strong physical information-processing primitives. However, they do not by themselves establish autonomous computation.

To claim transition-preserving computation, one must specify an abstract state space \(A\), a coarse-graining \(\Pi:X\to A\), an abstract transition \(G:A\to A\), and a domain \(U\subseteq X\) such that

\[
\Pi(F(x))=G(\Pi(x)).
\tag{43}
\]

The walker experiment does not provide such a symbolic transition system. One may construct limited finite-time abstractions, such as a coarse-grained reversal map on recent trajectory segments, but these are externally controlled and finite-horizon.

To claim robust symbolic computation, one must further identify separated physical basins \(B_i\) satisfying

\[
\dist(B_i,B_j)>2\varepsilon_X
\]

and transition stability

\[
K(x,B_{G(a_i)})\geq 1-\delta.
\tag{44}
\]

Such basins are not demonstrated as a symbolic alphabet or transition table.

Finally, to claim autonomous physical computation, one must identify an internal physical readout state \(y_n\) and an operation-selection map \(C:X_Y\to\calO\) such that

\[
x_{n+1}=F_{C(y_n)}(x_n).
\]

In the walker experiment, the \(\pi\)-shift is imposed externally. There is no internal readout state satisfying

\[
C(y_n)=o_\pi.
\]

Therefore, the walker realizes a wave-memory physical machine with writing, finite-time storage, local reading, feedback, and externally triggered erasure. It does not, as experimentally described, realize a closed autonomous physical computer.

We summarize the classification as

\[
\boxed{
\text{wave-memory physical machine with Turing-like primitives}
}
\]

but not

\[
\boxed{
\text{closed autonomous Turing machine}.
}
\]

This distinction preserves the physical insight of the walker while sharpening the computational claim.

\section{Constructive extension: closing the wave-memory loop}
\label{sec:construct}

The previous classification suggests a constructive route. To turn a wave-memory walker into a closed physical computer, one must add a physical readout-control layer:

\[
H_n,z_n
\longrightarrow
y_n
\longrightarrow
o_n
\longrightarrow
F_{o_n},
\tag{45}
\]

where

\[
z_n=(\mathbf r_n,\mathbf v_n,\sigma_n),
\]

\[
y_n\in X_Y
\]

is a physical readout state, and

\[
o_n=C(y_n)
\]

is the selected physical operation.

The extended state is


\[
\tilde x_n=
(\mathbf r_n,\mathbf v_n,\sigma_n,H_n,y_n),
\]

with


\[
\tilde X
=
\Omega\times\R^2\times\{+1,-1\}\times\calH\times X_Y.
\]

A generic closed-loop architecture has the form

\[
y_{n+1}
=
\mathcal R(y_n,H_n,z_n),
\tag{46}
\]

\[
o_n=C(y_n),
\tag{47}
\]

\[
(H_{n+1},z_{n+1})
=
F_{o_n}(H_n,z_n).
\tag{48}
\]

This realizes

\[
\tilde x_{n+1}
=
F_{C(y_n)}(\tilde x_n).
\tag{49}
\]

Several physical implementations are possible.

\subsection{Multistable readout}

A readout coordinate \(y\) may evolve in a bistable or multistable potential

\[
U_Y(y;q)
=
\frac{\alpha}{4}y^4
-
\frac{\beta}{2}y^2
-
\gamma qy,
\tag{50}
\]

where

\[
q_n
\]

is a physical scalar extracted from the wave state, such as

\[
q_n=\nabla H_n(\mathbf r_n)\cdot\hat{\mathbf e}
\]

or

\[
q_n=\int_\Omega \chi(\rho)H_n(\rho)\,d\rho.
\]

The readout basins

\[
Y_+,\;Y_-
\]

select operations:

\[
C(Y_+)=o_\pi,
\qquad
C(Y_-)=o_{\mathrm{normal}}.
\]

The \(\pi\)-shift then becomes internally triggered by the wave state.

\subsection{Thresholded wave detector}

A physical detector may compute

\[
q_n=\int_\Omega \chi(\rho)H_n(\rho)\,d\rho
\]

and update a hysteretic readout state

\[
y_{n+1}
=
\begin{cases}
1, & q_n>\theta_+,\\
0, & q_n<\theta_-,\\
y_n, & \theta_-\leq q_n\leq \theta_+.
\end{cases}
\tag{51}
\]

With

\[
C(0)=o_{\mathrm{normal}},
\qquad
C(1)=o_\pi,
\]

the system autonomously triggers erasure when its own wave field crosses a physical threshold.

\subsection{Boundary-controlled reservoir}

The readout state may also control bath parameters:

\[
o_n=
(\gamma_m^{(n)},\phi^{(n)},V^{(n)},\Omega^{(n)})
=
C(y_n).
\]

Then the wave update becomes state-dependent:

\[
H_{n+1}
=
\lambda(y_n)H_n
+
\sigma_n h_0(y_n)
J_0(k_F(y_n)|\rho-\mathbf r_n|).
\tag{52}
\]

This architecture allows the system to switch memory depth, confinement geometry, forcing phase, or routing conditions through its own physical readout.

\subsection{Coupled walkers}

A second walker may serve as a physical controller. Two walkers sharing a bath interact
through their superposed wave fields, exchanging energy and momentum via the memory field
\cite{Borghesi2014InteractionMemory,Protiere2006ParticleWave}. If walker \(B\) has basin
state \(y_n\), then

\[
o_n=C(y_n)
\]

can control the operation applied to walker \(A\). The shared field evolves as

\[
H_{n+1}
=
\lambda H_n
+
\sigma^A_n\psi_{\mathbf r^A_n}
+
\sigma^B_n\psi_{\mathbf r^B_n}.
\tag{53}
\]

This creates a physical finite-state controller embedded in the same wave-mediated substrate.

\subsection{Minimal target: an autonomous eraser}

The simplest decisive experiment is an autonomous eraser. Define

\[
\calO=\{o_{\mathrm{normal}},o_\pi\}.
\]

Let the readout detect wave energy,

\[
E_W[n]=
\frac{\|H_n\|_{L^2(\Omega)}^2}
{\|\psi_{\mathbf 0}\|_{L^2(\Omega)}^2}.
\]

Use a hysteretic threshold:

\[
y_{n+1}
=
\begin{cases}
1, & E_W[n]>\Theta_E,\\
0, & E_W[n]<\Theta_E-\Delta,\\
y_n, & \Theta_E-\Delta\leq E_W[n]\leq \Theta_E.
\end{cases}
\tag{54}
\]

Then

\[
C(0)=o_{\mathrm{normal}},
\qquad
C(1)=o_\pi.
\]

The loop is

\[
H_n
\to
E_W[n]
\to
y_n
\to
o_\pi
\to
\text{wave erasure}.
\]

This would close the gap in the original experiment: the erasure operation would be selected by the system's own physical state rather than by an externally scheduled perturbation.

\subsection{Reservoir-computing interpretation}

The same construction can be read as a physical reservoir computer \cite{Jaeger2001EchoState,JaegerHaas2004esn,Maass2002LiquidState,Tanaka2019PhysicalReservoir,Markovic2020NeuromorphicPhysics}. The wave field is the high-dimensional reservoir,

\[
H_n\in\calH.
\]

The droplet provides a localized probe,

\[
H_n\mapsto \nabla H_n(\mathbf r_n).
\]

The physical readout layer maps reservoir state into basins,

\[
(H_n,\mathbf r_n,\mathbf v_n)\mapsto y_n.
\]

The readout is fed back into the reservoir dynamics through operation selection,

\[
y_n\mapsto o_n\mapsto H_{n+1}.
\]

Thus the architecture is

\[
\begin{aligned}
\text{reservoir}
&\;\to\;
\text{physical readout}
\\
&\;\to\;
\text{operation selection}
\\
&\;\to\;
\text{reservoir}.
\end{aligned}
\]

This is the physical closure missing from standard externally read reservoir computation. The design principle is therefore:

\[
\boxed{
\begin{aligned}
\text{wave memory}
&+ \text{ intrinsic readout basins}
\\
&+ \text{ state-dependent operation selection}
\\[0.3em]
&= \text{ closed physical computation}.
\end{aligned}
}
\]

\section{Discussion}

The analysis developed here began from a concrete physical observation: a walking droplet can store information about its past trajectory in a self-generated wave field, can read this field through local slope coupling, and, after an imposed \(\pi\)-phase shift, can transiently retrace its previous path while erasing the wave memory that enabled the reversal\cite{Perrard2016WaveTuring}. This is a remarkable physical phenomenon. The point of the present paper is not to weaken that claim, but to locate it precisely in a hierarchy of physical information processing and computation.

The central conclusion is that wave memory, even when dynamically read and physically erased, is not by itself sufficient for autonomous computation. The walker realizes writing, finite-time storage, local reading, feedback, and externally triggered erasure. These are genuine physical primitives. However, autonomous physical computation requires an additional closure condition: the system must contain an internal physical readout state whose value selects the next physical operation. In the notation used above, the transition must have the form

\[
x_{n+1}
=
F_{C(p_Y(x_n))}(x_n),
\]

rather than

\[
x_{n+1}
=
F_{o_n}(x_n)
\]

with \(o_n\) supplied externally. This distinction is the core of the closure criterion.

\subsection{What the walker demonstrates}

The Perrard--Fort--Couder walker demonstrates that memory can be physically embodied in a continuous wave field. The wave field is not a passive record. It is a dynamical reservoir that influences the future motion of the droplet. Each impact writes a localized wave source into the bath. The wave field stores these sources over a finite memory time. 

The droplet reads this stored information through the local gradient \(\nabla H_n(\mathbf r_n)\), and this readout alters its subsequent motion \cite{Eddi2011InformationStored,Bush2015PilotWaveHydrodynamics}. The \(\pi\)-shift experiment \cite{Perrard2016WaveTuring} adds a further primitive: \textit{physical erasure}. After the phase shift, new wave sources are written with opposite phase relative to the old field. When the droplet backtracks, these new sources overlap with the previous trajectory trace and destructively interfere with the old memory. The erasure is therefore not merely epistemic. It is not simply that an observer loses track of the past. The physical memory field itself is modified.

This gives the walker a privileged status among physical memory systems. It is not simply a medium that stores information, nor merely a particle whose state evolves. It is a coupled wave--particle system with a recurrent loop:

\[
\begin{aligned}
\text{trajectory} 
&\rightarrow\; \text{wave memory} \\
&\rightarrow\; \text{local readout} \\
&\rightarrow\; \text{future trajectory}.
\end{aligned}
\]


The system therefore gives a concrete hydrodynamic example of how memory, motion, and physical feedback can be inseparable.

\subsection{What the walker does not yet demonstrate}

The same analysis also clarifies what has not been demonstrated. The walker does not, as experimentally described, implement a robust symbolic transition system. To make such a claim, one must specify an abstract state space \(A\), a coarse-graining \(\Pi:X\to A\), an abstract transition rule \(G:A\to A\), and a physical domain \(U\subseteq X\) on which

\[
\Pi\circ F=G\circ \Pi.
\]

For a real physical implementation, one must further identify separated physical basins \(B_i\) realizing abstract states \(a_i\), and show that the dynamics maps each basin into the correct successor basin with sufficient reliability:

\[
K(x,B_{G(a_i)})\geq 1-\delta.
\]

The walker may support limited finite-time abstractions. For example, under a \(\pi\)-shift, one may define a coarse reversal map on recent trajectory segments. But this is a finite-horizon, externally triggered primitive. It is not a general symbolic machine, and it does not establish Turing universality. 

This distinction matters because the phrase ``Turing machine'' can be read in two ways. In a loose operational sense, the walker exhibits operations reminiscent of a machine with memory: writing, storing, reading, and erasing. In the strict computability-theoretic sense, a Turing machine requires a finite alphabet, internal states, a transition table, and a robust physical encoding of configurations and updates \cite{Turing1937Computable,Church1936Entscheidungsproblem}---the symbolic architecture
that symbolic dynamics and computational mechanics make precise \cite{ShaliziCrutchfield2001ComputationalMechanics,Crutchfield2012OrderChaos}. The walker supplies the physical memory substrate and several primitive operations, but not that full symbolic architecture, and the classification of Sec.~\ref{sec:classification} therefore stands.

Thus the more precise classification is: \emph{``wave-memory physical machine with Turing-like primitives''} rather than ``autonomous universal Turing machine''.

\subsection{The closure criterion}
The closure criterion introduced here separates physical feedback from autonomous physical computation. Feedback is ubiquitous in many physical systems. A pendulum coupled to a driving field, a chemical oscillator in a reaction bath, a neural population coupled to its own recurrent activity, and a walking droplet coupled to its wave field all exhibit physical feedback. But computation in the stronger autonomous sense requires that some physical state of the system function as a readout-control variable \cite{Markovic2020NeuromorphicPhysics}. In biological systems, effective control or adaptation demands that the internal controller possess a variety at least equal to that of the system–environment complex \cite{Ashby1991,Baryam2004,Dehghani2018netcontrol}; at the cellular scale, ion channels provide this by acting simultaneously as the readout and the control variable for adaptation \cite{Dehghani2024hilbert}.

The closure condition can be written as

\[
X_R
\longrightarrow
X_Y
\longrightarrow
\mathcal O
\longrightarrow
X,
\]

where \(X_R\) is a reservoir or memory subsystem, \(X_Y\) is a physical readout-control subsystem, and \(\mathcal O\) is a set of physical operations. The crucial step is

\[
X_Y\longrightarrow\mathcal O.
\]

The readout state must select the next operation. A detector that is merely observed by an external camera is not enough. A threshold computed offline is not enough. A classification label assigned by an observer is not enough. The readout state must be physically coupled to the system so that it changes what the system does next.

In the walker, the missing operation is clear. The \(\pi\)-shift is the operation that triggers reversal and erasure. In the experiment, this operation is imposed externally. A closed autonomous version would require a physical readout state \(y_n\) such that

\[
C(y_n)=o_\pi
\]

under specified physical conditions. This could be realized by a multistable droplet trap, a thresholded wave detector, a coupled-walker controller, or a boundary-control element whose state changes the forcing phase or confinement geometry.

\subsection{Why memory is not enough}
A key lesson is that memory and computation must not be conflated. Physical systems often contain traces of their past. A magnetic material ``remembers'' its field history, and a broad class of nanoscale and complex materials exhibit memristive and related memory effects \cite{Chua1971Memristor,PershinDiVentra2011MemoryMaterials}; a fluid vortex field ``records'' previous forcing; a deformed solid retains a record of its loading; a neural circuit carries traces of previous activity \cite{Maass2002LiquidState}. Such memory may be rich and dynamically consequential. But memory becomes computation only when there is an abstraction under which physical transitions preserve an abstract transition structure. 

This is why the coarse-graining condition is central. The map

\[
\Pi:X\to A
\]

is not arbitrary labeling. It becomes computationally meaningful only when physical evolution descends to an abstract transition:

\[
\Pi(F(x))=G(\Pi(x)).
\]

The robust version requires that abstract states correspond to separated physical basins and that transitions among those basins be reliable. This avoids two extremes. It avoids the overly weak view that any physical process computes whatever an observer can map onto it. It also avoids the overly strong view that physical computation requires microscopic one-to-one identity between physical and symbolic states.

\subsection{Relation to physical reservoir computing}

The closure criterion is especially relevant for reservoir computing and neuromorphic systems. In standard reservoir computing, a high-dimensional dynamical system transforms input histories into a rich state space, and an external readout is trained to extract outputs \cite{Jaeger2001EchoState,JaegerHaas2004esn,Maass2002LiquidState}. The reservoir is physical or dynamical, but the readout is often mathematically or electronically external to the reservoir itself \cite{Tanaka2019PhysicalReservoir,Markovic2020NeuromorphicPhysics}. This holds even when the recurrent substrate itself is placed under selection rather than fixed: optimizing reservoir architecture for a predictive task reorganizes the substrate's internal structure while the readout remains an externally trained decoder\cite{Dehghani2026EvoRes}.

The walker makes this separation visible. The wave field is a high-dimensional reservoir. The droplet samples the reservoir locally. But the operation that produces erasure is selected externally. The constructive extension above internalizes the readout, closing the
reservoir--readout--operation-selection loop of Sec.~\ref{sec:construct}: the wave
field drives a physical readout basin, the basin selects an operation, and the operation
changes the subsequent reservoir dynamics. The decoded state is then not merely read off
after the fact but participates in the system's future physical evolution---\emph{a physically
closed reservoir computer} rather than the externally read reservoir of standard practice
\cite{Jaeger2001EchoState,JaegerHaas2004esn,Maass2002LiquidState,Tanaka2019PhysicalReservoir}. 




This distinction may be useful for neuromorphic computing more broadly. Many neuromorphic devices are described as computing because their dynamics transforms inputs into useful output states. The closure criterion asks a sharper question: are the output states merely measured, or do they form internal physical variables that select subsequent operations? In biological recurrent systems, this distinction is also relevant. Neural activity does not simply represent states for an external observer. It acts on downstream circuits, gates future processing, and changes the physical conditions under which later activity unfolds \cite{Dehghani2018netcontrol,Muller2018wave}. The closure criterion may therefore provide a language for comparing engineered reservoirs, physical unconventional computers, and biological recurrent systems without collapsing them into a single vague notion of ``information processing.''

\subsection{Cortical traveling waves and the risk of premature computational interpretation}

The same caution applies to cortical traveling waves. Waves of neural activity can propagate across cortical tissue, modulate excitability, organize spike timing, and shape spatiotemporal patterns of activity. They may therefore participate in neural computation \cite{ErmentroutKleinfeld2001wave,LeVanQuyen2016highfrq,Muller2018wave,KellerWelling2023wavemachine}. But the presence of a traveling wave is not, by itself, a demonstration of computation. Under the present criterion, a cortical wave becomes computationally relevant only when one can identify the physical variables it transforms, the coarse-grained states it helps stabilize or route, and the downstream readout/control pathways through which it changes later neural dynamics.

This distinction avoids a bad analogy. The claim is not that cortex is a hydrodynamic walker, nor that all traveling waves compute. The analogy is structural and limited: both systems contain distributed wave-like activity, local readout, and recurrent physical coupling. In the walker, the local readout is the droplet sampling \(\nabla H(\mathbf r)\). In cortex, the readout may be downstream spiking, synaptic integration, dendritic nonlinearities, local circuit thresholds, or long-range recurrent coupling. The computational question is whether these readouts form robust transition-preserving and closed loops, not whether the activity pattern visually resembles a wave.

This framing also suggests a concrete research program for cortical waves. One should ask whether wave variables define robust basins or phase states; whether those states predict or control transitions in downstream neural populations; whether perturbing wave phase, direction, or speed changes the inferred transition map; and whether the relevant readout is internal to the circuit rather than imposed by an external decoder. In this sense, cortical traveling waves provide an important biological test case for the closure criterion developed here.

\subsection{Category theory as organization, not ornament}

The category-theoretic formulation is not intended as formal decoration. It serves three roles that the dynamical-systems statement alone leaves implicit.

First, it clarifies that computation concerns processes, not isolated states. A physical-to-abstract map becomes computational only when it maps physical morphisms to abstract morphisms; treating the representation as a functor, rather than a correspondence of states, is what enforces this \cite{DehghaniCaterina2024PhysicalComputing}. Therefore, the relevant object is not merely a representation of a state, but the compatibility of state transitions in this mapping.

Second, it replaces bijective implementation with quotient functoriality. That allows accommodating the many-to-one character of physical realization. Abstract states are physical realization classes---basins, not microstates---and the operative demand is
that physical processes respect the quotient. In this sense, many microstates may realize the same abstract state. What matters is that the physical process respects this quotient. Allowing the representation pair
$(\mathcal{R}_T,\widetilde{\mathcal{R}}_T)$ to be an adjunction rather than a strict
inverse \cite{DehghaniCaterina2024PhysicalComputing} is what licenses this basin-based
realization, steering between one-to-one physical--symbol identity and arbitrary mapping.

Third, it expresses closure compositionally and makes closure a statement about the morphism itself. The closed-loop map
\[
\Phi
=
\mathrm{Act}\circ
\langle C\circ p_Y,\id_X\rangle
\]
folds operation selection into the physical process itself. Where the functorial picture takes the morphism relating $p$ to $p'$ as given, closure asks where that morphism originates and answers: from the system’s own state. Autonomy, in this view, is not a vague cognitive or interpretative overlay—of the sort easily ascribed to biological systems—but a structural property of how the process is generated. The system produces its own transitions rather than having them assigned from outside; morphism selection is internal. The same compositional machinery lets reservoir, readout, actuator, and abstraction sit at different scales and still compose through functors and natural transformations.

This is also why the framework naturally extends to multiscale systems \cite{DehghaniCaterina2024PhysicalComputing}. A physical reservoir, a readout layer, an actuator, and a symbolic abstraction may each live at different physical scales. Category theory provides a way to compose these levels without demanding that they be identical.

\subsection{A philosophical clarification: why the rock does not compute its trajectory}

The present framework also addresses a recurrent philosophical worry: if every physical system evolves according to physical law, does every physical system compute its own evolution? Does a rock compute its trajectory \cite{Chalmers1996RockFSA}? Does the Solar System compute Newton's laws \cite{CampbellYang2021SolarSystem}? Does any object with state transitions implement arbitrary finite-state machines?

The answer offered here is ``\textbf{\textit{NO!}}'', not in the relevant sense. A propelled rock follows a trajectory whether or not anyone models it. Its motion may be simulated by a computer, and an observer may map its positions onto symbols after the fact. But this does not make the rock a computer for all such symbolic descriptions. What is missing is a physically constrained abstraction whose states are robustly realized and whose transitions are preserved by the physical dynamics in a way that supports the intended abstract process.

The rock may instantiate physical information in the weak sense that its state carries consequences of forces, initial conditions, and environmental interactions. But unless there is a physically organized readout/write-in structure, a robust coarse-graining, and transition preservation relative to an abstract process, the claim that the rock computes its trajectory is an observer-side projection. The distinction is not that digital computers are magical or that natural systems cannot compute. The distinction is that computing systems are physically organized so that some degrees of freedom serve as realizers of abstract states and some physical processes preserve the corresponding abstract transitions.

The billiard-ball computer illustrates the contrast \cite{FredkinToffoli1982ConservativeLogic,Margolus1984PhysicsLikeComputation,DurandLose2002BilliardBall}. There, mechanical trajectories are constrained so that collisions implement logical operations. The physical process is organized to preserve an abstract transition structure. A generic moving ball is not thereby a logic gate; a ball in a constrained collision architecture can be. Similarly, the walking droplet is not a Turing machine merely because it has a history-dependent trajectory. But it can become part of a physical computing architecture if its wave memory is coupled to robust readout basins and internally selected operations.

This clarification avoids pancomputationalism without denying physical computation. It rejects the idea that arbitrary mappings are sufficient, while preserving the possibility that unconventional, analog, biological, quantum, or wave-based systems can compute when their physical organization satisfies the relevant structural criteria.

\subsection{Limitations}

The closure criterion is intentionally minimal. It does not by itself provide a full theory of computational power, efficiency, thermodynamic cost, trainability, or universality. It answers a prior question: when does a physical memory system become an autonomous physical computer rather than an externally interpreted dynamical process?

Several limitations remain. First, the criterion depends on identifying physically meaningful basins and noise scales. In some systems, especially high-dimensional biological or fluid systems, the relevant state variables may be difficult to measure. The framework therefore provides a set of conditions, not an automatic procedure for discovering the correct abstraction.

Second, robust transition preservation may hold only on restricted domains. This is not a defect. Physical computers also operate only within valid ranges of initialization, temperature, voltage, noise, and control. But the domain of validity must be specified.

Third, the constructive extensions proposed here are design principles rather than completed experiments. An autonomous wave-memory eraser, a wave-controlled branch, or a finite-state coupled-walker controller would be the natural next steps. Such systems would not prove Turing universality, but they would test the closure criterion directly.

Fourth, the relation between this closure criterion and computational universality remains to be developed. Universal computation would require additional structure: a robust finite alphabet, programmable transition rules, memory addressing or an equivalent substrate, and scalable composition of operations \cite{Turing1937Computable,Deutsch1985UniversalQuantumComputer}. The present paper stops short of that claim by design.

\subsection{Outlook}

The most direct experimental test is an autonomous eraser. Instead of imposing the \(\pi\)-shift externally, one would couple a physical detector of wave-memory energy, local slope, or trajectory state to the forcing phase. When the readout crosses a threshold, the system would trigger its own phase shift and erase its own memory. The relevant comparison would be between three cases:

\[
\begin{aligned}
\text{external phase schedule}, \\
\text{observer-side readout without feedback}, \\
\text{physical readout with feedback}.
\end{aligned}
\]

Only the third satisfies closure.

A second direction is to construct small autonomous finite-state wave machines. One could define two or three robust basins, use wave-field features to drive a physical readout, and use that readout to switch confinement, phase, or routing. The goal would not initially be universality, but reliable closed transition structure.

A third direction is to transfer the criterion to neuromorphic reservoirs \cite{Tanaka2019PhysicalReservoir,Markovic2020NeuromorphicPhysics}. Physical reservoirs are often evaluated by how well an external readout can decode their state. The closure criterion suggests a stronger benchmark: whether the readout can be physically internalized so that decoded states control subsequent reservoir dynamics. This may help distinguish passive physical reservoirs from autonomous physical computers.

\subsection{Conclusion}

The wave--particle walker shows that memory can be written into matter, read by matter, and erased by matter. That is already a profound physical result. The present work adds a criterion for when such memory becomes autonomous computation.

The criterion is:

\[
\begin{aligned}
&\quad \text{physical memory} \\
&+ \text{robust coarse-grained transition preservation} \\
&+ \text{internal operation selection} \\
&\Rightarrow\; \textbf{closed physical computation}.
\end{aligned}
\]

Applied to the Perrard--Fort--Couder walker, this yields a precise classification. The walker is a wave-memory physical machine with Turing-like primitives. It is not, as experimentally described, a closed autonomous Turing machine. The missing layer is not mysterious: it is a physical readout-control subsystem that selects subsequent operations.

This reframes the wave-based Turing-machine idea as a constructive research program. The question is no longer whether wave memory is metaphorically computational. The question is how to build physical architectures in which wave memory, readout basins, and operation selection compose into closed autonomous computation.

\section*{Acknowledgments}
I thank Gianluca Caterina and Baktash Babadi for deep and thoughtful discussions and for their collaboration on physical computing formalism and recurrent computing systems.
\section*{References}
\bibliography{waveComp}
\section*{Appendices}

%
%
%
%
%


\setcounter{section}{0}
\setcounter{subsection}{0}
\setcounter{equation}{0}
\setcounter{figure}{0}
\setcounter{table}{0}
\renewcommand{\thesection}{A\arabic{section}}
\renewcommand{\thesubsection}{\thesection.\arabic{subsection}}
\renewcommand{\theequation}{A\arabic{equation}}
\renewcommand{\thefigure}{A\arabic{figure}}
\renewcommand{\thetable}{A\arabic{table}}

\section{Stroboscopic physical model of the wave--particle walker}
\label{sec:S1}

The physical system considered here is a wave--particle entity consisting of a droplet bouncing on a vertically vibrated bath and interacting with the surface waves generated by its previous impacts. The bath is driven at frequency $(f_0)$, while the Faraday waves and the bouncing motion occur subharmonically at

\[
f_F = f_W = \frac{f_0}{2}.
\]

This provides a natural stroboscopic clock. We index the droplet impacts by

\[
n \in \mathbb{Z}_{\geq 0},
\]

and denote the corresponding impact times by

\[
t_n = nT_F,
\qquad
T_F = \frac{1}{f_F}.
\]

The goal of this section is not to reproduce the full hydrodynamic model of the walker, but to write the minimal discrete-time physical model needed to analyze memory, reading, erasing, and later computation. The stroboscopic description makes explicit which variables belong to the physical state and how the wave memory enters the droplet dynamics.

\subsection{Physical state space}

Let

\[
\Omega \subset \mathbb{R}^2
\]

denote the horizontal domain of the fluid bath. At each bounce $n$, we define the physical state as

\begin{equation}
x_n =
\left(
\mathbf r_n,
\mathbf v_n,
\sigma_n,
H_n
\right)
\in X.    
\label{eq:S-state}
\end{equation}

The components are:

\[
\mathbf r_n \in \Omega
\]

the horizontal position of the droplet at the $n$-th impact;

\[
\mathbf v_n \in \mathbb{R}^2
\]

the horizontal velocity of the droplet just before or immediately after the impact, depending on the chosen convention;

\[
\sigma_n \in \{+1,-1\}
\]

the phase of the droplet bounce relative to a fixed stroboscopic phase of the Faraday wave field;

\[
H_n : \Omega \to \mathbb{R}
\]

the slowly varying envelope of the surface-wave memory field sampled at the same stroboscopic phase.

We take

\[
H_n \in \mathcal H,
\]

where $\mathcal H$ is a function space over the bath domain, for example $L^2(\Omega)$ if only wave energy is needed, or $H^1(\Omega)$ if the local wave slope $\nabla H_n$ is explicitly used. Since the droplet reads the wave field through its local gradient, $H^1(\Omega)$ is the more natural choice for the analysis below.

Thus the physical state space is

\[
X
=
\Omega
\times
\mathbb{R}^2
\times
\{+1,-1\}
\times
\mathcal H.
\]

This state space separates the localized degrees of freedom of the droplet,

\[
(\mathbf r_n,\mathbf v_n,\sigma_n),
\]

from the extended wave-memory degree of freedom,

\[
H_n.
\]

This distinction is central: the droplet is a localized probe and actuator, while the wave field is a spatially distributed memory reservoir.

\subsection{Single-impact wave source}

Each impact of the droplet generates a localized standing Faraday-wave contribution. In the idealized infinite-domain approximation, the spatial profile of the impact-generated source is modeled by the zeroth-order Bessel function

\[
J_0(k_F|\rho-\mathbf r_n|),
\]

where

\[
\rho \in \Omega
\]

is a point on the fluid surface,

\[
k_F = \frac{2\pi}{\lambda_F}
\]

is the Faraday wavenumber, and

\[
\lambda_F
\]

is the Faraday wavelength.

We define the source kernel centered at $\mathbf r$ by

\begin{equation}
\psi_{\mathbf r}(\rho)
= 
h_0 J_0(k_F|\rho-\mathbf r|).
\label{S-kernel}
\end{equation}

Here

\[
h_0
\]

is the effective stroboscopic amplitude of a single impact-generated wave. In a finite bath, boundary effects and finite-size corrections may alter the precise spatial Green's function. In that case one may replace the Bessel kernel by a bath-dependent response kernel

\[
G_\Omega(\rho,\mathbf r),
\]

so that

\[
\psi_{\mathbf r}(\rho)
=
h_0 G_\Omega(\rho,\mathbf r).
\]

For the present analysis, the Bessel approximation is sufficient because the key point is not the exact hydrodynamic Green's function, but the existence of a localized wave source whose contributions superpose and decay over a memory time.

\subsection{Wave-memory update}

The wave memory is sustained by the vertical forcing but decays over a finite time scale. Let

\[
\tau
\]

denote the decay time of the Faraday-wave memory. The dimensionless memory parameter is

\[
M_e = \frac{\tau}{T_F}.
\]

Equivalently, if we sample the system once per bounce, the wave-memory decay factor per stroboscopic step is

\begin{equation}
\lambda
=
\exp\!\left(-\frac{T_F}{\tau}\right)
=
\exp\!\left(-\frac{1}{M_e}\right),
\qquad
0 < \lambda < 1.
\label{eq:S-decay}
\end{equation}

The stroboscopic wave-memory update is then

\begin{equation}
\begin{aligned}
H_{n+1}(\rho)
&=
\lambda\, H_n(\rho)
+
\sigma_n\, \psi_{\mathbf r_n}(\rho),
\\[0.6em]
H_{n+1}(\rho)
&=
\lambda\, H_n(\rho)
+
\sigma_n\, h_0\,
J_0\!\left(k_F \lvert \rho - \mathbf r_n \rvert\right).
\end{aligned}
\label{eq:S-Hupdate}
\end{equation}

The factor

\[
\sigma_n \in \{+1,-1\}
\]

records the phase of the emitted wave relative to the chosen Faraday phase. In the unperturbed walker, the droplet remains locked to one bouncing phase, so that

\[
\sigma_{n+1}=\sigma_n.
\]

A phase-switching perturbation, such as the $\pi$-shift used in the experiment, will later be represented as an intervention changing $\sigma_n\to-\sigma_n$.

If $H_0=0$, iterating the wave-memory update gives

\begin{equation}
\begin{aligned}
H_n(\rho)
&=
\sum_{j=0}^{n-1}
\lambda^{\,n-1-j}\,
\sigma_j\, h_0\,
J_0\!\left(k_F \lvert \rho - \mathbf r_j \rvert\right),
\\[0.6em]
\lambda^{m}
&=
\exp\!\left(-\frac{m}{M_e}\right).
\end{aligned}
\label{eq:S-trace}
\end{equation}

An impact contribution $m$ bounces in the past is weighted by $e^{-m/M_e}$. Thus $M_e$ controls the effective number of past impacts retained in the wave field. In this sense, $H_n$ is a distributed physical memory of the recent trajectory

\[
\mathbf r_{n-1},
\mathbf r_{n-2},
\ldots,
\mathbf r_{n-M_e}.
\]

This is the mathematical expression of the wave-memory reservoir.

\subsection{Local readout by the droplet}

The droplet does not read the entire field $H_n$. It samples the local slope of the wave field at its impact location. Define the physical readout vector

\begin{equation}
\mathbf g_n
=
\nabla H_n(\mathbf r_n).
\label{eq:S-readout}
\end{equation}

The wave-mediated horizontal force is proportional to the negative local slope,


\[
\begin{aligned}
\mathbf F^{\mathrm{wave}}_n 
&= -C\,\sigma_n\nabla H_n(\mathbf r_n) \\
&= -C\,\sigma_n\,\mathbf g_n.
\end{aligned}
\]

Here,
$
C>0
$
is the wave--droplet coupling coefficient.

The factor $\sigma_n$ expresses that the droplet samples the field at its own bouncing phase; in the phase-locked regime $\sigma_n^2=1$ against the phase carried by $H_n$, and the standard guidance is recovered.

This local readout is the physical mechanism by which the stored wave memory influences future droplet motion. The field $H_n$ stores information globally, but the droplet accesses it only through the local differential quantity

\[
\nabla H_n(\mathbf r_n).
\]

Thus the walker implements a recurrent physical loop:

\[
\begin{aligned}
\text{impact position}
&\longrightarrow
\text{wave written into } H_n
\\
&\longrightarrow
\text{local slope read at next impacts}
\\
&\longrightarrow
\text{future droplet motion}.
\end{aligned}
\]

\subsection{Droplet impact map}

We now write a minimal stroboscopic map for the droplet motion. Let

\[
V:\Omega\to\mathbb{R}
\]

be an external confinement potential. In the harmonic trap used in the walker experiments, one may take

\[
V(\mathbf r)
=
\frac{1}{2}m\omega^2|\mathbf r|^2,
\]

so that

\[
-\nabla V(\mathbf r)
\]

is the confining force.

A discrete-time impact map for the horizontal velocity is

\begin{equation}
\mathbf v_{n+1}
= a\,\mathbf v_n
  - \kappa\,\sigma_n \nabla H_n(\mathbf r_n)
  - \chi\,\nabla V(\mathbf r_n)
  + \boldsymbol{\xi}_n.
  \label{eq:S-vupdate}
\end{equation}

Here:

\[
a \in [0,1]
\]

is the effective horizontal persistence factor after damping over one bounce period;

\[
\kappa>0
\]

is the effective wave-induced kick strength;

\[
\chi>0
\]

sets the strength of the external confinement term in the discrete map;

\[
\boldsymbol\xi_n
\]

represents experimental noise, unresolved hydrodynamic effects, or small fluctuations in the impact dynamics.

The position update is

\begin{equation}
\mathbf r_{n+1}
=
\mathbf r_n
+
T_F\mathbf v_{n+1}.
\label{eq:posupdate}
\end{equation}

If the finite domain boundary is relevant, this may be written as

\[
\mathbf r_{n+1}
= \mathcal P_\Omega
\left(
\mathbf r_n
+
T_F\mathbf v_{n+1}
\right),
\]

where

\[
\mathcal P_\Omega
\]

is a boundary or projection operator encoding reflection, confinement, or exclusion outside the bath domain. For most of the formal analysis, we assume that the trajectory remains sufficiently far from hard boundaries, so that $\mathcal P_\Omega$ can be suppressed.

\subsection{Full stroboscopic physical evolution}

Combining the droplet update and the wave-memory update gives the physical evolution map

\[
F:X\to X,
\]

defined by

\[
x_{n+1}
= F(x_n).
\]

Explicitly, for

\[
x_n=
(\mathbf r_n,\mathbf v_n,\sigma_n,H_n),
\]

we compute

\[
\begin{aligned}
\mathbf g_n &= \nabla H_n(\mathbf r_n), 
\\[0.4em]
\mathbf v_{n+1}
&= a\,\mathbf v_n
  - \kappa\,\sigma_n \mathbf g_n
  - \chi\,\nabla V(\mathbf r_n)
  + \boldsymbol{\xi}_n.
\end{aligned}
\]

\[
\begin{aligned}
\mathbf r_{n+1} &= \mathbf r_n + T_F\, \mathbf v_{n+1}, \\
\sigma_{n+1} &= \sigma_n.
\end{aligned}
\]

In the absence of an externally imposed phase perturbation, and

\[
H_{n+1}(\rho)
=
\lambda H_n(\rho)
+
\sigma_n h_0
J_0
\left(
k_F|\rho-\mathbf r_n|
\right).
\]

Therefore,

\[
F(\mathbf r_n,\mathbf v_n,\sigma_n,H_n)
=
\left(
\mathbf r_{n+1},
\mathbf v_{n+1},
\sigma_{n+1},
H_{n+1}
\right).
\]

This is a closed physical dynamical system once the parameters

\[
a,\kappa,\chi,\lambda,h_0,k_F
\]

and the potential $V$ are specified. In the deterministic idealization,

\[
\boldsymbol\xi_n=0.
\]

In the noisy case, $F$ is replaced by a stochastic transition kernel

\[
\mathbb P(x_{n+1}\mid x_n).
\]

The deterministic map is sufficient for defining memory, phase reversal, and coarse-grained transition preservation. The stochastic version will be useful later when discussing robustness and physical symbolization.

\subsection{Dimensionless form}

It is often useful to nondimensionalize the model using the Faraday wavelength and period. Define

\[
\begin{aligned}
\bar{\mathbf r}_n &= \frac{\mathbf r_n}{\lambda_F}, \\
\bar{\rho}        &= \frac{\rho}{\lambda_F}, \\
\bar{\mathbf v}_n &= \frac{T_F}{\lambda_F}\,\mathbf v_n, \\
\bar H_n          &= \frac{H_n}{h_0}.
\end{aligned}
\]

Since

\[
k_F\lambda_F = 2\pi,
\]

the wave-memory update becomes

\[
\bar H_{n+1}(\bar\rho)
=
\lambda \bar H_n(\bar\rho)
+
\sigma_n
J_0
\left(
2\pi|\bar\rho-\bar{\mathbf r}_n|
\right).
\]

The local readout becomes

\[
\bar{\mathbf g}_n
=
\nabla_{\bar\rho}\bar H_n(\bar{\mathbf r}_n),
\]

and the nondimensional droplet update may be written as

\[
\begin{aligned}
\bar{\mathbf v}_{n+1}
&= a\,\bar{\mathbf v}_n
   - \bar\kappa\,\sigma_n \nabla_{\bar\rho}\bar H_n(\bar{\mathbf r}_n)
   - \bar\chi\,\nabla_{\bar{\mathbf r}}\bar V(\bar{\mathbf r}_n)
   + \bar{\boldsymbol\xi}_n.
\end{aligned}
\]

\[
\bar{\mathbf r}_{n+1}
=
\bar{\mathbf r}_n
+
\bar{\mathbf v}_{n+1}.
\]

The dimensionless form makes clear that the essential control parameters are the memory parameter $M_e$, the wave--droplet coupling strength $\bar\kappa$, the confinement strength $\bar\chi$, and the noise scale.

\subsection{Interpretation of the stroboscopic model}

The stroboscopic model separates four physical operations:

\[
\mathbf r_n
\longmapsto
\sigma_n h_0J_0(k_F|\rho-\mathbf r_n|)
\]

is the writing operation, by which the droplet impact deposits a wave source into the field;

\[
H_n
\longmapsto
\lambda H_n
\]

is the storage operation, by which previous wave sources persist but decay;

\[
H_n
\longmapsto
\nabla H_n(\mathbf r_n)
\]

is the local reading operation, by which the droplet samples the stored wave memory;

\[
\nabla H_n(\mathbf r_n)
\longmapsto
\mathbf v_{n+1}
\]

is the physical feedback operation, by which the read wave slope alters the next droplet motion.

The resulting system is therefore a recurrent physical memory system:

\[
(\mathbf r_n,\mathbf v_n)
\rightarrow
H_{n+1}
\rightarrow
\nabla H_{n+1}(\mathbf r_{n+1})
\rightarrow
(\mathbf r_{n+2},\mathbf v_{n+2}).
\]

This formulation is the starting point for distinguishing physical memory from physical computation. At this stage, no symbolic computation has been assumed. The model only states that the walker has an extended wave-memory field, a localized probe, and a closed physical feedback loop. Whether this loop implements an autonomous computation requires additional structure: a robust abstraction or readout layer whose states preserve transition structure and causally select future physical operations.

\section{Memory and erasure in the stroboscopic wave field}
\label{sec:S2}

The stroboscopic model introduced above separates the localized droplet variables from the extended wave-memory field. We now make this memory structure explicit and define what it means, mathematically, for the walker to store, read, and erase trajectory information.

Recall the stroboscopic state

\[
x_n =
\left(
\mathbf r_n,
\mathbf v_n,
\sigma_n,
H_n
\right),
\]

with wave-memory update

\[
H_{n+1}(\rho)
=
\lambda H_n(\rho)
+
\sigma_n h_0
J_0
\left(
k_F|\rho-\mathbf r_n|
\right),
\]

where

\[
\lambda = e^{-1/M_e}.
\]

Here $M_e=\tau/T_F$ is the dimensionless memory parameter, $T_F$ is the Faraday period, $\tau$ is the wave-memory decay time, $h_0$ is the amplitude scale of a single impact-generated wave, $k_F=2\pi/\lambda_F$ is the Faraday wavenumber, and $\sigma_n\in\{+1,-1\}$ denotes the phase of the impact-generated wave relative to the chosen stroboscopic phase.

For compactness, define the single-impact source centered at $\mathbf r$ by

\[
\psi_{\mathbf r}(\rho)
=
h_0
J_0
\left(
k_F|\rho-\mathbf r|
\right).
\]

Then the wave update is

\[
H_{n+1}
=
\lambda H_n
+
\sigma_n\psi_{\mathbf r_n}.
\]

\subsection{The wave field as an exponentially weighted trajectory trace}

Iterating the wave update gives

\[
H_n
=
\sum_{j=0}^{n-1}
\lambda^{n-1-j}
\sigma_j
\psi_{\mathbf r_j}
+
\lambda^n H_0.
\]

If the initial field has decayed or if we take $H_0=0$, then

\[
H_n
=
\sum_{j=0}^{n-1}
\lambda^{n-1-j}
\sigma_j
\psi_{\mathbf r_j}.
\]

Equivalently, indexing the past by the lag $q=n-j$,

\begin{equation}
H_n
=
\sum_{q=1}^{n}
\lambda^{q-1}
\sigma_{n-q}
\psi_{\mathbf r_{n-q}}.
\label{eq:S2-trace}
\end{equation}

Thus the present wave field is an exponentially weighted spatial trace of previous droplet impacts. The contribution of an impact $q$ bounces in the past is weighted by

\[
\lambda^{q-1}
=
\exp\left(-\frac{q-1}{M_e}\right).
\]

Therefore, the memory parameter $M_e$ controls the effective temporal depth of the wave field. Large $M_e$ corresponds to long memory, while small $M_e$ corresponds to rapid forgetting.

\subsection{Finite-memory approximation}

Although the formal expression for $H_n$ contains all previous impacts, older impacts are exponentially suppressed. Define the $L$-step truncated memory field by

\[
H_n^{(L)}
=
\sum_{q=1}^{L}
\lambda^{q-1}
\sigma_{n-q}
\psi_{\mathbf r_{n-q}}.
\]

The residual field due to impacts older than $L$ bounces is

\[
R_n^{(L)}
=
H_n - H_n^{(L)}
=
\sum_{q=L+1}^{n}
\lambda^{\,q-1}\,
\sigma_{n-q}\,
\psi_{\mathbf r_{n-q}}.
\]

Assume that the single-impact source is uniformly bounded in the chosen norm:

\[
|\psi_{\mathbf r}|_{\mathcal H}
\leq
\Psi_0
\qquad
\text{for all }
\mathbf r\in \Omega.
\]

Then

\[
|R_n^{(L)}|_{\mathcal H}
\leq
\sum_{q=L+1}^{\infty}
\lambda^{q-1}
\Psi_0
=
\Psi_0
\frac{\lambda^L}{1-\lambda}.
\]

Hence, for a prescribed field-resolution tolerance $\varepsilon_H>0$, the field is effectively $L$-step Markovian whenever

\[
\Psi_0
\frac{\lambda^L}{1-\lambda}
\leq
\varepsilon_H.
\]

Solving for $L$ gives

\[
L
\geq
\frac{
\log\left[
\varepsilon_H(1-\lambda)/\Psi_0
\right]
}{
\log \lambda
}.
\]

Since

\[
\lambda = e^{-1/M_e},
\qquad
\log \lambda = -\frac{1}{M_e},
\]

this becomes

\begin{equation}
L
\geq
M_e
\log
\left[
\frac{\Psi_0}{\varepsilon_H(1-\lambda)}
\right].
\label{eq:S2-horizon}
\end{equation}

Thus, up to a logarithmic tolerance factor, the effective memory horizon scales as

\[
L_{\mathrm{eff}}
\sim
M_e.
\]

This makes precise the statement that $M_e$ is the number of past impacts effectively retained in the wave field.

\subsection{Trajectory memory}

The wave field stores a spatial trace of the recent trajectory. Let the recent $L$-step trajectory segment be

\[
\mathcal T_n^{(L)}
=
\left(
\mathbf r_{n-L},
\mathbf r_{n-L+1},
\ldots,
\mathbf r_{n-1}
\right)
\in
\Omega^L.
\]

We assume the bounce phase is constant over the interval, $\sigma_{n-q}=\sigma$ for $1\leq q\leq L$. This holds for the unperturbed walker, which remains locked to a single bouncing phase; the case of a phase intervention is treated separately below. Under this assumption the wave field is determined by the trajectory positions alone, so the writing map is well-defined on $\Omega^L$.

Define the wave-writing map

\[
\mathcal W_L:
\Omega^L
\to
\mathcal H
\]

by

\[
\mathcal W_L
\left(
\mathcal T_n^{(L)}
\right)
=
\sum_{q=1}^{L}
\lambda^{q-1}
\sigma_{n-q}
\psi_{\mathbf r_{n-q}}.
\]

With the phase constant, the map simplifies to

\[
\mathcal W_L
\left(
\mathcal T_n^{(L)}
\right)
=
\sigma
\sum_{q=1}^{L}
\lambda^{q-1}
\psi_{\mathbf r_{n-q}}.
\]

We say that the wave field contains $L$-step trajectory memory at spatial resolution $\varepsilon_r$ if there exists a decoder

\[
D_L:\mathcal H\to \Omega^L
\]

such that

\[
d_L
\left(
D_L(H_n),
\mathcal T_n^{(L)}
\right)
<
\varepsilon_r,
\]

where $d_L$ is a metric on trajectory segments, for example

\[
\begin{aligned}
d_L(\mathcal T,\mathcal T')
&=
\left[
\frac{1}{L}
\sum_{q=1}^{L}
\left\lvert
\mathbf r_{n-q}
-
\mathbf r'_{n-q}
\right\rvert^{2}
\right]^{1/2}.
\end{aligned}
\]

This definition deliberately separates physical storage from computational use. A wave field may contain recoverable information about a previous trajectory without yet implementing a symbolic computation. Memory is a property of the physical state $H_n$; computation will require an additional abstraction or readout structure that preserves transition relations.

\subsection{Dynamically readable memory}

The droplet does not decode the full trajectory explicitly. Instead, it reads the wave field locally through the slope at its present position. Define the local wave readout

\[
\mathbf g_n
=
\nabla H_n(\mathbf r_n).
\]

Using the memory expansion,

\begin{equation}
\mathbf g_n
=
\sum_{q=1}^{n}
\lambda^{q-1}
\sigma_{n-q}
\nabla
\psi_{\mathbf r_{n-q}}
(\mathbf r_n).
\label{eq:S2-readout}
\end{equation}

Thus the current horizontal kick depends on the entire recent history through a weighted sum of gradients generated by previous impact locations:


\[
\mathbf F^{\mathrm{wave}}_n 
= 
-C\,\sigma_n\mathbf g_n 
= 
-C\,\sigma_n
\sum_{q=1}^{n}
\lambda^{q-1}
\sigma_{n-q}
\nabla
\psi_{\mathbf r_{n-q}}(\mathbf r_n).
\]

This is the dynamically relevant form of memory. The field does not merely contain a passive record of the past; its local gradient at the droplet position enters the next physical transition,

\[
\begin{aligned}
\mathbf v_{n+1}
&= a\,\mathbf v_n \\
&\quad - \kappa\,\sigma_n \nabla H_n(\mathbf r_n) \\
&\quad - \chi\,\nabla V(\mathbf r_n) \\
&\quad + \boldsymbol{\xi}_n.
\end{aligned}
\]

Hence the walker realizes a physical memory-feedback loop:

\[
\mathcal T_n^{(L)}
\longrightarrow
H_n
\longrightarrow
\nabla H_n(\mathbf r_n)
\longrightarrow
(\mathbf r_{n+1},\mathbf v_{n+1}).
\]

The wave field stores the recent trajectory, and the droplet reads a local functional of that stored trajectory.

\subsection{Wave energy functional}

To quantify storage and erasure, define the normalized wave energy

\begin{equation}
E_W[n]
=
\frac{
|H_n|_{L^2(\Omega)}^2
}{
|\psi_{\mathbf 0}|_{L^2(\Omega)}^2
}.
\label{eq:S2-energy}
\end{equation}

Equivalently,

\[
E_W[n]
=
\frac{
\int_{\Omega}
H_n(\rho)^2\,d\rho
}{
\int_{\Omega}
h_0^2J_0(k_F|\rho|)^2\,d\rho
}.
\]

The denominator normalizes the wave energy by the energy of a single impact-generated source. This is the stroboscopic analogue of the normalized wave-energy measure used to quantify the erasing process.

The energy is not simply the sum of the energies of individual sources, because the sources interfere. Expanding the norm gives

\[
|H_n|_{L^2}^2
=
\sum_{q,q'}
\lambda^{q-1}
\lambda^{q'-1}
\sigma_{n-q}
\sigma_{n-q'}
\left\langle
\psi_{\mathbf r_{n-q}},
\psi_{\mathbf r_{n-q'}}
\right\rangle,
\]

where

\[
\left\langle f,g\right\rangle
=
\int_{\Omega} f(\rho)g(\rho)\,d\rho.
\]

The cross terms

\[
\left\langle
\psi_{\mathbf r_{n-q}},
\psi_{\mathbf r_{n-q'}}
\right\rangle
\]

encode constructive or destructive interference among stored wave sources. This interference is what allows the same physical reservoir to store trajectory information and later erase it through phase-opposed writing.

\subsection{The $\pi$-shift as a phase intervention}

Let $n_\pi$ denote the bounce at which the $\pi$-shift intervention is imposed. Before the intervention, assume the droplet is locked to phase

\[
\sigma_n = \sigma_+
\qquad
\text{for }
n<n_\pi.
\]

The phase intervention changes the subsequent impact phase to

\[
\sigma_n=-\sigma_+
\qquad
\text{for }
n\geq n_\pi.
\]

As a map on the stroboscopic state, the phase intervention may be written

\begin{equation}
P_\pi:
(\mathbf r,\mathbf v,\sigma,H)
\mapsto
(\mathbf r,\mathbf v,-\sigma,H).
\label{eq:S2-pishift}
\end{equation}

The stored field $H$ is not instantaneously destroyed by the perturbation. Rather, the droplet begins to interact with the preexisting field with the opposite phase relation. Since the wave-mediated force is

\[
-C\,\sigma_n\nabla H(\mathbf r),
\]

and the intervention reverses \(\sigma_n\) while leaving \(H\) unchanged, the kick exerted by the preexisting field reverses sign:

\[
-C\,\sigma_+\nabla H(\mathbf r)
\mapsto
+C\,\sigma_+\nabla H(\mathbf r).
\]





This accounts for the initial velocity reversal. The deeper effect is that subsequent impacts write new wave sources with opposite phase relative to the previously stored field.

\subsection{Old and new field decomposition after the phase shift}

For $m\geq 0$, define the wave field $m$ bounces after the phase shift:

\[
H_{n_\pi+m}.
\]

Using the wave update, decompose this field into an old component and a newly written component:

\[
H_{n_\pi+m}
=
H_{\mathrm{old}}^{(m)}
+
H_{\mathrm{new}}^{(m)}.
\]

The old component is simply the pre-shift field after $m$ steps of passive decay:

\[
H_{\mathrm{old}}^{(m)}
=
\lambda^m H_{n_\pi}.
\]

The new component is the contribution of post-shift impacts:

\[
H_{\mathrm{new}}^{(m)}
=
-\sigma_+
\sum_{\ell=0}^{m-1}
\lambda^{m-1-\ell}
\psi_{\mathbf r_{n_\pi+\ell}}.
\]

Therefore,

\begin{equation}
H_{n_\pi+m}
=
\underbrace{\lambda^{m} H_{n_\pi}}_{\text{decayed memory}}
-
\underbrace{
\sigma_{+}
\sum_{\ell=0}^{m-1}
\lambda^{\,m-1-\ell}\,
\psi_{\mathbf r_{n_\pi+\ell}}
}_{\text{new impacts}}.
\label{eq:S2-decomp}
\end{equation}

The negative sign is the mathematical expression of phase-opposed writing. The new sources are not merely additional memory traces; they are written with the opposite phase and therefore can destructively interfere with the old field.

\subsection{Backtracking and wave cancellation}

Suppose that, for a finite time after the phase shift, the droplet approximately retraces its previous path. Define the backtracking error

\[
d_{\mathrm{rev}}(m)
=
\left\lvert
\mathbf r_{n_\pi+m}
-
\mathbf r_{n_\pi-m}
\right\rvert.
\]

We say that the system backtracks over a horizon $m_\ast$ at spatial tolerance $\varepsilon_{\mathrm{rev}}$ if

\[
d_{\mathrm{rev}}(m)
<
\varepsilon_{\mathrm{rev}}
\qquad
\text{for }
1\leq m\leq m_\ast.
\]

During such a backtracking interval, the post-shift impacts occur near locations previously visited by the droplet. Since the post-shift sources have the opposite phase, the newly written field overlaps spatially with the old field but contributes with opposite sign. This produces destructive interference.

The energy after $m$ post-shift impacts is

\[
E_W[n_\pi+m]
=
\frac{
\left\lVert
H_{\mathrm{old}}^{(m)}
+
H_{\mathrm{new}}^{(m)}
\right\rVert_{L^2}^{2}
}{
\left\lVert \psi_{\mathbf 0} \right\rVert_{L^2}^{2}
}.
\]

Expanding,

\[
E_W[n_\pi+m]
=
\frac{
\left|
H_{\mathrm{old}}^{(m)}
\right|^2
+
\left|
H_{\mathrm{new}}^{(m)}
\right|^2
+
2
\left\langle
H_{\mathrm{old}}^{(m)},
H_{\mathrm{new}}^{(m)}
\right\rangle
}{
|\psi_{\mathbf 0}|^2
}.
\]

Wave erasure occurs when the interference term is sufficiently negative:

\[
2
\left\langle
H_{\mathrm{old}}^{(m)},
H_{\mathrm{new}}^{(m)}
\right\rangle
+
\left|
H_{\mathrm{new}}^{(m)}
\right|^2
<
0.
\]

Equivalently,

\begin{equation}
\left|
H_{\mathrm{old}}^{(m)}
+
H_{\mathrm{new}}^{(m)}
\right|^2
<
\left|
H_{\mathrm{old}}^{(m)}
\right|^2.
\label{eq:S2-erase}
\end{equation}

This inequality gives a precise mathematical meaning to wave erasure: the newly written, phase-opposed field reduces the energy of the preexisting memory field more than would be expected from passive decay alone.

\subsection{Erasure efficiency}

Define the erasure efficiency over $m$ post-shift impacts by

\begin{equation}
\eta_{\mathrm{erase}}(m)
=
1
-
\frac{
\left\lVert
H_{\mathrm{old}}^{(m)}
+
H_{\mathrm{new}}^{(m)}
\right\rVert_{L^2}^{2}
}{
\left\lVert
H_{\mathrm{old}}^{(m)}
\right\rVert_{L^2}^{2}
}.
\label{eq:S2-eta}
\end{equation}

Thus

\[
\eta_{\mathrm{erase}}(m)>0
\]

means that post-shift writing has removed wave energy relative to passive decay of the old field. If

\[
\eta_{\mathrm{erase}}(m)=1,
\]

the old memory has been perfectly canceled in the $L^2$-energy sense. In practice, exact cancellation is not expected because the trajectory is only approximately retraced, the source amplitudes are not exactly matched, and the hydrodynamic system is noisy.

One may also normalize by the wave energy at the moment of the phase shift:

\[
\widetilde E_W(m)
=
\frac{
E_W[n_\pi+m]
}{
E_W[n_\pi]
}.
\]

The experimentally observed erasing interval corresponds to

\[
\widetilde E_W(m)<1
\]

for a finite range of $m$, with a minimum near the time over which the droplet backtracks.

\subsection{Finite-time read--erase reversibility}

The phase shift does not make the dissipative system globally time reversible. Rather, it produces finite-time trajectory reversal by exploiting the physical memory stored in the wave field. We can define this as follows.

Let

\[
\pi_{\mathbf r}:X\to \Omega
\]

be the projection from the full physical state to droplet position. The $\pi$-shift produces an approximate trajectory-level reversal over horizon $m_\ast$ if

\begin{equation}
\left|
\pi_{\mathbf r}\!\left(F^{m}(P_\pi x_{n_\pi})\right)
-
\pi_{\mathbf r}\!\left(x_{n_\pi-m}\right)
\right|
<
\varepsilon_{\mathrm{rev}}.
\label{eq:S2-reversal}
\end{equation}



for

\[
1\leq m\leq m_\ast
\]

This is weaker than global reversibility. The map $F$ need not be invertible on the full physical state space $X$. The condition only states that, along a particular memory-endowed trajectory, the phase-shifted dynamics causes the droplet to revisit previous positions for a finite time.

The reversal and erasure are coupled. The same stored wave field that guides the droplet backward is progressively canceled by the post-shift impacts. Thus the system reads its memory backward while destroying that memory:

\[
\begin{aligned}
\text{stored wave memory}
&\longrightarrow
\text{backward guidance}
\\
&\longrightarrow
\text{phase-opposed rewriting}
\\
&\longrightarrow
\text{memory erasure}.
\end{aligned}
\]

This coupling explains why the reversal is transient. Once the old field has been sufficiently erased, the droplet is no longer guided by the previous trajectory. The system then begins to form a new wave field, and the path diverges from the reversed trajectory.

\subsection{Memory and erasure are physical, but not yet computation}

The formal structure above establishes two points.

First, the wave field is a genuine physical memory reservoir. It stores a finite-resolution, exponentially weighted trace of previous droplet impacts:

\[
\mathcal T_n^{(L)}
\longmapsto
H_n.
\]

Second, the $\pi$-shift induces physical erasure by causing subsequent impacts to write phase-opposed wave sources into the same spatial reservoir:

\[
H_{\mathrm{old}}^{(m)}
+
H_{\mathrm{new}}^{(m)}
\approx
0
\]

over the portion of the field corresponding to the retraced trajectory.

However, neither memory nor erasure alone is sufficient to establish autonomous computation. At this stage, we have defined a physical memory field, a local readout through the wave slope, and an externally triggered erasing operation. We have not yet introduced an abstract state space, a robust coarse-graining, a symbolic transition map, or a physical readout layer whose states causally select future operations.

Those additional structures will be introduced next. The purpose of the present section is therefore to anchor the analysis in the physics: the walker stores recent trajectory information in an extended wave field and can erase that information through phase-opposed writing during finite-time backtracking.

\section{Computation as coarse-grained transition preservation}
\label{sec:S3}

The previous sections defined the walker as a stroboscopic physical dynamical system and showed that its wave field stores and can erase a finite-memory trace of previous droplet impacts. We now introduce the first formal criterion for when a physical process may be interpreted as implementing a computation.

At this stage we do not yet require autonomy. That is, we do not require the system to contain an internal readout layer that selects its own future operations. We only ask a more basic question: given a physical dynamical system and an abstract description, when does the physical evolution preserve the transition structure of the abstract system?

The answer is not a bijection between physical microstates and computational states. Physical computation generally does not require one physical state for one abstract symbol. Many physical microstates may instantiate the same abstract state. What is required is that the physical dynamics be compatible with the abstract dynamics under a coarse-graining.

\subsection{Physical and abstract state spaces}

Let

\[
X
\]

denote the physical state space of the system. For the stroboscopic walker model,

\[
X
=
\Omega
\times
\mathbb{R}^2
\times
\{+1,-1\}
\times
\mathcal H,
\]

with physical states

\[
x_n =
(\mathbf r_n,\mathbf v_n,\sigma_n,H_n).
\]

The physical dynamics is a map

\[
F:X\to X,
\]

so that

\[
x_{n+1}=F(x_n).
\]

For the deterministic walker model, $F$ is the stroboscopic update consisting of the droplet position update, velocity update, phase update, and wave-memory update. In a stochastic description, $F$ may be replaced by a transition kernel, but for now we work with the deterministic case.

Let

\[
A
\]

denote an abstract state space. Depending on the intended abstraction, $A$ may be finite, countable, or continuous. For example, $A$ may be:

\[
A=\{0,1\}
\]

for a binary abstraction,

\[
A=\{a_1,\ldots,a_K\}
\]

for a finite-state abstraction, or

\[
A\subseteq \mathbb{R}^d
\]

for a low-dimensional continuous abstraction.

An abstract computation or abstract transition system is specified by a map

\[
G:A\to A,
\]

so that

\[
a_{n+1}=G(a_n).
\]

The central question is whether the physical dynamics $F$ implements the abstract dynamics $G$ under a physically meaningful abstraction.

\subsection{Coarse-graining map}

A coarse-graining is a map

\[
\Pi:X\to A.
\]

For a physical state $x\in X$, the abstract state represented by $x$ is

\[
a=\Pi(x).
\]

The map $\Pi$ is generally many-to-one. That is, several distinct physical states may instantiate the same abstract state. The preimage of an abstract state $a\in A$,

\[
\Pi^{-1}(a)
=
\{x\in X:\Pi(x)=a\},
\]

is the physical realization class of $a$. These realization classes may correspond to macrostates, equivalence classes, experimentally indistinguishable states, or regions of physical state space that are treated as the same abstract state.

The coarse-graining $\Pi$ induces an equivalence relation on $X$:

\[
x\sim_\Pi x'
\quad
\Longleftrightarrow
\quad
\Pi(x)=\Pi(x').
\]

Thus $A$ can be regarded as a quotient of the physical state space,

\[
A \simeq X/{\sim_\Pi},
\]

at least on the part of $X$ where the abstraction is defined.

This quotient perspective is important. It makes clear that physical computation does not require a one-to-one mapping between physical and abstract states. Instead, computation is possible when the physical dynamics descends consistently to the quotient.

\subsection{Transition preservation}

The physical system implements the abstract transition $G$ on a domain

\[
U\subseteq X
\]

if the following diagram commutes for all $x\in U$:

\[
\Pi(F(x))
=
G(\Pi(x)).
\]

Equivalently,

\begin{equation}
\Pi\circ F
=
G\circ \Pi
\qquad
\text{on } U.
\label{eq:S3-commute}
\end{equation}

Diagrammatically,

\[
\begin{tikzcd}
X \arrow[r,"F"] \arrow[d,"\Pi"'] & X \arrow[d,"\Pi"] \\
A \arrow[r,"G"'] & A
\end{tikzcd}
\]


commutes on $U$.

This condition says that evolving physically and then abstracting gives the same result as abstracting first and then applying the abstract transition. In words:

\[
\begin{aligned}
\text{physical evolution followed by abstraction} \\
= \text{abstract transition applied to the abstraction}.
\end{aligned}
\]

This is the minimal transition-preservation criterion.

\subsection{Well-defined quotient dynamics}

The transition-preservation condition can also be expressed in terms of equivalence classes. Suppose $A=X/{\sim_\Pi}$. We would like to define an abstract dynamics on equivalence classes by

\[
G([x])=[F(x)],
\]

where

\[
[x]=\{x'\in X:x'\sim_\Pi x\}.
\]

For this definition to be well-defined, the choice of representative $x$ must not matter. Therefore, we require

\[
x\sim_\Pi x'
\quad
\Longrightarrow
\quad
F(x)\sim_\Pi F(x')
\]

for all relevant $x,x'\in U$.

Equivalently,

\begin{equation}
\Pi(x)=\Pi(x')
\quad
\Longrightarrow
\quad
\Pi(F(x))=\Pi(F(x')).
\label{S3-welldef}
\end{equation}

If this condition holds, then $F$ respects the coarse-graining $\Pi$, and an abstract map $G$ is induced on the quotient. This gives a concise mathematical criterion:

\[
\left\{
\begin{aligned}
F \text{ implements an abstract transition on } X/{\sim_\Pi} \\
\text{iff}\quad F \text{ preserves } \Pi\text{-equivalence classes.}
\end{aligned}
\right.
\]


This is the precise replacement for a bijective physical-symbol mapping.

\subsection{Proposition: transition preservation induces abstract dynamics}

\paragraph{Proposition.}
Let $F:X\to X$ be a physical transition map and let $\Pi:X\to A$ be a surjective coarse-graining. If

\[
\Pi(x)=\Pi(x')
\quad
\Longrightarrow
\quad
\Pi(F(x))=\Pi(F(x'))
\]

for all $x,x'\in U\subseteq X$, then there exists a unique map

\[
G:\Pi(U)\to A
\]

such that

\begin{equation}
\Pi(F(x))=G(\Pi(x))
\label{eq:S3-induced}
\end{equation}

for all $x\in U$.

\paragraph{Proof.}
For any $a\in \Pi(U)$, choose $x\in U$ such that

\[
\Pi(x)=a.
\]

Define

\[
G(a)=\Pi(F(x)).
\]

We must show that this definition does not depend on the choice of $x$. Suppose $x'\in U$ is another physical state with

\[
\Pi(x')=a.
\]

Then

\[
\Pi(x)=\Pi(x'),
\]

and by the assumed equivalence-preservation condition,

\[
\Pi(F(x))=\Pi(F(x')).
\]

Therefore $G(a)$ is independent of the chosen representative. Hence $G$ is well-defined.

Uniqueness follows immediately. If another map $G'$ satisfies

\[
\Pi(F(x))=G'(\Pi(x))
\]

for all $x\in U$, then for any $a=\Pi(x)\in\Pi(U)$,

\[
G'(a)
=
G'(\Pi(x))
=
\Pi(F(x))
=
G(\Pi(x))
=
G(a).
\]

Thus $G'=G$. $\square$

This proposition clarifies the role of the coarse-graining. The abstract computation is not arbitrarily imposed on the physical system. It is induced by the physical dynamics only when the physical dynamics respects the equivalence classes defined by the abstraction.

\subsection{Exact and approximate transition preservation}

The exact condition

\[
\Pi(F(x))=G(\Pi(x))
\]

is appropriate for idealized deterministic systems. Physical systems, however, often contain noise, unresolved degrees of freedom, and finite measurement precision. We therefore distinguish exact transition preservation from approximate transition preservation.

Let $d_A$ be a metric on the abstract state space $A$. The system approximately implements $G$ on $U$ with error tolerance $\varepsilon_A$ if

\[
d_A
\left(
\Pi(F(x)),
G(\Pi(x))
\right)
\leq
\varepsilon_A
\]

for all $x\in U$.

For a finite-state abstraction, this metric may be replaced by a mismatch indicator:

\[
\mathbf 1\left[
\Pi(F(x))\neq G(\Pi(x))
\right].
\]

In that case exact transition preservation means zero mismatch. Approximate or noisy transition preservation will later be treated in terms of probabilistic basin stability. For the present stage, the key point is only that computation requires preservation of transition structure, not a one-to-one physical-symbol encoding.

\subsection{Finite-time transition preservation}

Some physical systems may preserve an abstract transition structure only over a finite time horizon. This is especially relevant for dissipative memory systems such as the walker, where the wave field stores recent history only for a finite number of bounces.

For a horizon $T\in\mathbb{N}$, the physical system implements $G$ for $T$ steps on $U$ if

\begin{equation}
\Pi(F^m(x))
=
G^m(\Pi(x))
\label{eq:S3-finitetime}
\end{equation}

for all

\[
x\in U
\]

and all

\[
m=0,1,\ldots,T.
\]

Here $F^m$ denotes $m$-fold composition of the physical map, and $G^m$ denotes $m$-fold composition of the abstract map.

Approximate finite-time implementation is defined by

\begin{equation}
d_A
\left(
\Pi(F^m(x)),
G^m(\Pi(x))
\right)
\leq
\varepsilon_A
\label{eq:S3-approx}
\end{equation}

for

\[
m=0,1,\ldots,T.
\]

This distinction is important because a physical system may exhibit a transient computational structure without supporting that structure indefinitely. In the walker, the finite lifetime of the wave memory naturally limits any transition structure that depends on the stored wave field.

\subsection{Controlled physical transitions}

The physical system may also be subject to externally selected operations. Let

\[
\mathcal O
\]

be a set of physical operations or interventions. For each

\[
o\in \mathcal O,
\]

let

\[
F_o:X\to X
\]

be the corresponding physical transition. In the walker, examples include normal evolution, a phase-shift intervention, or a change in confinement.

Similarly, let

\[
\mathcal U
\]

be a set of abstract operations. For each

\[
u\in\mathcal U,
\]

let

\[
G_u:A\to A
\]

be the corresponding abstract transition.

A controlled physical system implements the abstract controlled transition system if there is a map

\[
\Theta:\mathcal O\to \mathcal U
\]

such that

\begin{equation}
\Pi(F_o(x))
=
G_{\Theta(o)}(\Pi(x))
\label{eq:S3-controlled}
\end{equation}

for all

\[
x\in U
\]

and all relevant operations

\[
o\in\mathcal O.
\]

This describes externally selected computation. The control operation is chosen from outside the system, but the physical transition still preserves the intended abstract transition under $\Pi$.

This is distinct from autonomous computation. In autonomous computation, the system itself must contain a physical readout or control state that selects which $F_o$ occurs next. That additional closure condition is not assumed here and will be introduced later.

\subsection{Application to the stroboscopic walker}

For the walker, the physical state is

\[
x_n =
(\mathbf r_n,\mathbf v_n,\sigma_n,H_n),
\]

and the physical transition is

\[
x_{n+1}=F(x_n).
\]

A coarse-graining

\[
\Pi:X\to A
\]

may extract different possible abstract descriptions from the walker. For example, $\Pi$ may assign abstract states according to:

\[
\Pi(x_n)
=
\text{region of droplet position},
\]

or

\[
\Pi(x_n)
=
\text{sign or magnitude class of } \nabla H_n(\mathbf r_n),
\]

or

\[
\Pi(x_n)
=
\text{coarse class of the wave-memory field } H_n.
\]

Each choice of $\Pi$ defines a different candidate abstraction. But the mere existence of such a map does not establish computation. The abstraction defines a computation only if there exists an abstract transition map

\[
G:A\to A
\]

such that

\[
\Pi(F(x))
=
G(\Pi(x))
\]

on the relevant physical domain.

Thus, for the walker, a claim of computation must identify:

\[
X
\]

the physical state space;

\[
F
\]

the physical transition map;

\[
A
\]

the abstract state space;

\[
\Pi:X\to A
\]

the coarse-graining from physical states to abstract states;

\[
G:A\to A
\]

the abstract transition rule;

and the domain

\[
U\subseteq X
\]

on which the transition-preservation condition holds.

Without these ingredients, a proposed computational interpretation remains only an observer-side description of physical dynamics.

\subsection{Memory does not imply transition preservation}

The wave field may store a recoverable trajectory trace,

\[
\mathcal T_n^{(L)}
\longmapsto
H_n,
\]

and the droplet may dynamically read that trace through

\[
\nabla H_n(\mathbf r_n).
\]

However, this does not by itself define an abstract computation. To obtain a computation, one must specify a coarse-graining $\Pi$ and show that the induced abstract state transitions are preserved by the physical dynamics.

For example, suppose we define an abstraction that assigns to each physical state the coarse region of the bath occupied by the droplet:

\[
\Pi(x_n)=a_i
\quad
\text{if}
\quad
\mathbf r_n\in R_i.
\]

This becomes a computation only if there is a transition rule

\[
G(a_i)=a_j
\]

such that whenever

\[
\mathbf r_n\in R_i,
\]

the physical dynamics reliably carries the system to a state satisfying

\[
\mathbf r_{n+1}\in R_j.
\]

If two physical states have the same abstract state but their next physical states map to different abstract states, then the abstraction fails to support a well-defined computation.

In quotient terms, if

\[
\Pi(x)=\Pi(x')
\]

but

\[
\Pi(F(x))\neq \Pi(F(x')),
\]

then the physical dynamics does not descend to a well-defined abstract transition on the quotient $X/{\sim_\Pi}$.

\subsection{Relation to physical computation}

This coarse-grained criterion separates three notions that are often conflated.

First, a system may possess physical memory: its current state contains information about its past.

Second, a system may admit an observer-defined abstraction: an external observer may label regions of its state space as abstract symbols.

Third, a system may implement an abstract transition structure: the physical dynamics may preserve an abstract transition map under the coarse-graining.

Only the third condition is computation in the transition-preserving sense used here.

Thus, at this stage, we define physical implementation of an abstract computation as:

\begin{equation}
\boxed{
\Pi\circ F
=
G\circ \Pi
\quad
\text{on } U\subseteq X.
}
\label{S3-criterion}
\end{equation}

This is a dynamical version of physical-to-abstract compatibility. It does not require bijection. It does not yet require autonomy. It only requires that the physical transition be compatible with the abstract transition under the chosen coarse-graining.

The next step is to strengthen this criterion for real physical systems by adding robustness: abstract states must correspond to distinguishable physical regions, and physical evolution must carry those regions to the appropriate successor regions despite noise and perturbations.

\section{Robust physical symbolization: basins, noise, and failure modes}
\label{sec:S4}

The transition-preservation criterion,

\[
\Pi\circ F = G\circ \Pi ,
\]

defines when a physical dynamics $F:X\to X$ induces a well-defined abstract dynamics $G:A\to A$ under a coarse-graining $\Pi:X\to A$. This criterion is exact and structural. However, real physical systems are noisy, finite-resolution, and subject to perturbations. A physically meaningful computation therefore cannot rely on arbitrarily thin partitions of state space or infinitely precise distinctions between physical states.

We now strengthen the coarse-grained criterion by introducing robust physical symbolization. The central idea is that abstract states must correspond to separated physical regions, and the physical dynamics must carry those regions into the appropriate successor regions with high probability. This section does not yet introduce autonomy. The operation $F$ may still be externally selected. The purpose here is only to define what it means for an abstract transition structure to be physically robust.

\subsection{Physical metric and finite-resolution distinguishability}

Let

\[
X
\]

be the physical state space. We assume that $X$ is equipped with a metric

\[
d_X:X\times X\to \mathbb{R}_{\geq 0},
\]

or, more generally, a physically meaningful notion of distance or distinguishability. For the walker,

\[
x=
(\mathbf r,\mathbf v,\sigma,H),
\]

so a natural metric may take the form

\[
d_X(x,x')^2
=
\alpha_r|\mathbf r-\mathbf r'|^2
+
\alpha_v|\mathbf v-\mathbf v'|^2
+
\alpha_\sigma |\sigma-\sigma'|^2
+
\alpha_H|H-H'|_{\mathcal H}^2,
\]

where the positive coefficients

\[
\alpha_r,\alpha_v,\alpha_\sigma,\alpha_H
\]

set the relative physical scales of position, velocity, phase, and wave-field differences. The exact metric is not unique; what matters is that it reflects experimentally or dynamically meaningful resolution.

Let

\[
\varepsilon_X>0
\]

denote the physical uncertainty scale. This may represent measurement precision, thermal or hydrodynamic noise, unresolved degrees of freedom, or the smallest perturbation scale that cannot be reliably controlled.

For a subset

\[
B\subset X,
\]

define its $\varepsilon_X$-neighborhood by

\[
N_{\varepsilon_X}(B)
=
\{x\in X:\operatorname{dist}(x,B)\leq \varepsilon_X\},
\]

where

\[
\operatorname{dist}(x,B)
=
\inf_{y\in B} d_X(x,y).
\]

Two physical regions $B_i,B_j\subset X$ are distinguishable at resolution $\varepsilon_X$ if their $\varepsilon_X$-neighborhoods do not overlap:

\[
N_{\varepsilon_X}(B_i)
\cap
N_{\varepsilon_X}(B_j)
=
\varnothing .
\]

Equivalently, a sufficient condition is

\[
\operatorname{dist}(B_i,B_j)>2\varepsilon_X,
\]

where

\[
\operatorname{dist}(B_i,B_j)
=
\inf_{x\in B_i,\,y\in B_j}
d_X(x,y).
\]

This replaces a purely dimensional requirement with a physically meaningful separability requirement. A high-dimensional state space does not by itself provide usable abstract states. What matters is the number of reliably distinguishable regions available at the relevant noise scale.

\subsection{Basins as physical realizers of abstract states}

Let the abstract state space be finite,

\[
A=\{a_1,\ldots,a_K\}.
\]

A robust physical realization of $A$ is a collection of physical regions

\[
\mathcal B
=
\{B_1,\ldots,B_K\},
\qquad
B_i\subset X,
\]

such that

\[
B_i
\]

realizes abstract state

\[
a_i.
\]

The coarse-graining map

\[
\Pi:X\to A
\]

is then defined on the union

\[
U=\bigcup_{i=1}^K B_i
\]

by

\[
\Pi(x)=a_i
\qquad
\text{if}
\qquad
x\in B_i.
\]

In this formulation,

\[
B_i=\Pi^{-1}(a_i)
\]

on the relevant domain. The regions $B_i$ may be attractor basins, metastable states, thresholded readout regions, symbolic memory states, or other physically stable macrostates.

The regions $B_i$ should not be understood as exact mathematical points. A physical symbol is not a microstate; it is a finite-volume region of state space. This is what gives the abstraction physical robustness.

\subsection{Separation condition}

A necessary condition for robust symbolization is that the basins representing distinct abstract states are separated:

\[
\Delta_{ij}
=
\operatorname{dist}(B_i,B_j)
>
2\varepsilon_X
\qquad
i\neq j.
\]

Define the minimal basin separation

\[
\Delta_{\min}
=
\min_{i\neq j}
\Delta_{ij}.
\]

Then robust distinguishability requires

\[
\Delta_{\min}>2\varepsilon_X.
\]

The margin

\[
m_{\mathrm{sep}}
=
\frac{1}{2}\Delta_{\min}-\varepsilon_X
\]

measures how much separation remains after accounting for physical uncertainty. A positive margin,

\[
m_{\mathrm{sep}}>0,
\]

means that the abstract states are distinguishable under perturbations of size $\varepsilon_X$.

This condition makes explicit why the old dimensionality argument is insufficient. The relevant quantity is not the dimension of a readout manifold alone, but the number of separated, stable, physically accessible regions at the relevant noise scale. A one-dimensional physical readout can support many symbols if it has many well-separated basins; a high-dimensional readout can support no reliable symbols if its candidate regions are not robustly separable.

\subsection{Robust symbolic capacity}

Let

\[
N_{\varepsilon_X}(U)
\]

denote the maximum number of pairwise $\varepsilon_X$-distinguishable physical regions that can be embedded in a domain $U\subset X$. This is a coarse capacity of the physical readout domain.

A finite abstraction

\[
A=\{a_1,\ldots,a_K\}
\]

can be physically symbolized in $U$ only if

\begin{equation}
N_{\varepsilon_X}(U)\geq K.
\label{eq:S4-capacity}
\end{equation}

Equivalently, if one wants to physically realize an alphabet $\Gamma$, one needs

\[
N_{\varepsilon_X}(U)\geq |\Gamma|.
\]

This is a finite-resolution replacement for dimension-based claims. The number of available symbols is controlled by physically distinguishable basins, not simply by manifold dimension.

One may also express this in information-theoretic form. Let $A$ be the intended abstract state and let $Y$ be the noisy physical readout. Reliable symbolization requires

\[
I(A;Y)
\approx
H(A),
\]

or, for a uniform alphabet,

\[
I(A;Y)
\geq
\log_2 K-\eta,
\]

where $\eta$ is the allowed information loss. This information criterion is optional, but it captures the same idea: the physical readout must preserve enough information to discriminate the intended abstract states.

\subsection{Deterministic basin transition stability}

Let

\[
G:A\to A
\]

be the intended abstract transition rule. Suppose

\[
G(a_i)=a_{g(i)},
\]

where

\[
g:\{1,\ldots,K\}\to\{1,\ldots,K\}
\]

is the index map induced by $G$.

In the deterministic case, the strongest basin-level transition condition is

\[
F(B_i)\subseteq B_{g(i)}
\qquad
\text{for all } i.
\]

This means that every physical state realizing $a_i$ is carried by the physical dynamics into the basin realizing the correct successor abstract state.

A more robust version accounts for perturbations before the transition. We require

\begin{equation}
F\left(N_{\varepsilon_X}(B_i)\right)
\subseteq
B_{g(i)}
\label{eq:S4-basintrans}
\end{equation}

or, more weakly,

\[
F\left(N_{\varepsilon_X}(B_i)\right)
\subseteq
N_{\varepsilon_X}(B_{g(i)}).
\]

The first condition says that perturbations around the source basin are corrected into the target basin. The second condition says that the transition remains correct up to the same physical resolution.

This gives a physically meaningful strengthening of exact commutation:

\[
\Pi(F(x))=G(\Pi(x)).
\]

Instead of requiring this only for ideal points $x\in B_i$, robust implementation requires that it hold in a neighborhood around each basin.

\subsection{Stochastic transition stability}

Real physical systems are often better modeled by a stochastic transition kernel

\[
K(x,dy),
\]

where

\[
K(x,S)
\]

is the probability that the next physical state lies in the set $S\subseteq X$, given current state $x$.

For the noisy walker, this corresponds to replacing

\[
x_{n+1}=F(x_n)
\]

by

\[
x_{n+1}=F(x_n,\xi_n),
\]

where $\xi_n$ represents noise in the impact, phase, wave field, or unresolved hydrodynamic variables.

The abstraction is $(\delta,\varepsilon_X)$-robust for one step if

\[
K(x,B_{g(i)})
\geq
1-\delta
\]

for all

\[
x\in N_{\varepsilon_X}(B_i)
\]

and all

\[
i=1,\ldots,K.
\]

Here

\[
\delta\in[0,1]
\]

is the allowed one-step failure probability. Equivalently,

\[
\mathbb P
\left[
x_{n+1}\in B_{G(a_i)}
\,\middle|\,
x_n=x
\right]
\geq
1-\delta
\qquad
\text{for all } x\in N_{\varepsilon_X}(B_i).
\]

This is the stochastic analogue of basin transition preservation.

\subsection{Confusion matrix of the physical implementation}

A useful empirical diagnostic is the transition confusion matrix. Let

\[
Q_{ij}
=
\mathbb P
\left[
\Pi(x_{n+1})=a_j
\,\middle|\,
\Pi(x_n)=a_i
\right].
\]

If $G(a_i)=a_{g(i)}$, then ideal implementation gives

\[
Q_{i,g(i)}=1
\]

and

\[
Q_{ij}=0
\qquad
j\neq g(i).
\]

A noisy but robust implementation satisfies

\[
Q_{i,g(i)}\geq 1-\delta
\]

for all $i$. The off-target probabilities

\[
Q_{ij},
\qquad
j\neq g(i),
\]

measure transition errors.

This matrix makes the criterion experimentally testable. One can initialize the physical system in each basin $B_i$, repeatedly apply the physical operation, and estimate the probability that the resulting state lands in the correct target basin.

\subsection{Multi-step reliability}

If the one-step transition error is at most $\delta$, then the probability of remaining correct for $T$ steps can be bounded. Under a union-bound estimate,

\[
\mathbb P
\left[
\Pi(x_m)=G^m(\Pi(x_0))
\text{ for all } m=1,\ldots,T
\right]
\geq
1-T\delta.
\]

A sharper bound may be obtained if the transition errors are independent:

\[
\mathbb P
\left[
\Pi(x_T)=G^T(\Pi(x_0))
\right]
\geq
(1-\delta)^T.
\]

For small $\delta$,

\[
(1-\delta)^T
\approx
e^{-T\delta}.
\]

Thus even small one-step errors accumulate over time. This matters for wave-memory systems because memory decay, phase noise, and trajectory divergence naturally limit the time over which any abstract transition structure can remain reliable.

A finite-time robust implementation over horizon $T$ therefore requires

\[
T\delta \ll 1
\]

or, equivalently,

\[
\delta \ll \frac{1}{T}.
\]

\subsection{Basin stability}

For continuous physical systems, it may be unrealistic to require that every point in $B_i$ transition correctly. Some points may lie near basin boundaries or represent rare states. We therefore define a measure-based version.

Let

\[
\mu_i
\]

be a probability measure over basin $B_i$, representing the physically relevant distribution of initial states that realize $a_i$. The one-step basin stability of state $a_i$ is

\[
S_i
=
\int_{B_i}
K(x,B_{g(i)})\,d\mu_i(x).
\]

The abstraction is $\delta$-stable with respect to the basin measures ${\mu_i}$ if

\[
S_i\geq 1-\delta
\qquad
\text{for all } i.
\]

This measure-based condition is often the most practical one experimentally. It does not require controlling every possible microstate in $B_i$. Instead, it asks whether typical physical realizations of the abstract state are carried to the correct successor basin.

\subsection{Failure mode 1: aliasing}

Aliasing occurs when physically distinct states that should correspond to different abstract states are assigned the same readout state, or when distinct computationally relevant states cannot be separated at the physical noise scale.

Formally, aliasing occurs if there exist

\[
x,x'\in X
\]

such that the intended abstraction requires

\[
\Pi_{\mathrm{intended}}(x)\neq
\Pi_{\mathrm{intended}}(x'),
\]

but the physical readout or available partition gives

\[
\Pi_{\mathrm{phys}}(x)=
\Pi_{\mathrm{phys}}(x').
\]

In basin language, aliasing means that two intended basins overlap at the relevant resolution:

\[
N_{\varepsilon_X}(B_i)
\cap
N_{\varepsilon_X}(B_j)
\neq
\varnothing
\qquad
i\neq j.
\]

In such a case, no physical process can reliably distinguish the two abstract states without additional degrees of freedom, improved measurement, stronger attractor separation, or a different physical encoding.

\subsection{Failure mode 2: representative dependence}

A coarse-graining fails to define a computation if two physical states in the same abstract basin evolve into different abstract basins. That is, for some

\[
x,x'\in B_i,
\]

we have

\[
F(x)\in B_j,
\qquad
F(x')\in B_k,
\qquad
j\neq k.
\]

Equivalently,

\[
\Pi(x)=\Pi(x')
\]

but

\begin{equation}
\Pi(F(x))\neq \Pi(F(x')).
\label{eq:S4-repdep}
\end{equation}

This is the failure of quotient dynamics. The abstract transition is not well-defined because the next abstract state depends on physical details that were discarded by the coarse-graining.

Representative dependence is especially important in high-dimensional physical reservoirs. A coarse label may appear meaningful, but if unresolved reservoir degrees of freedom alter the next transition, the abstraction does not support a well-defined computation.

\subsection{Failure mode 3: noise-induced basin crossing}

Even if the deterministic dynamics maps

\[
F(B_i)\subseteq B_{g(i)},
\]

noise may push trajectories across basin boundaries. Let the noisy transition be

\[
x_{n+1}=F(x_n)+\xi_n
\]

in local coordinates. If the distance from $F(x)$ to the boundary of the target basin is smaller than the typical noise amplitude, then the abstract transition is unreliable.

Define the target-basin safety margin

\[
m_i(x)
=
\operatorname{dist}
\left(
F(x),
X\setminus B_{g(i)}
\right).
\]

A sufficient condition for robustness under bounded noise

\[
|\xi_n|\leq \varepsilon_X
\]

is

\[
m_i(x)>\varepsilon_X
\qquad
\text{for all } x\in B_i.
\]

If

\[
m_i(x)\leq \varepsilon_X,
\]

then noise can produce an incorrect abstract transition.

\subsection{Failure mode 4: memory decay}

In memory-based physical systems, the basin structure may depend on an internal memory variable that decays. For the walker, the relevant memory variable is the wave field $H_n$, whose past contributions decay as

\[
\lambda^q=e^{-q/M_e}.
\]

If an abstract state depends on distinctions in the wave field that decay below the noise scale, then the corresponding basins merge.

Let two wave-memory states be

\[
H
\quad\text{and}\quad
H'.
\]

They are distinguishable only if

\[
|H-H'|_{\mathcal H}>2\varepsilon_H.
\]

If

\[
|F_H^m(H)-F_H^m(H')|_{\mathcal H}
\leq
2\varepsilon_H
\]

after $m$ steps, where $F_H$ is the wave-memory update, then the physical distinction between the two abstract memory states is lost.

For exponentially decaying memory, this imposes a finite reliable horizon. If the initial separation is

\[
D_0=|H-H'|_{\mathcal H},
\]

then after passive decay,

\[
D_m\approx \lambda^m D_0.
\]

The states become indistinguishable when

\[
\lambda^mD_0\leq 2\varepsilon_H.
\]

Solving for $m$ gives

\begin{equation}
m
\geq
M_e
\log
\left(
\frac{D_0}{2\varepsilon_H}
\right).
\label{eq:S4-decayhorizon}
\end{equation}

Thus the memory parameter $M_e$ sets not only the physical memory horizon, but also the time horizon over which memory-dependent abstract distinctions can remain robust.

\subsection{Failure mode 5: transition deformation}

A physical transition may preserve the intended abstract map for one region of state space but fail elsewhere. This occurs when

\[
\Pi(F(x))=G(\Pi(x))
\]

holds on a restricted domain

\[
U_0\subset X
\]

but not on a larger domain

\[
U\supset U_0.
\]

The implementation is then domain-limited. For chaotic or strongly nonlinear systems, small deviations from the intended region may lead to transition deformation:

\[
\Pi(F(x))\neq G(\Pi(x))
\qquad
x\in U\setminus U_0.
\]

This is not necessarily a defect if the system is only claimed to compute on $U_0$. But it must be stated explicitly. Physical computation is always relative to a domain of valid initialization, control, and operation.

\subsection{Application to the walker}

In the stroboscopic walker, the full physical state is

\[
x_n=
(\mathbf r_n,\mathbf v_n,\sigma_n,H_n).
\]

Different candidate abstractions may choose basins based on droplet position, velocity, wave-field features, local wave slope, orbit class, or phase state. For example, one might define position-region states by partitioning the bath into regions

\[
R_1,\ldots,R_K\subset\Omega
\]

and setting

\[
B_i
=
\{x\in X:\mathbf r\in R_i\}.
\]

Alternatively, one might define wave-readout states by partitioning the possible local slopes

\[
\nabla H(\mathbf r)
\]

into separated regions. Or one might define orbit-level states corresponding to stable orbital modes in a confined walker.

However, none of these abstractions is physically valid merely because it can be written down. Each must satisfy the robustness requirements:

\begin{equation}
\operatorname{dist}(B_i,B_j)>2\varepsilon_X,
\label{eq:S4-separation}
\end{equation}

and

\begin{equation}
K(x,B_{g(i)})\geq 1-\delta
\qquad
x\in N_{\varepsilon_X}(B_i).
\label{eq:S4-kernel}
\end{equation}

For the walker, likely sources of failure include:

\[
\text{hydrodynamic noise},
\]

\[
\text{finite precision of phase control},
\]

\[
\text{chaotic sensitivity to initial conditions},
\]

\[
\text{decay of the wave-memory field},
\]

and

\[
\text{overlap of wave-field configurations under coarse readout}.
\]

Therefore, a claim that the walker implements an abstract transition system must specify not only a coarse-graining $\Pi$, but also the basin separation, noise tolerance, and time horizon over which the transition structure remains valid.

\subsection{Robust transition preservation}

We can now state the robust version of the Stage 3 criterion.

\paragraph{Definition.}
Let $A=\{a_1,\ldots,a_K\}$ be a finite abstract state space, let $G:A\to A$ be an abstract transition map, and let $B_i\subset X$ be physical basins realizing $a_i$. Let $K(x,dy)$ be the physical stochastic transition kernel. The physical system robustly implements $G$ with tolerance $(\varepsilon_X,\delta)$ on $U=\bigcup_iB_i$ if:

\begin{equation}
N_{\varepsilon_X}(B_i)
\cap
N_{\varepsilon_X}(B_j)
=
\varnothing
\qquad
i\neq j,
\label{eq:S4-robustsep}
\end{equation}

and

\begin{equation}
K(x,B_{g(i)})
\geq
1-\delta
\qquad
\text{for all }
x\in N_{\varepsilon_X}(B_i),
\label{eq:S4-robustkernel}
\end{equation}

where

\[
G(a_i)=a_{g(i)}.
\]

In the deterministic limit, this reduces to

\[
F(N_{\varepsilon_X}(B_i))
\subseteq
B_{g(i)}
\]

or, at finite resolution,

\[
F(N_{\varepsilon_X}(B_i))
\subseteq
N_{\varepsilon_X}(B_{g(i)}).
\]

This definition expresses the physical content of computation in the presence of noise. Abstract states must be separated physical basins, and physical transitions must carry each basin into the basin representing the correct abstract successor state.

\subsection{Summary}

The exact condition

\[
\Pi\circ F=G\circ \Pi
\]

is the structural criterion for coarse-grained computation. The robust criterion adds the physical requirements needed for real systems:

\[
\begin{aligned}
&\text{separated basins}
+
\text{noise tolerance}\\
&\quad+
\text{transition stability}
+
\text{finite-time reliability}.
\end{aligned}
\]

This separates robust physical computation from fragile observer-side labeling. A physical system does not compute merely because an observer can assign symbols to its states. It computes, in the present transition-preserving sense, only when those symbols correspond to physically distinguishable basins and the physical dynamics reliably maps those basins according to the abstract transition rule.

The next stage adds a further requirement: autonomy. Robust transition preservation can still be externally controlled. A closed physical computer must also contain an internal physical readout or control variable whose state selects the subsequent physical transition.

\section{Closure and autonomy: physically internalized transition selection}
\label{sec:S5}

The previous two sections defined computation as coarse-grained transition preservation and then strengthened this criterion by requiring robust basins, noise tolerance, and transition stability. These conditions are necessary for physical computation, but they do not yet distinguish externally controlled computation from autonomous physical computation.

A physical system may robustly implement an abstract transition map only because an external agent selects the relevant physical operation at each step. For example, an external experimenter may impose a phase shift, change a boundary condition, alter a potential landscape, or switch the forcing protocol. Such externally selected operations may still preserve an abstract transition structure, but the transition selection is not physically internal to the system.

The aim of this section is to define the missing closure condition. A physical computation is closed, in the sense used here, when the physical state of the system contains a readout or control variable whose state causally selects the next physical operation. The readout is not merely measured by an observer. It is a physical subsystem coupled back into the dynamics.

\subsection{From external control to internal control}

Let

\[
X
\]

be the physical state space. Let

\[
\mathcal O
\]

be a set of physically realizable operations. For each operation

\[
o\in\mathcal O,
\]

let

\[
F_o:X\to X
\]

be the physical transition induced by applying operation $o$.

In an externally controlled physical system, the operation sequence

\[
o_0,o_1,o_2,\ldots
\]

is supplied from outside the system. The physical evolution is

\[
x_{n+1}
=
F_{o_n}(x_n).
\]

The operation $o_n$ may be chosen by an experimenter, an external controller, or an externally programmed schedule. Even if each $F_{o_n}$ robustly preserves an abstract transition, the system is not closed unless the selection of $o_n$ is itself physically generated by the system.

In a closed physical system, the operation is selected by a physical variable internal to the state. We therefore decompose the state as

\begin{equation}
X
=
X_R
\times
X_Y
\times
X_Z.
\label{eq:S5-decomp}
\end{equation}

Here, $X_R$ is the reservoir or memory part of the system, $X_Y$ is the physical readout/control subsystem and $X_Z$ contains the remaining physical degrees of freedom, such as position, velocity, phase, boundary variables, actuator states, or other internal variables.

For the walker, $X_R$ may include the wave field $H_n$, while $X_Z$ may include the droplet position, velocity, and phase. The key additional ingredient is $X_Y$: a physical readout layer whose state can select subsequent operations.

Let

\[
p_Y:X\to X_Y
\]

be the projection onto the readout subsystem. The readout state at time $n$ is

\[
y_n=p_Y(x_n).
\]

A closed physical controller is a map

\[
C:X_Y\to \mathcal O
\]

that assigns a physical operation to each readout state. The closed-loop physical dynamics is then

\begin{equation}
x_{n+1}
=
F_{C(y_n)}(x_n),
\qquad
y_n=p_Y(x_n).
\label{eq:S5-closedloop}
\end{equation}

Equivalently,

\[
x_{n+1}
=
\Phi(x_n),
\]

where the closed physical evolution map is

\begin{equation}
\Phi:X\to X,
\qquad
\Phi(x)
=
F_{C(p_Y(x))}(x).
\label{eq:S5-Phi}
\end{equation}

This equation is the basic closure condition. The operation applied at the next step is not supplied externally. It is selected by the system's own physical readout state.

\subsection{Readout as a physical state, not an observer label}

A coarse-graining

\[
\Pi:X\to A
\]

assigns abstract states to physical states. Such a map may be defined by an observer. By itself, it does not imply that the system contains a physical readout. Closure requires more.

A physical readout state must satisfy three conditions.

First, it must be part of the physical state:

\[
y_n\in X_Y
\subseteq X.
\]

Second, it must be dynamically produced from the system's physical variables. This may occur through an internal readout map

\[
D:X_R\times X_Z\to X_Y,
\]

or more generally through a readout update

\[
y_{n+1}
=
R(x_n,y_n).
\]

Third, it must causally influence the next physical transition by selecting or modulating the operation:

\[
x_{n+1}
=
F_{C(y_n)}(x_n).
\]

Thus a camera measurement, an external classification algorithm, or a post hoc symbolic labeling does not constitute closure unless its result is physically coupled back into the system and changes subsequent dynamics.

\subsection{Autonomous closed computation}

Let

\[
A
\]

be an abstract state space and let

\[
\Pi:X\to A
\]

be a coarse-graining. Let

\[
\mathcal U
\]

be a set of abstract operations, with

\[
G_u:A\to A
\]

for each

\[
u\in\mathcal U.
\]

Let

\[
\Theta:\mathcal O\to \mathcal U
\]

map physical operations to the abstract operations they implement.

A closed physical system implements the abstract controlled transition system if

\[
\Pi(F_{C(p_Y(x))}(x))
=
G_{\Theta(C(p_Y(x)))}(\Pi(x))
\]

for all

\[
x\in U\subseteq X.
\]

This condition combines transition preservation with closure. The physical readout state $p_Y(x)$ selects a physical operation $C(p_Y(x))$, and that physical operation preserves the corresponding abstract transition.

For a fixed abstract transition map

\[
G:A\to A,
\]

the closed system implements $G$ if

\[
\Pi(\Phi(x))
=
G(\Pi(x)),
\qquad
x\in U,
\]

where

\[
\Phi(x)
=
F_{C(p_Y(x))}(x).
\]

The difference from Stage 3 is that the physical transition map is no longer an externally chosen $F$. It is an internally selected closed-loop map $\Phi$.

\subsection{Closed finite-state realization}

For a finite-state abstraction, let

\[
A=\{a_1,\ldots,a_K\}
\]

and let the physical basins realizing these abstract states be

\[
B_1,\ldots,B_K\subset X.
\]

Let the readout/control subsystem have robust basins

\[
Y_1,\ldots,Y_M\subset X_Y.
\]

Each readout basin

\[
Y_\ell
\]

selects an operation

\[
o_\ell=C(Y_\ell)\in\mathcal O.
\]

The physical operation $F_{o_\ell}$ induces an abstract transition

\[
G_\ell:A\to A.
\]

The closed-loop system evolves by

\[
x_{n+1}
=
F_{o_\ell}(x_n)
\qquad
\text{whenever}
\qquad
p_Y(x_n)\in Y_\ell.
\]

The corresponding abstract transition is

\[
a_{n+1}
=
G_\ell(a_n)
\qquad
\text{whenever}
\qquad
p_Y(x_n)\in Y_\ell.
\]

Thus the readout state determines which abstract transition is applied. In deterministic form, robust closure requires

\[
F_{o_\ell}(B_i\cap p_Y^{-1}(Y_\ell))
\subseteq
B_{G_\ell(a_i)}
\]

for all abstract basins $B_i$ and all readout basins $Y_\ell$ for which the intersection is physically reachable.

At finite resolution, one requires

\[
F_{o_\ell}
\left(
N_{\varepsilon_X}(B_i\cap p_Y^{-1}(Y_\ell))
\right)
\subseteq
N_{\varepsilon_X}(B_{G_\ell(a_i)}).
\]

This is the basin-level form of closed transition preservation.

\subsection{Stochastic closed-loop criterion}

For noisy systems, let

\[
K_o(x,dz)
\]

be the stochastic transition kernel associated with physical operation $o$. The closed-loop transition kernel is

\[
K_{\mathrm{cl}}(x,dz)
=
K_{C(p_Y(x))}(x,dz).
\]

The closed-loop system robustly implements the abstract transition structure with one-step error probability $\delta$ if

\[
K_{\mathrm{cl}}(x,B_{G_\ell(a_i)})
\geq
1-\delta
\]

for all

\[
x\in N_{\varepsilon_X}(B_i\cap p_Y^{-1}(Y_\ell)).
\]

Equivalently,

\[
\mathbb P
\left[
x_{n+1}\in B_{G_\ell(a_i)}
\,\middle|\,
x_n=x,\;
p_Y(x_n)\in Y_\ell
\right]
\geq
1-\delta.
\]

This says that, conditional on the physical readout being in basin $Y_\ell$, the resulting operation must carry the current abstract state into the correct successor basin with high probability.

\subsection{Closure of operation selection}

The preceding condition assumes that the readout state reliably selects the intended operation. This also requires robustness.

Let the actual operation applied at step $n$ be a random variable

\[
O_n\in\mathcal O.
\]

The system has $\delta_C$-reliable operation closure if

\begin{equation}
\mathbb P
\left[
O_n=C(y_n)
\,\middle|\,
y_n\in Y_\ell
\right]
\geq
1-\delta_C
\label{eq:S5-opclosure}
\end{equation}

for all readout basins $Y_\ell$.

Equivalently, the operation-selection confusion matrix

\[
M_{\ell o}
=
\mathbb P
\left[
O_n=o
\,\middle|\,
y_n\in Y_\ell
\right]
\]

should satisfy

\[
M_{\ell,C(Y_\ell)}
\geq
1-\delta_C.
\]

If operation selection is deterministic, then

\[
M_{\ell,C(Y_\ell)}=1.
\]

If the operation is externally imposed and not determined by the physical readout state, then $O_n$ is not a function of $y_n$. In that case closure fails, even if the externally imposed operation produces a correct transition.

An information-theoretic quantity that accompanies closure is

\[
H(O_n\mid Y_n).
\]

For deterministic closed operation selection, $O_n=C(Y_n)$ implies

\begin{equation}
H(O_n\mid Y_n)=0.
\label{eq:S5-diagnostic}
\end{equation}

The converse fails, so this quantity is a one-sided diagnostic rather than a definition of closure. An externally scheduled deterministic operation can also satisfy $H(O_n\mid Y_n)=0$: at any fixed step $n$ a deterministic schedule makes $O_n$ constant, and even averaged over a run the readout state may be statistically predictive of an external schedule---for instance because the scheduled interventions themselves shape the dynamics that produce $y_n$---without playing any causal role in selecting the operation. Closure is therefore the causal, physical condition $o_n=C(y_n)$ with $y_n$ physically coupled to the actuation pathway, as defined above; it cannot be certified by a conditional entropy alone. An information-theoretic formulation would need to condition on time and on all external control variables, $H(O_n\mid Y_n,n,E_n)$, or replace the conditional entropy by an interventional quantity testing whether setting $y_n$ changes the distribution of $O_n$. The diagnostic retains one valid direction: if

\[
H(O_n\mid Y_n)>0
\]

then the readout does not determine the operation, and closed deterministic operation selection is ruled out.

\subsection{Operational closure versus dynamical feedback}

The walker already has ordinary dynamical feedback: the droplet writes waves, the waves influence the droplet through the local slope, and the droplet writes new waves. This is a closed physical feedback loop. However, not every feedback loop is a closed computation.

Operational closure is stronger. It requires that a physical readout state select among distinct physically realizable operations. In symbols, one needs a nontrivial operation-selection map

\[
C:X_Y\to \mathcal O
\]

with

\begin{equation}
|\operatorname{Im}(C)|\geq 2
\label{eq:S5-nontrivial}
\end{equation}

for a genuinely state-dependent operation selection.

If

\[
|\operatorname{Im}(C)|=1,
\]

then the system follows a single autonomous physical evolution law. It may still be a dynamical system with memory, and it may still admit a coarse-grained abstract description. But it does not internally select among operations. For computation in the stronger controlled sense, the physical readout must be able to choose between alternatives such as writing, erasing, branching, changing confinement, switching phase, or modifying coupling.

\subsection{The closure diagram}

The distinction can be summarized by the following physical chain:

\[
X_R
\longrightarrow
X_Y
\longrightarrow
\mathcal O
\longrightarrow
X.
\]

The reservoir state contributes to a physical readout state:

\[
y_n=D(x_n).
\]

The readout state selects an operation:

\[
o_n=C(y_n).
\]

The selected operation updates the full physical state:

\[
x_{n+1}=F_{o_n}(x_n).
\]

The updated physical state changes the reservoir, closing the loop:

\[
x_{n+1}
\longrightarrow
X_R.
\]

Thus the closed physical computation has the form

\[
X_R
\rightarrow
X_Y
\rightarrow
\mathcal O
\rightarrow
X_R.
\]

When coupled to an abstraction

\[
\Pi:X\to A,
\]

the corresponding abstract transition is

\[
a_{n+1}
=
G_{\Theta(o_n)}(a_n),
\qquad
a_n=\Pi(x_n).
\]

Closure requires that $o_n$ be produced by $y_n$, and $y_n$ be part of the physical system.

\subsection{Closed computation as state-dependent transition preservation}

We can now state the central definition.

\paragraph{Definition.}
Let $X$ be a physical state space, $A$ an abstract state space, $\Pi:X\to A$ a coarse-graining, $\mathcal O$ a set of physical operations, $F_o:X\to X$ the physical transition associated with $o\in\mathcal O$, $X_Y$ a physical readout state space, $p_Y:X\to X_Y$ the physical readout projection, and $C:X_Y\to\mathcal O$ an operation-selection map.

The system implements a closed physical computation on $U\subseteq X$ if the closed-loop map

\[
\Phi(x)=F_{C(p_Y(x))}(x)
\]

satisfies

\[
\Pi(\Phi(x))
=
G(\Pi(x))
\]

for a fixed abstract transition $G:A\to A$, or more generally

\[
\Pi(F_{C(p_Y(x))}(x))
=
G_{\Theta(C(p_Y(x)))}(\Pi(x))
\]

for an abstract controlled transition family $\{G_u\}_{u\in\mathcal U}$.

In the robust finite-state case, abstract states and readout states must be realized by separated physical basins, and the closed-loop stochastic transition must carry each source basin into the correct successor basin with probability at least $1-\delta$.

\subsection{Proposition: closure converts external control into internal transition selection}

\paragraph{Proposition.}
Suppose a family of physical operations $\{F_o\}_{o\in\mathcal O}$ robustly implements a family of abstract transitions $\{G_{\Theta(o)}\}_{o\in\mathcal O}$ under a coarse-graining $\Pi:X\to A$. Suppose further that the physical system contains a readout subsystem $X_Y$ and a robust operation-selection map $C:X_Y\to\mathcal O$. Then the closed-loop physical map

\[
\Phi(x)=F_{C(p_Y(x))}(x)
\]

implements the state-dependent abstract transition

\[
a_{n+1}
=
G_{\Theta(C(p_Y(x_n)))}(a_n),
\qquad
a_n=\Pi(x_n).
\]

\paragraph{Proof.}
By assumption, for every physical operation $o\in\mathcal O$,

\[
\Pi(F_o(x))
=
G_{\Theta(o)}(\Pi(x))
\]

on the domain where operation $o$ is valid. In the closed-loop system, the applied operation is

\[
o=C(p_Y(x)).
\]

Substituting this operation into the transition-preservation condition gives

\[
\Pi(F_{C(p_Y(x))}(x))
=
G_{\Theta(C(p_Y(x)))}(\Pi(x)).
\]

Since

\[
\Phi(x)=F_{C(p_Y(x))}(x),
\]

we obtain

\[
\Pi(\Phi(x))
=
G_{\Theta(C(p_Y(x)))}(\Pi(x)).
\]

Thus the abstract transition is selected by the physical readout state. $\square$

This proposition is simple but important. It says that closure is not an additional metaphorical property. It is the replacement of an externally supplied operation sequence by a physically internal operation-selection map.

\subsection{Memory is insufficient without closure}

The distinction between memory and closed computation can now be stated precisely.

A system may have physical memory if its state contains recoverable information about previous states:

\[
\mathcal T_n^{(L)}
\longmapsto
H_n.
\]

It may have transition-preserving computation if there exists a coarse-graining

\[
\Pi:X\to A
\]

such that

\[
\Pi(F(x))=G(\Pi(x)).
\]

But it has closed physical computation only if the transition map itself is selected or modulated by an internal physical readout state:

\[
x_{n+1}
=
F_{C(p_Y(x_n))}(x_n).
\]

Thus physical memory, transition preservation, and closure are distinct.

\[
\begin{aligned}
\text{memory}
&\not\Rightarrow\;
\text{transition-preserving computation}
\\
&\not\Rightarrow\;
\text{closed autonomous computation}.
\end{aligned}
\]

The first implication fails because a memory trace may never be mapped to a robust abstract transition. The second implication fails because the transition may be externally selected rather than internally selected.

\subsection{Relation to the wave--particle walker}

In the stroboscopic walker, the wave field $H_n$ is a reservoir and the droplet position $\mathbf r_n$ is a localized probe. The local wave slope

\[
\nabla H_n(\mathbf r_n)
\]

affects the next droplet velocity and therefore participates in a closed physical feedback loop.

However, the $\pi$-shift operation is externally imposed in the experiment. In the notation above, it corresponds to a physical operation

\[
P_\pi
\]

chosen by an external perturbation of the forcing. Unless the walker contains an internal physical readout state

\[
y_n\in X_Y
\]

that triggers

\[
C(y_n)=o_\pi,
\]

the erasing operation is not autonomously selected by the walker.

Therefore, the walker as experimentally described realizes physical wave memory, local reading, feedback, and externally triggered erasure. To become a closed autonomous physical computer in the stronger sense defined here, it would require an additional physical readout-control layer:

\[
(H_n,\mathbf r_n,\mathbf v_n,\sigma_n)
\longrightarrow
y_n
\longrightarrow
o_n
\longrightarrow
F_{o_n}.
\]

The readout state $y_n$ would need to select whether the system continues normal propagation, flips phase, changes confinement, branches into another trajectory class, writes, or erases. Without such internal operation selection, the system remains a wave-memory dynamical machine with externally controlled computational primitives.

\subsection{Summary}

The closure criterion can be summarized as

\begin{equation}
\boxed{
x_{n+1}
=
F_{C(p_Y(x_n))}(x_n)
}
\label{eq:S5-box-closure}
\end{equation}

together with transition preservation,

\begin{equation}
\boxed{
\Pi(F_{C(p_Y(x))}(x))
=
G_{\Theta(C(p_Y(x)))}(\Pi(x)).
}
\label{eq:S5-box-pres}
\end{equation}

The first equation states that the next physical operation is selected by an internal physical readout state. The second states that the resulting physical transition preserves the corresponding abstract transition.

This is the mathematical criterion that separates externally interpreted or externally controlled physical information processing from closed autonomous physical computation.

\section{Categorical formulation of coarse-grained physical computation}
\label{sec:S6}

The preceding sections developed the mathematical criterion in dynamical language. We began with a physical state space $X$, a physical transition map $F:X\to X$, an abstract state space $A$, and a coarse-graining

\[
\Pi:X\to A.
\]

The physical system implements the abstract transition

\[
G:A\to A
\]

on a domain $U\subseteq X$ when

\[
\Pi\circ F = G\circ \Pi
\qquad
\text{on } U.
\]

We then strengthened this condition by requiring robust physical basins, noise tolerance, and finally closure: the next physical operation must be selected by an internal physical readout state rather than externally imposed.

We now translate the same structure into category-theoretic language. The purpose is not to replace the physics with formalism, but to make the compositional structure explicit. Category theory is useful here because it allows us to distinguish physical processes, abstract transitions, coarse-grainings, refinements, and closed-loop compositions in a single language.

\subsection{Physical process category}

Let

\[
\mathbf{PhysProc}
\]

denote a category of physical processes.

For the present paper, we use a concrete version adapted to dynamical systems. The objects are physically meaningful regions of state space,

\[
U,V,W\subseteq X.
\]

These may be full physical state spaces, domains of valid initialization, robust basins, metastable regions, or experimentally accessible regions. A morphism

\[
f:U\to V
\]

is a physically realizable process that carries states in $U$ into $V$. In the deterministic case this means

\[
f(U)\subseteq V.
\]

For example, the stroboscopic walker map restricted to a valid domain gives a morphism

\[
F:U\to F(U).
\]

A controlled physical operation

\[
F_o:X\to X
\]

also defines a morphism

\[
F_o:U\to V
\]

whenever

\[
F_o(U)\subseteq V.
\]

Composition in $\mathbf{PhysProc}$ is ordinary sequential composition of physical processes. If

\[
f:U\to V
\]

and

\[
g:V\to W,
\]

then

\[
g\circ f:U\to W
\]

is the physical process obtained by first applying $f$ and then applying $g$. The identity morphism on $U$ is the physical process that leaves states in $U$ unchanged:

\[
\mathrm{id}_U:U\to U.
\]

This category captures the physically realizable state transitions of the system. In the walker, the morphisms include normal stroboscopic evolution, phase-shift interventions, changes of confinement, and any other physically implementable operation.

\subsection{Abstract process category}

Let

\[
\mathbf{AbsProc}
\]

denote a category of abstract processes.

For a finite-state abstraction, objects are sets of abstract states or subsets of an abstract state space,

\[
S,T\subseteq A.
\]

A morphism

\[
g:S\to T
\]

is an abstract transition rule carrying states in $S$ to states in $T$. For a deterministic finite-state system, this is simply a function

\[
g:S\to T.
\]

For example, if

\[
A=\{a_1,\dots,a_K\}
\]

and

\[
G:A\to A
\]

is an abstract transition map, then $G$ is a morphism in $\mathbf{AbsProc}$.

Composition in $\mathbf{AbsProc}$ is ordinary composition of abstract transitions. If

\[
g:S\to T
\]

and

\[
h:T\to R,
\]

then

\[
h\circ g:S\to R
\]

is the abstract transition obtained by applying $g$ and then $h$.

Thus $\mathbf{PhysProc}$ contains physically realizable transformations, while $\mathbf{AbsProc}$ contains abstract transition rules.

\subsection{Coarse-graining as a quotient functor}

A coarse-graining

\[
\Pi:X\to A
\]

assigns abstract states to physical states. It also maps physical regions to abstract regions. For

\[
U\subseteq X,
\]

define

\[
\Pi(U)
=
\{\Pi(x):x\in U\}
\subseteq A.
\]

Thus $\Pi$ maps physical objects $U$ in $\mathbf{PhysProc}$ to abstract objects $\Pi(U)$ in $\mathbf{AbsProc}$.

The nontrivial question is whether $\Pi$ maps physical morphisms to abstract morphisms. Given a physical process

\[
f:U\to V,
\]

we would like to define an abstract process

\[
\Pi(f):\Pi(U)\to \Pi(V)
\]

by

\[
\Pi(f)(\Pi(x))=\Pi(f(x)).
\]

This is well-defined only if the result is independent of the physical representative $x$. Therefore, we require

\[
\Pi(x)=\Pi(x')
\quad
\Longrightarrow
\quad
\Pi(f(x))=\Pi(f(x'))
\]

for all

\[
x,x'\in U.
\]

This is exactly the quotient condition introduced earlier. If it holds, the physical process $f$ descends to an abstract process

\[
\bar f:\Pi(U)\to \Pi(V),
\]

where

\[
\bar f(\Pi(x))=\Pi(f(x)).
\]

In this case, the coarse-graining defines a local quotient functor

\[
Q_\Pi:\mathbf{PhysProc}_\Pi\to \mathbf{AbsProc},
\]

where

\[
\mathbf{PhysProc}_\Pi
\]

is the subcategory of physical regions and physical processes that preserve the equivalence relation induced by $\Pi$.

The functor acts on objects by

\[
Q_\Pi(U)=\Pi(U),
\]

and on morphisms by

\begin{equation}
Q_\Pi(f)=\bar f.
\label{eq:S6-functor}
\end{equation}

Functoriality follows from preservation of composition:

\begin{equation}
Q_\Pi(g\circ f)
=
Q_\Pi(g)\circ Q_\Pi(f),
\label{eq:S6-functoriality}
\end{equation}

because

\[
\begin{aligned}
Q_\Pi(g\!\circ\! f)(\Pi(x))
&= \Pi(g(f(x)))
\\
&= Q_\Pi(g)(\Pi(f(x)))
\\
&= (Q_\Pi(g)\circ Q_\Pi(f))(\Pi(x)).
\end{aligned}
\]

It also preserves identities:

\[
Q_\Pi(\mathrm{id}_U)
=
\mathrm{id}_{\Pi(U)}.
\]

Thus the dynamical commutative diagram

\begin{equation}
\Pi\circ F=G\circ \Pi
\label{eq:S6-commute}
\\ \text{see:} \eqref{eq:S3-commute}
\end{equation}

is the one-morphism case of a quotient functor from physical processes to abstract processes.

\subsection{Why this is weaker than bijection}

The quotient functor does not require a one-to-one mapping between physical states and abstract states. The physical realization of an abstract state

\[
a\in A
\]

is the fiber

\[
\Pi^{-1}(a)\subseteq X.
\]

This fiber may contain many physical microstates. What matters is not bijection, but preservation of transition structure. If all physical states in the same fiber are carried into the fiber of the same abstract successor state, then the physical dynamics descends to the abstraction.

In categorical terms, the physical process $F$ need not be isomorphic to the abstract process $G$. Rather, $G$ is the image of $F$ under the quotient functor $Q_\Pi$:

\[
Q_\Pi(F)=G.
\]

This gives a precise formulation of physical-to-abstract compatibility without requiring physical microstates and abstract symbols to be identical.

\subsection{Abstraction and concretization as an adjoint pair}

The quotient map $\Pi:X\to A$ also induces a pair of maps between regions of physical state space and regions of abstract state space.

Let

\[
\mathcal P(X)
\]

denote the powerset of $X$, partially ordered by inclusion, and let

\[
\mathcal P(A)
\]

denote the powerset of $A$, also ordered by inclusion. These partially ordered sets may be regarded as categories: there is a unique morphism

\[
U\to V
\]

whenever

\[
U\subseteq V.
\]

Define the abstraction map

\[
\alpha:\mathcal P(X)\to \mathcal P(A)
\]

by

\[
\alpha(U)=\Pi(U).
\]

Define the concretization map

\[
\gamma:\mathcal P(A)\to \mathcal P(X)
\]

by

\[
\gamma(S)=\Pi^{-1}(S).
\]

Then $\alpha$ and $\gamma$ form a Galois connection:

\[
\alpha(U)\subseteq S
\quad
\Longleftrightarrow
\quad
U\subseteq \gamma(S).
\]

This says that all physical states in $U$ abstract into $S$ if and only if $U$ lies inside the physical realization of $S$.

For a single abstract state $a\in A$,

\[
\gamma(\{a\})
=
\Pi^{-1}(a)
\]

is the physical basin or realization class of $a$.

This abstraction--concretization pair is useful because it avoids requiring an inverse map from abstract states to unique physical states. Instead, an abstract state is concretized as a region of physical state space.

\subsection{Transition preservation in adjoint form}

Using $\alpha$ and $\gamma$, the transition-preservation condition can be written regionally.

Let

\[
G:A\to A
\]

be an abstract transition map. It acts on abstract regions by direct image:

\[
G(S)=\{G(a):a\in S\}.
\]

A physical process

\[
F:X\to X
\]

implements $G$ if, for every abstract region $S\subseteq A$,

\[
F(\gamma(S))
\subseteq
\gamma(G(S)).
\]

In words: if the physical state begins in the physical realization of an abstract region $S$, then after physical evolution it lands in the physical realization of the abstract successor region $G(S)$.

For a single abstract state $a$, this becomes

\[
F(\gamma(\{a\}))
\subseteq
\gamma(\{G(a)\}).
\]

If we write

\[
B_a=\gamma(\{a\}),
\]

then

\[
F(B_a)\subseteq B_{G(a)}.
\]

This is exactly the basin-transition condition introduced earlier.

Equivalently, applying abstraction after physical evolution gives

\[
\alpha(F(\gamma(S)))\subseteq G(S).
\]

For exact deterministic implementation, this inclusion may be an equality on the reachable domain. In robust or approximate implementations, inclusion is the more natural condition because physical dynamics may contract regions, discard irrelevant degrees of freedom, or map only a subset of the physical basin into the target basin.

\subsection{Robust implementation as a lax condition}

In real physical systems, noise and finite resolution weaken strict functoriality. Instead of requiring

\[
F(B_a)\subseteq B_{G(a)},
\]

we require

\[
F(N_{\varepsilon_X}(B_a))
\subseteq
N_{\varepsilon_X}(B_{G(a)})
\]

in the deterministic finite-resolution case, or

\[
K(x,B_{G(a)})
\geq
1-\delta
\qquad
x\in N_{\varepsilon_X}(B_a)
\]

in the stochastic case.

Categorically, this can be viewed as a lax or approximate version of the quotient functor. The physical morphism does not map exactly into the target realization class for every perturbation, but it maps into it up to a specified tolerance.

Thus the exact functorial statement

\[
Q_\Pi(F)=G
\]

becomes an approximate statement

\begin{equation}
Q_\Pi(F)\approx_{\varepsilon,\delta}G.
\label{eq:S6-lax}
\end{equation}

Here

\[
\varepsilon
\]

controls physical indistinguishability or basin separation, while

\[
\delta
\]

controls transition failure probability.

This formulation is important for physics. It allows computation to be defined for noisy physical systems without falling back into arbitrary observer labeling. The abstraction must still be physically robust: its basins must be separated, and its transitions must be stable under perturbations.

\subsection{Controlled operations as morphism families}

Now suppose the system has a set of physical operations

\[
\mathcal O.
\]

Each operation

\[
o\in\mathcal O
\]

defines a physical morphism

\[
F_o:X\to X.
\]

On the abstract side, suppose there is a set of abstract operations

\[
\mathcal U,
\]

with

\[
G_u:A\to A
\]

for each

\[
u\in\mathcal U.
\]

A map

\[
\Theta:\mathcal O\to\mathcal U
\]

assigns to each physical operation the abstract operation it implements.

The controlled physical system implements the controlled abstract system under $\Pi$ if

\[
Q_\Pi(F_o)=G_{\Theta(o)}
\]

for all physically relevant operations

\[
o\in\mathcal O.
\]

Equivalently,

\[
\Pi(F_o(x))
=
G_{\Theta(o)}(\Pi(x))
\]

for all relevant physical states $x$.

In basin form,

\[
F_o(B_a)
\subseteq
B_{G_{\Theta(o)}(a)}.
\]

This is the categorical version of externally controlled physical computation. The family of physical morphisms $\{F_o\}$ maps, under the quotient functor, to a family of abstract morphisms $\{G_{\Theta(o)}\}$.

\subsection{Closure as internalization of morphism selection}

The closure condition introduced earlier can now be phrased categorically.

Let

\[
X_Y
\]

be the physical readout/control state space, and let

\[
p_Y:X\to X_Y
\]

be the physical projection onto the readout subsystem. Let

\[
C:X_Y\to\mathcal O
\]

select a physical operation from the readout state.

The closed-loop physical morphism is

\[
\Phi:X\to X,
\]

defined by

\[
\Phi(x)=F_{C(p_Y(x))}(x).
\]

This is not simply one externally chosen morphism. It is a composite construction:

\[
X
\xrightarrow{\langle C\circ p_Y,\mathrm{id}_X\rangle}
\mathcal O\times X
\xrightarrow{\mathrm{Act}}
X,
\]

where

\[
\mathrm{Act}(o,x)=F_o(x).
\]

Thus

\[
\Phi
=
\mathrm{Act}
\circ
\langle C\circ p_Y,\mathrm{id}_X\rangle.
\]

This expression makes closure explicit. The system first maps its own physical state to a readout state, then maps that readout state to an operation, and then applies that operation to itself.

The corresponding abstract transition is

\[
Q_\Pi(\Phi).
\]

If the controlled physical operations implement the abstract operations under $\Theta$, then

\[
Q_\Pi(\Phi)(\Pi(x))
=
G_{\Theta(C(p_Y(x)))}(\Pi(x)).
\]

Thus morphism selection has been internalized: the abstract transition is selected by a physical readout state within the system.

\subsection{Closed computation diagram}

The closed computation diagram has two coupled layers.

On the physical side:

\[
X
\xrightarrow{p_Y}
X_Y
\xrightarrow{C}
\mathcal O,
\]

and

\[
\mathcal O\times X
\xrightarrow{\mathrm{Act}}
X.
\]

On the abstract side:

\[
A
\xrightarrow{G_u}
A
\]

for each abstract operation

\[
u\in\mathcal U.
\]

The compatibility condition is

\[
\Pi\left(
\mathrm{Act}(C(p_Y(x)),x)
\right)
=
G_{\Theta(C(p_Y(x)))}(\Pi(x)).
\]

Equivalently,

\[
\Pi(\Phi(x))
=
G_{\Theta(C(p_Y(x)))}(\Pi(x)).
\]

The categorical point is that the physical-to-abstract map does not merely connect isolated states. It connects a closed composite process on the physical side to a corresponding transition on the abstract side.

\subsection{Refinement as natural transformation}

Different coarse-grainings may describe the same physical system at different levels of abstraction. Let

\[
\Pi:X\to A
\]

and

\[
\Pi':X\to A'
\]

be two coarse-grainings. Suppose there exists a map

\[
\rho:A\to A'
\]

such that

\[
\Pi'=\rho\circ \Pi.
\]

Then $\Pi'$ is a coarser abstraction than $\Pi$, and $\rho$ is a refinement-forgetting map.

If both coarse-grainings preserve transition structure, with

\[
\Pi\circ F=G\circ \Pi
\]

and

\[
\Pi'\circ F=G'\circ \Pi',
\]

then substituting

\[
\Pi'=\rho\circ\Pi
\]

gives

\[
\rho\circ \Pi\circ F
=
G'\circ \rho\circ \Pi.
\]

Using

\[
\Pi\circ F=G\circ \Pi,
\]

we obtain

\[
\rho\circ G\circ \Pi
=
G'\circ \rho\circ \Pi.
\]

On the image of $\Pi$, this implies

\[
\rho\circ G
=
G'\circ \rho.
\]

This is the usual commuting condition for refinement:

\[
\begin{array}{ccc}
A & \xrightarrow{G} & A \\
\downarrow \rho &  & \downarrow \rho \\
A' & \xrightarrow{G'} & A'
\end{array}
\]

In categorical language, $\rho$ is a natural transformation-like relation between two abstraction functors. It expresses the fact that two descriptions of the same physical process are compatible across levels of abstraction.

This is useful for physical computation because a system may have multiple valid descriptions: a fine-grained physical readout, a coarser symbolic state, and an even coarser task-level description. These levels should not be arbitrary; their transitions should commute under refinement maps.

\subsection{Why category theory is useful here}

The categorical formulation adds four useful clarifications.

First, it replaces arbitrary representation with structure preservation. A physical-to-abstract map becomes computational only when it maps physical morphisms to abstract morphisms.

Second, it replaces bijection with quotient functoriality. Abstract states correspond to physical realization classes or basins, not unique microstates.

Third, it represents robustness through basin-level or approximate functoriality. Physical computation need not be exact at the microstate level, but it must be stable at the level of distinguishable physical regions.

Fourth, it represents autonomy as internal morphism selection. A closed physical computer is not merely a physical system to which an external sequence of operations is applied. It is a system whose own physical readout state selects the next physical morphism.

\subsection{Summary}

The dynamical criterion

\[
\Pi\circ F=G\circ \Pi
\]

is the one-step expression of a quotient functor

\[
Q_\Pi:\mathbf{PhysProc}_\Pi\to\mathbf{AbsProc}.
\]

The robust basin condition is the finite-resolution version of this functorial relation:

\[
F(B_a)\subseteq B_{G(a)}
\]

or, in noisy form,

\[
K(x,B_{G(a)})\geq 1-\delta.
\]

The closure condition internalizes operation selection:

\[
\Phi(x)=F_{C(p_Y(x))}(x),
\]

and the categorical compatibility condition becomes

\[
\Pi(\Phi(x))
=
G_{\Theta(C(p_Y(x)))}(\Pi(x)).
\]

Thus the category-theoretic lift does not add a new physical assumption. It organizes the previous stages into a compositional framework:

\[
\begin{aligned}
\text{physical process}
&\rightarrow\;
\text{coarse-grained abstract process}
\\
&\rightarrow\;
\text{robust basin realization}
\\
&\rightarrow\;
\text{closed internal operation selection}.
\end{aligned}
\]

This is the categorical form of the closure criterion for autonomous physical computation.

\section{Classification of the Perrard--Fort--Couder walker}
\label{sec:S7}

We can now classify the wave--particle walker using the criteria developed above. The purpose of this classification is not to diminish the physical significance of the experiment. On the contrary, the walker provides an unusually clean physical system in which writing, storage, reading, finite-time reversal, and erasure are all realized in one coupled wave--particle dynamics. The point is to distinguish these demonstrated physical primitives from stronger claims about autonomous computation or Turing universality.

The classification proceeds through the hierarchy developed in the previous sections:

\begin{equation}
\begin{aligned}
\text{physical dynamics}
&\rightarrow\;
\text{physical memory}
\\
&\rightarrow\;
\text{physical read/write feedback}
\\
&\rightarrow\;
\text{controlled erasure}
\\
&\rightarrow\;
\text{coarse-grained transition preservation}
\\
&\rightarrow\;
\text{closed autonomous computation}.
\end{aligned}
\label{eq:S7-hierarchy}
\end{equation}

The walker satisfies the early levels of this hierarchy in a strong and physically explicit sense. It does not, as experimentally described, satisfy the later requirements of robust symbolic transition preservation and autonomous operation selection.

\subsection{Physical state and stroboscopic dynamics}

In the notation used above, the stroboscopic state of the walker is

\[
x_n=
(\mathbf r_n,\mathbf v_n,\sigma_n,H_n)
\in X,
\]

where

\[
\mathbf r_n\in\Omega
\]

is the horizontal droplet position at the $n$-th impact,

\[
\mathbf v_n\in\mathbb R^2
\]

is the horizontal velocity,

\[
\sigma_n\in\{+1,-1\}
\]

is the bounce phase relative to the chosen Faraday phase, and

\[
H_n\in\mathcal H
\]

is the wave-memory field.

The wave-memory update is

\[
H_{n+1}(\rho)
=
\lambda H_n(\rho)
+
\sigma_n h_0J_0(k_F|\rho-\mathbf r_n|),
\]

with

\[
\lambda=e^{-1/M_e}.
\]

The droplet reads the local slope

\[
\nabla H_n(\mathbf r_n)
\]

and its horizontal dynamics can be written schematically as

\[
\begin{aligned}
\mathbf v_{n+1}
&= a\,\mathbf v_n
\\
&\quad - \kappa\,\sigma_n \nabla H_n(\mathbf r_n)
\\
&\quad - \chi\,\nabla V(\mathbf r_n)
\\
&\quad + \boldsymbol\xi_n .
\end{aligned}
\]

\[
\mathbf r_{n+1}
=
\mathbf r_n+T_F\mathbf v_{n+1}.
\]

Thus the walker is a recurrent physical system:

\[
\mathbf r_n
\rightarrow
H_{n+1}
\rightarrow
\nabla H_{n+1}(\mathbf r_{n+1})
\rightarrow
\mathbf r_{n+2}.
\]

This is the physical substrate on which the classification rests.

\subsection{Criterion 1: physical memory}

The first criterion is whether the system contains a physical memory variable. The answer is yes.

The field

\[
H_n
\]

stores an exponentially weighted trace of previous impact positions:

\[
H_n
=
\sum_{q=1}^{n}
\lambda^{q-1}
\sigma_{n-q}
h_0J_0(k_F|\rho-\mathbf r_{n-q}|).
\]

The contribution from an impact $q$ bounces in the past is weighted by

\[
\lambda^{q-1}
=
e^{-(q-1)/M_e}.
\]

Therefore, the effective memory horizon scales as

\[
L_{\mathrm{eff}}\sim M_e.
\]

In the language of the previous section, the walker realizes a memory map

\[
\mathcal W_L:\Omega^L\to\mathcal H,
\]

where

\[
\mathcal W_L
\left(
\mathbf r_{n-L},\ldots,\mathbf r_{n-1}
\right)
=
\sum_{q=1}^{L}
\lambda^{q-1}
\sigma_{n-q}
\psi_{\mathbf r_{n-q}}.
\]

Thus the walker clearly satisfies the physical-memory criterion:

\[
\boxed{
\text{Perrard walker has physical trajectory memory.}
}
\]

However, memory alone is not computation. The field $H_n$ may contain recoverable or dynamically usable information about the past without yet defining an abstract state space, a transition rule, or an autonomous control architecture.

\subsection{Criterion 2: physical writing}

The second criterion is whether the system physically writes information into the memory variable. The answer is also yes.

At each bounce, the droplet adds a localized wave source to the field:

\[
H_{n+1}
=
\lambda H_n
+
\sigma_n\psi_{\mathbf r_n}.
\]

The writing operation is therefore

\[
W_{\mathbf r_n,\sigma_n}:
H_n
\mapsto
\lambda H_n+\sigma_n\psi_{\mathbf r_n}.
\]

The written quantity is not a discrete symbol. It is a spatially extended wave source centered at the droplet impact position. Nevertheless, it is a genuine physical write operation: the droplet changes the state of the bath, and that changed bath state persists long enough to affect later dynamics.

Thus:

\[
\boxed{
\text{The droplet writes positional information into the wave field.}
}
\]

\subsection{Criterion 3: physical storage}

The third criterion is whether the written information persists. The answer is yes, but only for a finite memory time.

The storage operation is the passive decay

\[
H_n\mapsto \lambda H_n,
\]

with

\[
\lambda=e^{-1/M_e}.
\]

The memory is neither permanent nor symbolic. It is analog, distributed, and exponentially decaying. A source written $q$ impacts in the past contributes with weight

\[
e^{-q/M_e}.
\]

Thus:

\[
\boxed{
\begin{array}{c}
\text{The walker stores information}\\
\text{as a finite-lifetime wave-memory trace.}
\end{array}
}
\]

The storage is physically real, but it is not equivalent to a stable symbolic tape. Its reliability is limited by $M_e$, noise, interference, and the finite resolution with which wave states can be distinguished.

\subsection{Criterion 4: physical reading}

The fourth criterion is whether the stored memory is physically read by the system. The answer is yes.

The droplet reads the local slope of the wave field:

\[
\mathbf g_n=\nabla H_n(\mathbf r_n).
\]

This quantity enters the next velocity update:

\[
\begin{aligned}
\mathbf v_{n+1}
&= a\,\mathbf v_n
\\
&\quad - \kappa\,\sigma_n \mathbf g_n
\\
&\quad - \chi\,\nabla V(\mathbf r_n)
\\
&\quad + \boldsymbol\xi_n .
\end{aligned}
\]

Thus the wave field is not a passive record. It is causally active. The droplet samples the stored wave memory locally, and the sampled gradient changes subsequent motion.

Therefore:

\[
\boxed{
\begin{array}{c}
\text{The walker physically reads its wave memory}\\
\text{through local slope sampling.}
\end{array}
}
\]

This is stronger than observer-side reconstruction. The reading is performed by the droplet itself through the physical force law.

\subsection{Criterion 5: physical feedback}

The fifth criterion is whether reading feeds back into future writing. The answer is yes.

The system has the loop

\[
\mathbf r_n
\rightarrow
H_{n+1}
\rightarrow
\nabla H_{n+1}(\mathbf r_{n+1})
\rightarrow
\mathbf v_{n+2}
\rightarrow
\mathbf r_{n+2}
\rightarrow
H_{n+3}.
\]

Therefore, the walker is a closed analog feedback system. The droplet writes the wave field, the wave field guides the droplet, and the guided droplet writes the next wave field.

This justifies calling the walker a physical memory-feedback machine:

\[
\boxed{
\text{The walker is a closed wave--particle feedback system.}
}
\]

However, this should not yet be confused with closed autonomous computation. Feedback means the physical variables influence one another. Closed computation requires a stronger condition: a physically realized readout state must select among distinct operations.

\subsection{Criterion 6: externally triggered erasure}

The sixth criterion is whether the system can erase memory. The answer is yes, under an externally imposed phase operation.

Let

\[
P_\pi:X\to X
\]

be the phase-shift intervention

\[
P_\pi:
(\mathbf r,\mathbf v,\sigma,H)
\mapsto
(\mathbf r,\mathbf v,-\sigma,H).
\]

After this phase shift, newly written sources have the opposite phase relative to the preexisting wave field. If the droplet approximately backtracks, post-shift impacts occur near previous impact locations, but the newly written sources have opposite sign. Thus

\[
H_{n_\pi+m}
=
H_{\mathrm{old}}^{(m)}
+
H_{\mathrm{new}}^{(m)},
\]

with

\[
H_{\mathrm{old}}^{(m)}
=
\lambda^m H_{n_\pi},
\]

and

\[
H_{\mathrm{new}}^{(m)}
=
-\sigma_+
\sum_{\ell=0}^{m-1}
\lambda^{m-1-\ell}
\psi_{\mathbf r_{n_\pi+\ell}}.
\]

The wave energy after the phase shift is

\[
E_W[n_\pi+m]
=
\frac{
\left|
H_{\mathrm{old}}^{(m)}
+
H_{\mathrm{new}}^{(m)}
\right|_{L^2}^2
}{
|\psi_{\mathbf 0}|_{L^2}^2
}.
\]

Erasure occurs when

\[
\left|
H_{\mathrm{old}}^{(m)}
+
H_{\mathrm{new}}^{(m)}
\right|_{L^2}^2
<
\left|
H_{\mathrm{old}}^{(m)}
\right|_{L^2}^2.
\]

Equivalently, the interference term is sufficiently negative:

\[
2
\left\langle
H_{\mathrm{old}}^{(m)},
H_{\mathrm{new}}^{(m)}
\right\rangle
+
\left|
H_{\mathrm{new}}^{(m)}
\right|^2
<
0.
\]

Thus:

\[
\boxed{
\begin{array}{c}
\text{The walker realizes physical erasure}\\
\text{by phase-opposed wave writing.}
\end{array}
}
\]

But this erasure is controlled by an externally imposed phase perturbation. In the experiment, the $\pi$-shift is not selected by an internal readout state of the walker. It is supplied by a controlled disturbance of the forcing. Therefore, the erasure primitive is real, but externally triggered.

\subsection{Criterion 7: finite-time trajectory reversal}

The walker also realizes finite-time trajectory reversal. After the $\pi$-shift, the effective wave force changes sign:

\[
-C\,\sigma_+\nabla H(\mathbf r)
\mapsto
+C\,\sigma_+\nabla H(\mathbf r).
\]

For a finite time, the droplet retraces its recent trajectory. Define the reversal error

\[
d_{\mathrm{rev}}(m)
=
|\mathbf r_{n_\pi+m}-\mathbf r_{n_\pi-m}|.
\]

The system exhibits finite-time reversal over horizon $m_\ast$ if

\[
d_{\mathrm{rev}}(m)<\varepsilon_{\mathrm{rev}}
\]

for

\[
1\leq m\leq m_\ast.
\]

The relevant horizon is controlled by the memory lifetime:

\[
m_\ast \sim \frac{M_e}{2}.
\]

This reversal is not global time-reversal invariance. The system is dissipative, and the full physical map is not inverted. Instead, the system exploits its own stored wave memory to guide the droplet backward while simultaneously erasing that memory.

Thus:

\[
\boxed{
\begin{array}{c}
\text{The walker realizes finite-time memory-mediated reversal,}\\
\text{not global dynamical reversibility.}
\end{array}
}
\]

\subsection{Criterion 8: coarse-grained transition preservation}

The next question is whether the walker implements a computation in the coarse-grained transition-preserving sense.

For this, one must identify:

\[
A
\]

an abstract state space,

\[
\Pi:X\to A
\]

a coarse-graining,

\[
G:A\to A
\]

an abstract transition rule, and

\[
U\subseteq X
\]

a physical domain on which

\[
\Pi(F(x))=G(\Pi(x)).
\]

The Perrard experiment does not explicitly define such an abstract state space or transition rule. It demonstrates a physical memory-feedback process and an externally controlled reversal/erasure operation. Therefore, the experiment by itself does not establish a symbolic computation in the transition-preserving sense.

However, one can define limited abstractions under which the walker realizes finite-time controlled transitions. For example, let an abstract state encode a recent trajectory segment:

\[
\begin{aligned}
a_n
&= \Pi(x_n)
\\
&= [\,\mathbf r_{n-L},\ldots,\mathbf r_{n-1}\,] .
\end{aligned}
\]

up to finite spatial resolution. Under the externally applied phase shift $P_\pi$, one may define an approximate reversal transition

\[
G_{\mathrm{rev}}:
[\mathbf r_{n-L},\ldots,\mathbf r_{n-1}]
\mapsto
[\mathbf r_{n-1},\ldots,\mathbf r_{n-L}],
\]

valid over a finite horizon and tolerance. Then, for $m\lesssim M_e/2$,

\[
\Pi(F_{P_\pi}^m(x_n))
\approx
G_{\mathrm{rev}}^m(\Pi(x_n)).
\]

This is a legitimate finite-time, externally controlled, coarse-grained physical transition. But it is not a general-purpose symbolic computation. It is a specific memory-mediated reversal primitive.

Thus:

\[
\boxed{
\begin{array}{c}
\text{The walker can support limited externally}\\
\text{controlled transition preservation,}
\end{array}
}
\]

but

\[
\boxed{
\begin{array}{c}
\text{a full abstract transition system}\\
\text{is not specified or demonstrated.}
\end{array}
}
\]

\subsection{Criterion 9: robust symbolic basins}

A stronger computational claim would require robust symbolic basins

\[
B_1,\ldots,B_K\subset X
\]

corresponding to abstract states

\[
a_1,\ldots,a_K\in A.
\]

These basins would need to satisfy separation:

\[
\operatorname{dist}(B_i,B_j)>2\varepsilon_X
\qquad
i\neq j,
\]

and transition stability:

\[
F(B_i)\subseteq B_{G(a_i)}
\]

or, in the noisy case,

\[
K(x,B_{G(a_i)})
\geq 1-\delta.
\]

The Perrard experiment does not identify such symbolic basins. Stable orbit classes, such as circular, lemniscate, or other quantized orbits, may provide candidates for coarse dynamical states. But to count as robust symbolic states, one would need to show that these states are reliably distinguishable, that they can be initialized and read out, and that physical operations carry each basin into the correct successor basin.

Therefore:

\[
\boxed{
\text{Robust symbolic basin computation is not demonstrated.}
}
\]

This is not a criticism of the experiment. The experiment was designed to study memory-mediated reversal and erasure, not to implement a finite-state symbolic machine.

\subsection{Criterion 10: closure and autonomous operation selection}

The central criterion of this paper is closure. A closed physical computer must contain an internal physical readout state

\[
y_n\in X_Y
\]

and an operation-selection map

\[
C:X_Y\to\mathcal O
\]

such that

\[
x_{n+1}
=
F_{C(y_n)}(x_n).
\]

For the walker, the relevant operation set includes at least


\[
\mathcal O
=
\{
o_{\mathrm{normal}},
o_\pi,
\ldots
\},
\qquad
F_{o_{\mathrm{normal}}}=F,
\quad
F_{o_\pi}=P_\pi.
\]

The $\pi$-shift operation is

\[
P_\pi:
(\mathbf r,\mathbf v,\sigma,H)
\mapsto
(\mathbf r,\mathbf v,-\sigma,H).
\]

In the experiment, this operation is imposed externally. The system does not contain a physical readout state $y_n$ such that

\[
C(y_n)=o_\pi.
\]

Therefore, the walker does not satisfy the closure criterion for autonomous operation selection.

The distinction is important. The walker does have physical feedback:

\[
H_n
\rightarrow
\nabla H_n(\mathbf r_n)
\rightarrow
\mathbf v_{n+1}.
\]

But this feedback does not select among computational operations such as normal propagation versus erasure. The erasure operation is selected from outside the walker.

Thus:

\[
\boxed{
\text{The Perrard walker is a closed analog feedback system,}
}
\]

but

\[
\boxed{
\begin{array}{c}
\text{it is not, as described,}\\
\text{a closed autonomous physical computer.}
\end{array}
}
\]

\subsection{Criterion 11: universal Turing computation}

Finally, universal Turing computation would require substantially more structure. At minimum, one would need:

\[
\Gamma
\]

a finite alphabet,

\[
Q
\]

a finite set of internal states,

\[
\delta:Q\times\Gamma\to Q\times\Gamma\times\{L,R\}
\]

a transition rule or equivalent programmable operation set,




\[
\Pi:X\to Q\times\Gamma^{\mathbb{Z}}\times\mathbb{Z}
\]

a robust physical-to-symbolic encoding of machine configurations (internal state, tape contents, and head position),

and physical operations that preserve the transition table, where $\Delta_\delta$ denotes the configuration-update map induced by $\delta$ (apply the local rule at the head, write, and move):

\[
\Pi(F(x))=\Delta_\delta(\Pi(x)),
\]

The Perrard walker does not provide such a construction. It has a wave-memory repository and physical read/write/erase primitives, but no demonstrated finite alphabet, no symbolic transition table, no internally selected instruction set, and no proof of universal simulation.

Therefore:

\[
\boxed{
\begin{array}{c}
\text{The Perrard walker does not}\\
\text{demonstrate Turing universality.}
\end{array}
}
\]

It is better described as a wave-memory physical machine with Turing-like operational primitives.

\subsection{Classification table}

The classification can be summarized as follows:

\begin{table*}[t]
\centering
\begin{tabular*}{\textwidth}{@{\extracolsep{\fill}} l l l}
\hline
\textbf{Criterion} & \textbf{Status} & \textbf{Reason} \\
\hline

Physical wave dynamics & Yes &
Droplet and wave field form a coupled physical system. \\

Internal clock & Yes &
The Faraday/bounce period provides a stroboscopic clock. \\

Writing & Yes &
Each impact writes a localized wave source. \\

Storage & Yes &
The wave field persists for a memory time $M_eT_F$. \\

Reading & Yes &
The droplet reads $\nabla H(\mathbf r)$. \\

Feedback & Yes &
The read slope changes future droplet motion. \\

Erasure & Yes, externally triggered &
A controlled $\pi$-shift causes phase-opposed writing. \\

Finite-time reversal & Yes &
The droplet backtracks while erasing the old wave field. \\

Coarse-grained computation & Limited/not specified &
No explicit $A,\Pi,G$ are defined in the experiment. \\

Robust symbolic basins & Not demonstrated &
No finite alphabet or basin transition table is provided. \\

Closed autonomous operation selection & No &
The $\pi$-shift is externally imposed, not internally selected. \\

Turing universality & No &
No universal symbolic machine construction is demonstrated. \\
\hline
\end{tabular*}
\end{table*}













\subsection{Main classification result}

The classification can be stated as a proposition.

\paragraph{Proposition.}
The Perrard--Fort--Couder walker realizes a physical wave-memory machine with writing, finite-time storage, local reading, feedback, and externally triggered erasure. It does not, as experimentally described, realize a closed autonomous physical computer, because the operation that produces erasure and reversal is selected externally rather than by an internal physical readout state.

\paragraph{Argument.}
The wave field $H_n$ stores a recent trajectory trace through exponentially weighted impact-generated sources. The droplet reads this memory through the local slope $\nabla H_n(\mathbf r_n)$, and this readout changes the next physical state. Therefore, the system realizes physical memory, reading, writing, and feedback.

The $\pi$-shift operation changes the phase relation between the droplet and the preexisting wave field. It reverses the effective wave-induced kick and causes subsequent impacts to write phase-opposed sources. During backtracking, these new sources destructively interfere with the old field, producing erasure. Therefore, the system realizes a physical erasure primitive.

However, the phase shift is imposed externally. There is no internal readout state $y_n$ and no operation-selection map $C:X_Y\to\mathcal O$ such that $C(y_n)=o_\pi$. Hence the system does not satisfy

\[
x_{n+1}=F_{C(p_Y(x_n))}(x_n)
\]

for the erasure operation. It satisfies physical feedback, but not autonomous operation selection.

Therefore, the walker is best classified as:

\[
\boxed{
\text{a wave-memory physical machine with Turing-like primitives,}
}
\]

not as

\[
\boxed{
\text{a closed autonomous Turing machine.}
}
\]

\subsection{Interpretive conclusion}

This classification preserves the strongest physical insight of the experiment while avoiding overstatement. The walker is remarkable because its past is physically stored in the wave field, and because a phase operation allows the droplet to read that memory backward while erasing it. This is a genuine physical information-processing phenomenon.

But autonomous computation requires more than memory and externally triggered erasure. It requires robust symbolic or coarse-grained states whose transitions are physically preserved, and it requires internal operation selection by a physical readout state. The Perrard walker supplies the reservoir, probe, write, read, and erase primitives. The missing ingredient is an intrinsic readout-control layer.

Thus the result should be read not as a refutation of the wave-based Turing-machine language, but as its precise sharpening:

\[
\begin{aligned}
\text{wave memory }
&\\+\text{read/write/erase primitives}
&\\\neq
\text{autonomous computation},
\end{aligned}
\]

unless

\[
\begin{aligned}
\text{physical readout}
&\rightarrow\;
\text{operation selection}
\\
&\rightarrow\;
\text{transition-preserving feedback}
\end{aligned}
\]

is also realized inside the system.

\section{Constructive extensions: from wave memory to closed physical computation}
\label{sec:S8}

The previous section classified the Perrard--Fort--Couder walker as a wave-memory physical machine with writing, storage, reading, feedback, finite-time reversal, and externally triggered erasure. The missing ingredient for closed autonomous computation was an intrinsic readout-control layer. In this section we make that statement constructive. We ask what physical modifications would turn a wave-memory walker into a closed physical computer in the sense defined above.

The general design principle is

\[
\begin{aligned}
\text{wave reservoir}
&\longrightarrow\;
\text{physical readout basin}
\\
&\longrightarrow\;
\text{operation selection}
\\
&\longrightarrow\;
\text{new wave evolution}.
\end{aligned}
\]

In symbols, one must add a physically realized readout variable

\[
y_n\in X_Y
\]

and an operation-selection map

\[
C:X_Y\to\mathcal O
\]

so that the next physical transition is selected internally:

\begin{equation}
x_{n+1}
=
F_{C(y_n)}(x_n).
\label{eq:S8-closedloop}
\end{equation}

This transforms the walker from an externally manipulated wave-memory system into a closed-loop physical computing architecture.

\subsection{Extended state space}

Let the original stroboscopic walker state be

\[
x_n=
(\mathbf r_n,\mathbf v_n,\sigma_n,H_n),
\]

where $H_n$ is the wave-memory field, $\mathbf r_n$ and $\mathbf v_n$ are the droplet position and velocity, and $\sigma_n\in\{+1,-1\}$ is the bounce phase.

To construct a closed physical computer, extend the state to


\[
\tilde x_n=
(\mathbf r_n,\mathbf v_n,\sigma_n,H_n,y_n),
\]

where

\[
y_n\in X_Y
\]

is a physical readout/control state. The readout variable $y_n$ must be a physical degree of freedom, not an observer-side label.
It may be the state of a bistable trap, a threshold detector, a second droplet, a mechanical switch, an electronic feedback element, or a controllable boundary mode.

The extended physical state space is


\[
\Omega\times\R^2\times\{+1,-1\}\times\calH\times X_Y.
\]

The original walker dynamics is recovered if $X_Y$ is absent or dynamically inert. The constructive goal is to couple them to the wave field and to the subsequent physical operation.

\subsection{Generic closed-loop architecture}

Let

\[
z_n=(\mathbf r_n,\mathbf v_n,\sigma_n)
\]

denote the localized droplet variables. The wave-memory field is

\[
H_n\in\mathcal H.
\]

A physical readout map extracts a finite-dimensional physical state from the wave--droplet configuration:

\[
y_{n+1}
=
\mathcal R(y_n,H_n,z_n).
\]

In a simple memoryless readout,

\[
y_{n+1}
=
D(H_n,z_n),
\]

where

\[
D:\mathcal H\times Z\to X_Y.
\]

In a more realistic physical readout, $y_n$ has its own dynamics:

\[
y_{n+1}
=
R_y(y_n,H_n,z_n)+\eta^Y_n,
\]

where $\eta^Y_n$ denotes readout noise.

The readout selects an operation

\[
o_n=C(y_n)\in\mathcal O.
\]

The operation set may include normal propagation, phase flipping, changing memory depth, changing confinement, changing boundary conditions, routing to another spatial region, or coupling to another walker:

\[
\mathcal O
=
{
o_{\mathrm{normal}},
o_{\pi},
o_{\mathrm{confine}},
o_{\mathrm{route}},
o_{\mathrm{erase}},
o_{\mathrm{write}},
\ldots
}.
\]

The closed-loop physical update is then

\[
\tilde x_{n+1}
=
F_{o_n}(\tilde x_n),
\qquad
o_n=C(y_n).
\]

Equivalently,

\[
\tilde x_{n+1}
=
F_{C(y_n)}(\tilde x_n).
\]

This is the constructive version of the closure criterion.

\subsection{Physical requirements}

A physical implementation of this architecture must satisfy four conditions.

First, the readout states must be physically distinguishable. If

\[
Y_1,\ldots,Y_K\subset X_Y
\]

are readout basins, then

\[
\operatorname{dist}(Y_i,Y_j)>2\varepsilon_Y
\qquad
i\neq j,
\]

where $\varepsilon_Y$ is the readout noise scale.

Second, readout must be driven by the reservoir. The wave field must bias or determine the readout state:

\[
H_n,z_n
\longrightarrow
y_n.
\]

Third, the readout must select the next physical operation:

\[
y_n
\longrightarrow
o_n=C(y_n).
\]

Fourth, the selected operation must feed back into the wave--particle dynamics:

\[
o_n
\longrightarrow
F_{o_n}
\longrightarrow
H_{n+1},\mathbf r_{n+1},\mathbf v_{n+1},\sigma_{n+1}.
\]

Without the last step, the readout is merely a measurement. With the last step, it becomes part of a closed physical computing loop.

\subsection{Architecture I: multistable droplet trap as intrinsic readout}

A direct physical route is to add a multistable readout element. This could be a secondary droplet, a trapped interface mode, a movable boundary element, or another bistable mechanical degree of freedom. Let

\[
y\in\mathbb R
\]

be the readout coordinate. Suppose $y$ evolves in a multistable potential

\[
U_Y(y;q_n),
\]

where

\[
q_n
\]

is a scalar quantity extracted physically from the wave--particle state. Examples include

\[
q_n=
\nabla H_n(\mathbf r_n)\cdot \hat{\mathbf e},
\]

the local wave slope along direction $\hat{\mathbf e}$, or

\[
q_n=
\int_\Omega \chi(\rho)H_n(\rho)\,d\rho,
\]

a spatially localized wave amplitude measured by a detector profile $\chi$.

For a binary readout, take a double-well potential

\[
\begin{aligned}
U_Y(y;q)
&= \frac{\alpha}{4} y^4
\\
&\quad - \frac{\beta}{2} y^2
\\
&\quad - \gamma q y .
\end{aligned}
\]

with

\[
\alpha,\beta>0.
\]

For $q=0$, the readout has two stable basins near

\[
y_\pm=\pm\sqrt{\frac{\beta}{\alpha}}.
\]

The wave-dependent term

\[
-\gamma qy
\]

tilts the potential. The readout dynamics may be modeled as

\[
\zeta\dot y
=
-\frac{\partial U_Y}{\partial y}
+
\eta_Y(t),
\]

or stroboscopically,

\[
y_{n+1}
=
R_y(y_n,q_n)+\eta^Y_n.
\]

The readout basins are

\[
Y_+={y>0},
\qquad
Y_-={y<0}.
\]

Define the operation-selection map

\[
C(Y_+)=o_\pi,
\qquad
C(Y_-)=o_{\mathrm{normal}}.
\]

Then the system autonomously flips phase when the wave state drives the readout into the $Y_+$ basin:

\[
\sigma_{n+1}
=
\begin{cases}
-\sigma_n, & y_n\in Y_+,\\
\sigma_n, & y_n\in Y_-.
\end{cases}
\]

This turns the externally imposed $\pi$-shift into an internally selected operation.

For reliable operation, the readout barrier

\[
\Delta U
\]

must exceed the effective noise scale:

\[
\Delta U \gg k_BT_{\mathrm{eff}},
\]

or, more generally,

\[
\mathbb P[y_n\text{ crosses basin boundary by noise alone}]\ll 1.
\]

At the same time, the wave-induced tilt must be strong enough to switch the readout when the intended condition is met:

\[
|\gamma q_n|>q_{\mathrm{switch}}.
\]

This gives an experimentally testable design principle: tune the barrier and coupling so that the readout is stable against noise but switchable by the wave-memory signal.

\subsection{Architecture II: thresholded wave detector coupled to phase control}

A second route is to couple a physical wave detector directly to the forcing phase. Let

\[
q_n=
\int_\Omega \chi(\rho)H_n(\rho)\,d\rho
\]

be the detector signal, where $\chi$ is localized near a chosen region of the bath. The detector has hysteretic states

\[
y_n\in\{0,1\}
\]

with update rule

\[
y_{n+1}
=
\begin{cases}
1, & q_n>\theta_+,\\
0, & q_n<\theta_-,\\
y_n, & \theta_-\le q_n\le \theta_+,
\end{cases}
\]

where

\[
\theta_-<\theta_+
\]

define a hysteresis band. The hysteresis prevents rapid noise-induced switching.

The operation-selection map is

\[
C(0)=o_{\mathrm{normal}},
\qquad
C(1)=o_\pi.
\]

The selected operation modifies the bounce phase:

\[
o_\pi:
\sigma_n\mapsto -\sigma_n.
\]

Then the full closed-loop dynamics is

\[
q_n=\int_\Omega \chi(\rho)H_n(\rho)\,d\rho,
\]

\[
y_{n+1}=R_{\theta_-,\theta_+}(y_n,q_n),
\]

\[
o_n=C(y_n),
\]

\[
\tilde x_{n+1}=F_{o_n}(\tilde x_n).
\]

This architecture implements a physically internalized version of the experimental phase perturbation. The $\pi$-shift is no longer scheduled externally. It is triggered when the wave-memory field reaches a thresholded physical condition.

The robust-computation criterion becomes

\[
\mathbb P
[
y_{n+1}=1\mid q_n>\theta_+
]
\ge 1-\delta_Y,
\]

\[
\mathbb P
[
y_{n+1}=0\mid q_n<\theta_-
]
\ge 1-\delta_Y,
\]

and

\[
\mathbb P
[
o_n=C(y_n)
]
\ge 1-\delta_C.
\]

The combined one-step failure probability is approximately

\[
\delta_{\mathrm{total}}
\lesssim
\delta_Y+\delta_C+\delta_F,
\]

where $\delta_F$ is the probability that the selected operation fails to carry the system to the intended successor basin.

\subsection{Architecture III: boundary-controlled Faraday bath}

A third extension is to allow the readout state to control the bath parameters. The walker already depends sensitively on forcing amplitude, memory parameter, confinement, and phase. These quantities can become operation variables.

Let the physical operation be parameterized by

\[
o_n=
(\gamma_m^{(n)},\phi^{(n)},V^{(n)},\Omega^{(n)}),
\]

where

\[
\gamma_m^{(n)}
\]

is the forcing amplitude,

\[
\phi^{(n)}
\]

is the forcing phase,

\[
V^{(n)}(\mathbf r)
\]

is the confinement potential, and

\[
\Omega^{(n)}
\]

is the effective boundary geometry or accessible domain.

The readout state selects these parameters:

\[
(\gamma_m^{(n)},\phi^{(n)},V^{(n)},\Omega^{(n)})
=
C(y_n).
\]

For example,

\[
C(y_1)=
(\gamma_1,\phi_1,V_1,\Omega_1),
\]

\[
C(y_2)=
(\gamma_2,\phi_2,V_2,\Omega_2).
\]

The memory parameter depends on the forcing amplitude through the distance to the Faraday threshold. Schematically,

\[
M_e(y_n)=M_e(C(y_n)).
\]

Thus the wave decay factor becomes readout-controlled:

\[
\lambda(y_n)
=
\exp\left[-\frac{1}{M_e(y_n)}\right].
\]

The wave update becomes

\[
H_{n+1}(\rho)
=
\lambda(y_n)H_n(\rho)
+
\sigma_n h_0(y_n)
J_0
\left(
k_F(y_n)|\rho-\mathbf r_n|
\right).
\]

The droplet update becomes

\[
\begin{aligned}
\mathbf v_{n+1}
&= a(y_n)\,\mathbf v_n
\\
&\quad - \kappa(y_n)\,\sigma_n \nabla H_n(\mathbf r_n)
\\
&\quad - \chi(y_n)\,\nabla V_{y_n}(\mathbf r_n)
\\
&\quad + \boldsymbol\xi_n .
\end{aligned}
\]

This architecture makes the physical law itself state-dependent through the internal readout. It can implement operations such as:

\[
\text{increase memory},
\]

\[
\text{decrease memory},
\]

\[
\text{switch confinement geometry},
\]

\[
\text{route the droplet to another region},
\]

\[
\text{trigger erasure},
\]

or

\[
\text{stabilize a particular orbit class}.
\]

In this case, computation is not encoded in a symbolic tape but in controlled transitions among robust physical dynamical regimes.

\subsection{Architecture IV: coupled walkers as readout and actuator}

A fourth route is to use more than one walker. Let walker $A$ be the primary wave-memory system and walker $B$ be a readout-control element. Let their positions be

\[
\mathbf r^A_n,\mathbf r^B_n,
\]

and let the shared or coupled wave field be

\[
H_n.
\]

The wave update may include both sources:

\[
H_{n+1}(\rho)
=
\lambda H_n(\rho)
+
\sigma^A_n\psi_{\mathbf r^A_n}(\rho)
+
\sigma^B_n\psi_{\mathbf r^B_n}(\rho).
\]

Walker $B$ is confined to a multistable readout region with basins

\[
Y_1,\ldots,Y_K.
\]

Its basin state is

\[
y_n=i
\qquad
\text{if}
\qquad
\mathbf r^B_n\in Y_i.
\]

This readout state controls the operation applied to walker $A$:

\[
o_n=C(y_n).
\]

For instance,

\[
C(1)=o_{\mathrm{normal}},
\]

\[
C(2)=o_\pi,
\]

\[
C(3)=o_{\mathrm{route}}.
\]

Then walker $B$ functions as a physical finite-state controller. It reads the shared wave field through its own wave-mediated dynamics, settles into a basin, and that basin controls the subsequent operation on walker $A$.

This creates a physically closed architecture:

\[
H_n
\longrightarrow
\mathbf r^B_n
\longrightarrow
y_n
\longrightarrow
o_n
\longrightarrow
(\mathbf r^A_{n+1},H_{n+1}).
\]

The important point is that the controller is itself a physical wave--particle subsystem, not an external symbolic computer.

\subsection{Architecture V: autonomous finite-state wave machine}

The previous architectures can be summarized as an autonomous finite-state wave machine. Let

\[
A=\{a_1,\ldots,a_K\}
\]

be an intended abstract finite-state space. Let physical basins

\[
B_1,\ldots,B_K\subset \tilde X
\]

realize these states:

\[
\Pi(\tilde x)=a_i
\qquad
\text{if}
\qquad
\tilde x\in B_i.
\]

Let readout basins

\[
Y_1,\ldots,Y_M\subset X_Y
\]

select operations

\[
o_\ell=C(Y_\ell).
\]

The target abstract controlled transition is

\[
G_\ell:A\to A
\]

for each operation $o_\ell$. The physical design must satisfy

\[
F_{o_\ell}(B_i\cap p_Y^{-1}(Y_\ell))
\subseteq
B_{G_\ell(a_i)}.
\]

In the stochastic case,

\[
K_{o_\ell}(x,B_{G_\ell(a_i)})
\geq
1-\delta
\]

for all

\[
x\in N_{\varepsilon_X}(B_i\cap p_Y^{-1}(Y_\ell)).
\]

This gives a direct experimental program. Choose a small transition graph, construct physical basins for its states, couple wave-memory readouts to operation selection, and measure whether the physical transition matrix matches the desired abstract transition table.

\subsection{Minimal demonstration: autonomous eraser}

The simplest constructive demonstration is not universal computation. It is an autonomous eraser.

Define two operations:

\[
\mathcal O=\{o_{\mathrm{normal}},o_{\pi}\}.
\]

Let

\[
o_{\mathrm{normal}}
\]

be ordinary walker evolution and

\[
o_\pi
\]

be a phase flip.

Let the readout detect when the wave-memory energy exceeds a threshold:

\[
E_W[n]
=
\frac{|H_n|_{L^2(\Omega)}^2}{|\psi_{\mathbf 0}|_{L^2(\Omega)}^2}.
\]

Define

\[
y_{n+1}
=
\begin{cases}
1, & E_W[n]>\Theta_E,\\
0, & E_W[n]<\Theta_E-\Delta,\\
y_n, & \Theta_E-\Delta\leq E_W[n]\leq \Theta_E,
\end{cases}
\]

where $\Delta>0$ is a hysteresis margin.

Let

\[
C(0)=o_{\mathrm{normal}},
\qquad
C(1)=o_\pi.
\]

Then the system autonomously triggers erasure when its own wave-memory energy crosses a physical threshold. The closed-loop map is

\[
\tilde x_{n+1}
=
F_{C(y_n)}(\tilde x_n).
\]

This would demonstrate the missing closure condition in the most direct form. The experiment would show:

\[
H_n
\longrightarrow
E_W[n]
\longrightarrow
y_n
\longrightarrow
o_\pi
\longrightarrow
\text{wave erasure}.
\]

The key point is not that this already gives universal computation. It gives an autonomous physical memory-erasure primitive. That would move the system beyond externally triggered erasure.

\subsection{Minimal demonstration: autonomous branch}

A second minimal demonstration is a wave-controlled branch. Let the abstract states be

\[
A=\{a_L,a_R\},
\]

corresponding to two spatial basins of the droplet:

\[
B_L=\{\mathbf r\in R_L\},
\qquad
B_R=\{\mathbf r\in R_R\}.
\]

Let the readout variable be determined by the sign of a local wave slope:

\[
q_n=\nabla H_n(\mathbf r_n)\cdot \hat{\mathbf e}.
\]

Define


\[
y_{n+1}=
\begin{cases}
L, & q_n<-\theta,\\
R, & q_n>\theta,\\
y_n, & |q_n|\leq\theta.
\end{cases}
\]
The hold case makes the readout hysteretic, preventing rapid switching near threshold.

The readout state selects a confinement or boundary operation:

\[
C(L)=o_L,
\qquad
C(R)=o_R.
\]



The operations route the walker from a common initialization region $B_0\subset\tilde X$ into the corresponding spatial basin:

\[
F_{o_L}(B_0)\subseteq B_L,
\qquad
F_{o_R}(B_0)\subseteq B_R,
\]

This realizes a physical conditional:

\[
\text{if wave readout is } L,\text{ route left;}
\]

\[
\text{if wave readout is } R,\text{ route right.}
\]

Such a device would demonstrate internally selected branching, which is a more computation-like primitive than passive memory or externally triggered reversal.

\subsection{Reservoir-computing interpretation}

The constructive extensions can also be viewed as a physical version of reservoir computing. The wave field is the high-dimensional reservoir:

\[
H_n\in\mathcal H.
\]

The droplet is a localized probe of the reservoir:

\[
H_n\mapsto \nabla H_n(\mathbf r_n).
\]

The readout layer maps the reservoir/probe state into a low-dimensional physical basin:

\[
(H_n,\mathbf r_n,\mathbf v_n)
\mapsto
y_n.
\]

The operation-selection map feeds the readout back into the reservoir dynamics:

\[
y_n
\mapsto
o_n
\mapsto
H_{n+1}.
\]

Thus the architecture is

\[
H_n
\rightarrow
y_n
\rightarrow
o_n
\rightarrow
H_{n+1}.
\]

Standard reservoir computing often uses an external trained readout. In the present framework, the question is when the readout becomes physically internalized. The constructive answer is:

\[
\begin{aligned}
\text{external readout}
&\quad\longrightarrow\quad
\text{physical attractor readout}
\\
&\quad\longrightarrow\quad
\text{closed-loop operation selection}.
\end{aligned}
\]

The design problem is therefore to construct a physical readout $y_n$ that is trainable or tunable, robustly separated into basins, and coupled back to the wave dynamics.

\subsection{Experimental validation}

A constructive extension should be evaluated by the same criteria developed above.

First, identify the physical basins:

\[
B_1,\ldots,B_K\subset \tilde X.
\]

Second, measure basin separation:

\[
\operatorname{dist}(B_i,B_j)>2\varepsilon_X.
\]

Third, estimate the operation-selection confusion matrix:

\[
M_{\ell o}
=
\mathbb P[O_n=o\mid y_n\in Y_\ell].
\]

Closure requires

\[
M_{\ell,C(Y_\ell)}\geq 1-\delta_C.
\]

Fourth, estimate the transition confusion matrix:

\[
Q_{ij}
=
\mathbb P[\Pi(\tilde x_{n+1})=a_j\mid \Pi(\tilde x_n)=a_i].
\]

If the intended transition is

\[
G(a_i)=a_{g(i)},
\]

then robust implementation requires

\[
Q_{i,g(i)}\geq 1-\delta_F.
\]

Fifth, measure multi-step reliability:

\[
\mathbb P[
\Pi(\tilde x_m)=G^m(\Pi(\tilde x_0))
\text{ for }m=1,\ldots,T
]
\]

as a function of time horizon $T$, memory parameter $M_e$, noise level, and basin separation.

Finally, test closure directly by comparing three conditions:

\[
\text{external operation schedule},
\]

\[
\text{observer-side readout with no feedback},
\]

\[
\text{physical readout with feedback}.
\]

Only the third condition satisfies the closure criterion.

\subsection{What these extensions would and would not show}

These constructive extensions would not by themselves prove Turing universality. That would require a robust finite alphabet, a programmable transition table, memory addressing or equivalent tape-like structure, and a demonstration that arbitrary computations can be embedded.

The more immediate goal is different. The goal is to show how a wave-memory system can be upgraded from

\[
\text{memory + externally triggered operation}
\]

to

\[
\text{memory + physical readout + internally selected operation}.
\]

This is the minimal transition from wave-memory dynamics to closed physical computation.

A successful demonstration of an autonomous eraser, autonomous branch, or small finite-state wave machine would already establish the central claim:

\[
\text{wave memory alone is insufficient,}
\]

but

\[
\text{wave memory plus intrinsic readout-control feedback}
\]

can realize closed physical computation.

\subsection{Design principle}

The constructive design principle can be stated compactly.

Let

\[
H_n
\]

be the wave reservoir, let

\[
z_n=(\mathbf r_n,\mathbf v_n,\sigma_n)
\]

be the probe state, let

\[
y_n
\]

be a physical readout basin, and let

\[
o_n=C(y_n)
\]

be the selected physical operation. Then a closed wave-based physical computer has the form

\[
y_{n+1}
=
\mathcal R(y_n,H_n,z_n),
\]

\[
o_n=C(y_n),
\]

\[
(H_{n+1},z_{n+1})
=
F_{o_n}(H_n,z_n).
\]

Together,

\[
(H_n,z_n,y_n)
\mapsto
(H_{n+1},z_{n+1},y_{n+1})
\]

defines a closed physical evolution. Under a robust abstraction

\[
\Pi:\tilde X\to A,
\]

it implements an abstract transition if

\[
\Pi(\tilde x_{n+1})
=
G(\Pi(\tilde x_n))
\]

or, for controlled transitions,

\[
\Pi(F_{C(y_n)}(\tilde x_n))
=
G_{\Theta(C(y_n))}(\Pi(\tilde x_n)).
\]

This is the constructive version of the closure criterion. It turns the critique into a design program: build physical reservoirs whose readout states are not external measurements but intrinsic dynamical variables that select subsequent physical operations.


\end{document}